\documentclass[journal]{IEEEtran}
\usepackage{amsmath,graphicx,cite,epsfig}
\usepackage{multirow}
\usepackage{booktabs}
\usepackage{float}
\usepackage{subfigure}
\usepackage{caption}
\usepackage{threeparttable}
\usepackage{diagbox}
\usepackage{url}
\usepackage{bm}
\usepackage{algorithm}
\usepackage{algorithmic}

\ifCLASSINFOpdf

\else

\fi

\begin{document}
\title{Blind Omnidirectional Image Quality \\
	Assessment with Viewport Oriented Graph Convolutional Networks}

\author{Jiahua Xu, Wei Zhou,~\IEEEmembership{Student~Member,~IEEE}, and Zhibo Chen,~\IEEEmembership{Senior~Member,~IEEE}% <-this % stops a space
\thanks{This work was supported in part by NSFC under Grant U1908209, 61632001 and the National Key Research and Development Program of China 2018AAA0101400 and Alibaba Corporate. (Jiahua Xu and Wei Zhou contributed equally to this work.) (Corresponding anthor: Zhibo Chen.)}
\thanks{J. Xu, W. Zhou and Z. Chen are with the CAS Key Laboratory of Technology in Geo-Spatial Information Processing and Application System, University of Science and Technology of China, Hefei 230027, China (e-mail: xujiahua@mail.ustc.edu.cn; weichou@mail.ustc.edu.cn; chenzhibo@ustc.edu.cn).}
}

\markboth{IEEE Transactions on Circuits and Systems for Video Technology}%
{Shell \MakeLowercase{\textit{et al.}}: Bare Demo of IEEEtran.cls for IEEE Journals}



\maketitle

\begin{abstract}
Quality assessment of omnidirectional images has become increasingly urgent due to the rapid growth of virtual reality applications. Different from traditional 2D images and videos, omnidirectional contents can provide consumers with freely changeable viewports and a larger field of view covering the $360^{\circ}\times180^{\circ}$ spherical surface, which makes the objective quality assessment of omnidirectional images more challenging. In this paper, motivated by the characteristics of the human vision system (HVS) and the viewing process of omnidirectional contents, we propose a novel Viewport oriented Graph Convolution Network (VGCN) for blind omnidirectional image quality assessment (IQA). Generally, observers tend to give the subjective rating of a 360-degree image after passing and aggregating different viewports information when browsing the spherical scenery. Therefore, in order to model the mutual dependency of viewports in the omnidirectional image, we build a spatial viewport graph. Specifically, the graph nodes are first defined with selected viewports with higher probabilities to be seen, which is inspired by the HVS that human beings are more sensitive to structural information. Then, these nodes are connected by spatial relations to capture interactions among them. Finally, reasoning on the proposed graph is performed via graph convolutional networks. Moreover, we simultaneously obtain global quality using the entire omnidirectional image without viewport sampling to boost the performance according to the viewing experience. Experimental results demonstrate that our proposed model outperforms state-of-the-art full-reference and no-reference IQA metrics on two public omnidirectional IQA databases.
\end{abstract}


% \fi
\IEEEpeerreviewmaketitle

\section{Introduction}
\IEEEPARstart{V}{irtual} reality (VR) is an immersive technology that can offer producers and consumers a new way to generate, use and interact with visual information \cite{background1}. It takes traditional media beyond conventional screen and provides a 360-degree view with the help of head mounted display (HMD). Thus, VR as a new technology, is becoming increasingly popular. According to \cite{background2}, HMD with 8K monocular resolution will be used in industrial applications after 2025. The ultra-high resolution and sphere representation of omnidirectional contents will bring difficulties to current image/video processing systems, e.g., acquisition, compression, transmission, restoration, etc. \cite{importance}. Therefore, it is crucial to research on the quality assessment of omnidirectional contents to guide the optimization of existing systems and algorithms \cite{overview360}.

When it comes to image quality assessment (IQA), it can be roughly classified into two categories, namely subjective IQA and objective IQA \cite{subjective&objective}. In the field of 360-degree contents, subjective IQA means the omnidirectional images are viewed by human beings in the HMD \cite{OIQADatabse,SOLIDDatabase}, and pair comparison methods cannot be adopted since observers can browse one omnidirectional content every time \cite{SAMPVIQ}. It can provide the most accurate results, but it is cumbersome, expensive and impractical in real-time applications \cite{exploration}. Thus, it is necessary to develop objective IQA metrics which can automatically predict the perceptual quality of given images \cite{IQA&VQAsurvey}. To this end, many researchers have proposed useful metrics for IQA to meet the requirements of both academia and industry in the past decades \cite{StatisticalEvaluation}. 

Among traditional objective IQA metrics, the availability of distortion-free images leads to three approaches to predict the quality of images with certain degradation, namely full-reference (FR), reduced-reference (RR) and no-reference (NR) IQA metrics \cite{WaDIQaM}. Note that NR IQA is also known as blind IQA (BIQA). FR and RR IQA metrics need full or part of the original images and their distorted versions while NR IQA (BIQA) metrics only require distorted images \cite{RR&NR}. The most famous FR metrics, e.g., mean square error (MSE) and peak signal-to-noise ratio (PSNR), are commonly used when evaluating the performance of coding and restoration technologies owing to their simplicity and mathematical convenience \cite{IFC}. But the low correlation with human judgements promotes researchers to develop new metrics based on the human vision system (HVS) \cite{HVS}, e.g. SSIM \cite{ssim} and its variants \cite{msssim,iwssim,cwssim,fsim}. 

\begin{figure}[t]
	\centerline{\includegraphics[width=8.8cm]{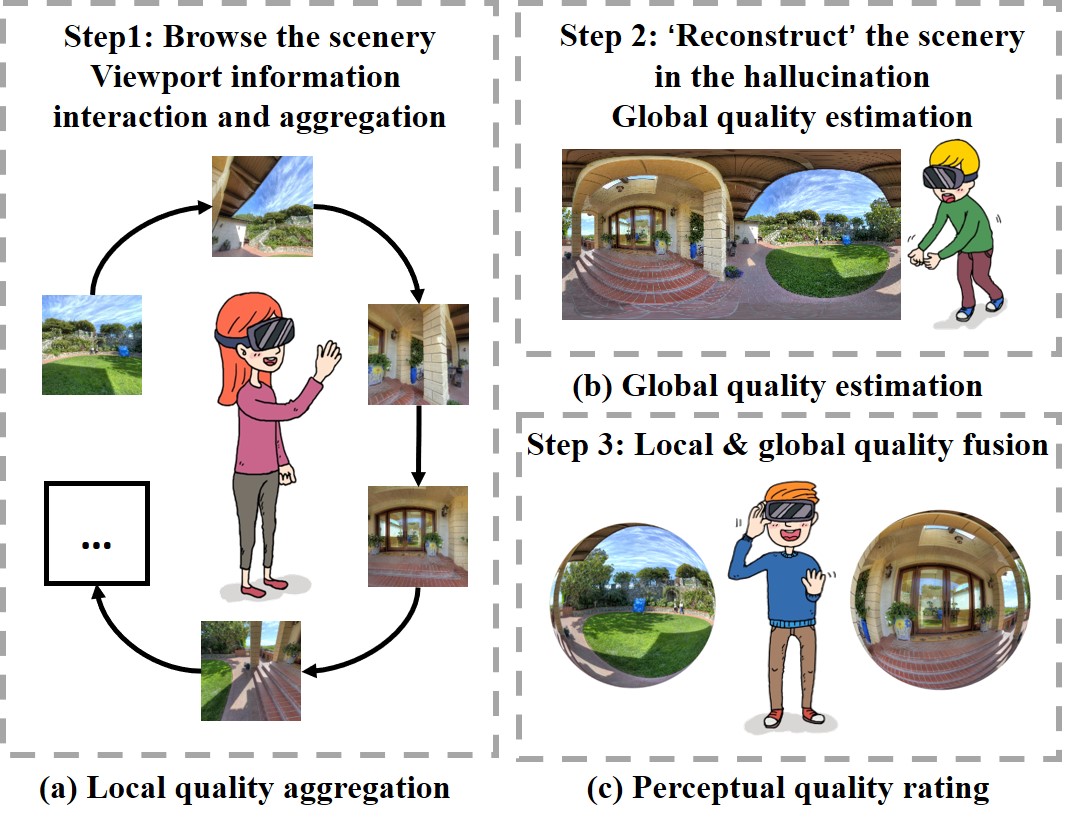}}
	\caption{The viewing process of omnidirectional contents. (a) Firstly, subjects browse the spherical scenery, and the visual information of different viewports is interacted and aggregated as local quality aggregation. (b) Then, they will `reconstruct' the whole scenery in the hallucination according to what they see, and have a general impression on the quality level as global quality estimation. (c) Finally, the perceptual quality is obtained considering both local quality aggregation and global quality estimation.}
	\centering
	\label{fig:fig1}
\end{figure}

Reference images are difficult to acquire in most real-world situations \cite{DIIVINE}. Thus, enormous efforts have been put into the NR IQA to predict the quality without clear pristine images \cite{WPDSE}. Firstly, BIQA mainly focuses on specific distortion types, including blurriness \cite{NRSharpness,NRBlur}, blockiness \cite{NRJPEG1,NRJPEG2}, ringing artifacts\cite{ringing}, etc. Although these metrics can achieve high performance for specific distortion, their generalization capability is limited. Then, the general-purpose NR IQA has been proposed to handle different distortion types and levels \cite{NIQE,DIIVINE,BRISQUE,BLINDS,wu2015blind,CORNIA,HOSA,wu2017blind}. In recent years, with the fast proliferation of neural network, many convolutional neural network (CNN)-based models have been proposed to predict the perceptual quality in an end-to-end manner \cite{NIMA,DIQA,CNNIQA,WaDIQaM,MEON,DBCNN}. 

However, the study for omnidirectional image quality assessment (OIQA) is not as mature as traditional IQA since VR has only become popular recently \cite{upenik2017performance}. The 360-degree contents cannot be transmitted as a sphere, thus, projection is necessary to convert the spherical source into 2D format. Therefore, it would introduce the geometry deformation. To tackle this problem, most existing works have focused on extending traditional FR IQA metrics to FR OIQA \cite{SPSNR,WSPSNR,CPPPSNR,CPPCNR,WSSSIM,SSSIM,ssim360}. Yu \textit{et al.} \cite{SPSNR} developed spherical PSNR (S-PSNR) and selected uniformly distributed points on the sphere to overcome the pixel redundancy issue in the projected omnidirectional image. Sun \textit{et al.} \cite{WSPSNR} proposed weighted-to-spherically-uniform PSNR (WS-PSNR) by multiplying the error map with a weight map to reduce the influence of stretched areas. Zakharchenko \textit{et al.} \cite{CPPPSNR} put forward craster parabolic projection PSNR (CPP-PSNR) and calculated PSNR on the craster parabolic projection plane. It guaranteed the uniform sampling density but would lower precision due to the interpolation. Xu \textit{et al.} \cite{CPPCNR} proposed non-content-based PSNR (NCP-PSNR) by weighting the pixel with location and content-based PSNR (CP-PSNR) by predicting the viewing direction. Considering the HVS, metrics based on SSIM have been proposed successively. Zhou \textit{et al.} \cite{WSSSIM} proposed weighted-to-spherically-uniform SSIM (WS-SSIM) which were similar to WS-PSNR. The location weight map was adopted to convert SSIM into WS-SSIM. To avoid the geometric distortion of projection, Chen \textit{et al.} \cite{SSSIM} designed spherical SSIM (S-SSIM) and computed similarity between reference and distorted 360-degree images on the sphere. The location weight map was also used in S-SSIM. SSIM360 \cite{ssim360} was proposed by Facebook to solve the warping problems of omnidirectional images. They put a weight on each re-sample SSIM and the weight is determined by how much the sampled area is stretched in the projection representation.

Compared with conventional 2D NR IQA, NR OIQA involves new challenges such as stitching artifacts, sphere representation and wide field of view (FoV). To measure the stitching artifacts, Ling \textit{et al.} \cite{ling2018no} applied convolutional sparse coding and compound feature selection to NR quality assessment of stitched panoramic images. Considering the sphere representation of 360-degree contents, Kim \textit{et al.} \cite{DeepVRIQA1,DeepVRIQA2} proposed a VR image quality assessment deep learning framework (DeepVR-IQA) with adversarial learning. It predicted the quality scores of sampled patches and weighted them with their position on the sphere. Based on the image patch again, Li \textit{et al.} \cite{VQA-ODV} predicted the head movement (HM) and eye movement (EM) maps, and adopted them to weight the quality score. However, patches sampled from the equirectangular projection (ERP) format of omnidirectional images contains heavy geometry deformation and cannot reflect the characteristics of actual viewing contents \cite{xu2019quality}. As a result, the concept of viewport is introduced which represents the visual information inside the viewing window. Li \textit{et al.} \cite{VCNN} proposed a viewport-based CNN (V-CNN) approach, it predicted the quality score of viewport rather than image patch and had auxiliary tasks for HM and EM map prediction. Sun \textit{et al.} \cite{mc360iqa1,mc360iqa2} developed a multi-channel convolution neural network for blind 360-degree image quality assessment (MC360IQA). It included six parallel hyper-ResNet-34 networks to process viewport images and an image quality regressor to fuse features for obtaining the final quality score.

However, the interactions among different viewports are both ignored in V-CNN and MC360IQA approaches. During the viewing process of omnidirectional contents, subjects first browse the sphere scenery. Meanwhile, the visual information of different viewports is interacted and aggregated as local quality aggregation. When finishing the browsing step, subjects will `reconstruct' the whole scenery in the hallucination according to what they see (pseudo reconstruction step), and have a general impression on the quality level which we named as global quality estimation. The entire process is illustrated in Fig. \ref{fig:fig1}, where the local quality aggregation and global quality estimation are both needed for the final decision of quality rating. 

\begin{figure*}[t]
	\centerline{\includegraphics[width=18cm]{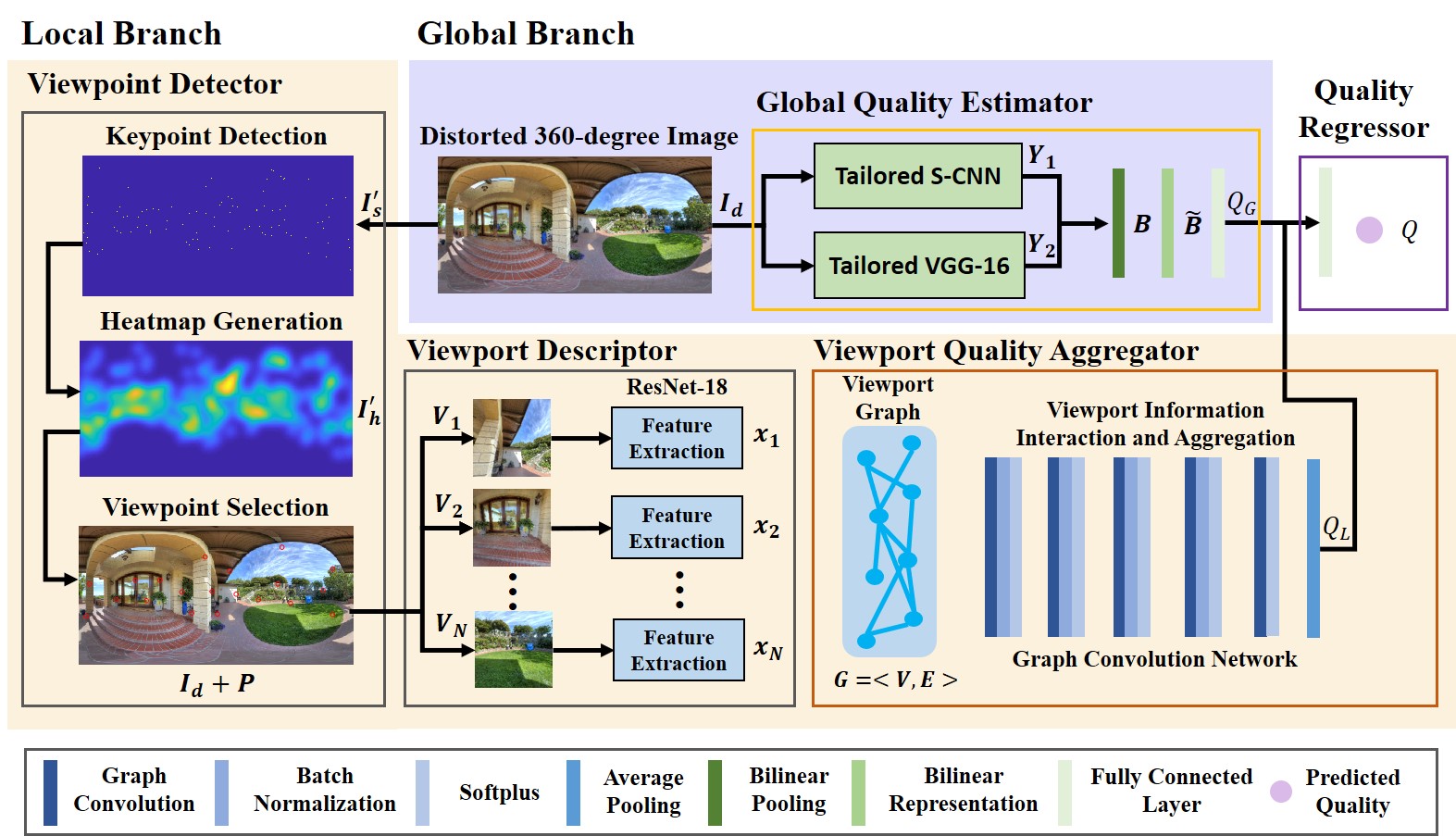}}
	\caption{The pipeline of our proposed VGCN for NR OIQA. The distorted omnidirectional image in ERP format is input into the network and processed through local and global branches. After that, the output of viewport quality aggregator and global quality estimator are regressed onto a scalar value as the perceptual quality prediction.}
	\centering
	\label{fig:fig2}
\end{figure*}

Based on these observations, we propose a Viewport oriented Graph Convolutional Network (VGCN) for blind omnidirectional image quality assessment. Firstly, we develop a viewpoint detector to select viewports with higher probabilities to be seen, the detector is designed according to the HVS which is sensitive to structural information \cite{ssim}. Secondly, we utilize the viewport descriptor to extract discriminative features for quality prediction. Thirdly, we build a spatial viewport graph to model the mutual dependency of viewports in the omnidirectional image. The graph nodes are defined with selected viewports and connected by spatial relations to capture interactions between each other. Finally, reasoning on the proposed graph is performed via graph convolutional networks. For simplicity, we omit the pseudo reconstruction step and directly use the entire omnidirectional image for obtaining global quality to boost the performance. Experimental results demonstrate that our proposed model outperforms state-of-the-art full-reference and no-reference IQA metrics on two public omnidirectional IQA databases. Besides, the generalization ability of the proposed VGCN is verified through cross database validation.

To sum up, our contributions are listed as follows:

\begin{itemize}
\item We develop a viewpoint detector motivated by the importance of structural information in the HVS \cite{ssim}. The viewports selected by the viewpoint detector have a great chance to be noticed by observers. Besides, we pre-train the ResNet-18 on the traditional 2D IQA database and adopt it as viewport descriptor to extract discriminative features for quality prediction. 
\item Inspired by the viewing process of omnidirectional contents, we build a spatial viewport graph and utilize graph convolutional networks to capture interactions between different viewports. The results of ablation study validate the effectiveness of the GCN architecture and the source code is available online for public research usage $\footnote{\url{http://staff.ustc.edu.cn/~chenzhibo/resources.html}}$.
\item We add a global branch for mimicking the global quality estimation step in the viewing experience to boost the performance. Compared with state-of-the-art IQA and OIQA metrics, the proposed VGCN predicts the perceptual quality more accurately under various distortion types and levels.
\end{itemize}

The rest of this paper is organized as follows. Section II introduces the proposed VGCN for blind OIQA in details. We present the experimental results and analysis in Section III and conclude the paper with an outlook on the future work in Section IV.

\section{Proposed Viewport Oriented Graph Convolutional Network}

Motivated by the viewing process of omnidirectional contents, we construct a framework with two branches, namely local branch and global branch corresponding to Fig. \ref{fig:fig1}. In the local branch, the viewpoint detector and viewport descriptor are first introduced to select specific viewports and extract features. Afterwards, the graph convolution network is adopted to aggregate information of different viewports, and acquire local quality. In the global branch, the entire omnidirectional image is utilized to obtain global quality. Then, the qualities inferred by these two branches are regressed onto the final perceptual quality prediction. Our proposed model does not require reference image to measure similarity, thus only the distorted image (in ERP format) is needed for obtaining the objective OIQA score.

\begin{algorithm}[tb] 
	\caption{ Viewpoint selection} 
	\label{Vselection} 
	\begin{algorithmic}[1] 
		\REQUIRE ~~\\ 
		The heatmap removing padding areas $\bm{I'_h}$\\
		The angular distance threshold $d_{th}$\\
		The number of viewpoints to be selected $N$
		\ENSURE ~~\\ 
		The set of selected viewpoints $\bm{P}$
		\STATE $k \gets 1$, $\bm{P} \gets \emptyset$
		\WHILE {$k \le N$}
		\STATE $(x,y) \gets argmax_{\{(x,y)|\bm{I'_h}(x,y)\in\bm{I'_h} \}}\bm{I'_h}$
		%\STATE Find the pixel $(x,y)$ with the largest value in $\bm{I'_h}$;
		\STATE $\bm{I'_h}(x,y) \gets 0$;
		\STATE $(x,y)_{2D} \to (\phi,\theta)_{Sphere}$
		%\STATE Convert the $(x,y)$ in the rectangular coordinate into $(\phi,\theta)$ in the spherical coordinate;
		\IF {$\bm{P}=\emptyset$}
		\STATE $\bm{P} \gets \bm{P} \cup (\phi,\theta)$
		\STATE $k \gets k+1$
		\ELSE
		\STATE $d_s \gets min(AngularDist[(\phi,\theta),\bm{P}])$
		%\STATE Compute the angular distance of $(\phi,\theta)$ and points in set $\bm{P}$;
		%\STATE Find the smallest angular distance $d_s$;
		\IF {$d_s>d_{th}$}
		\STATE $\bm{P} \gets \bm{P} \cup (\phi,\theta)$
		\STATE $k \gets k+1$
		\ENDIF
		\ENDIF 
		\ENDWHILE
		\RETURN $\bm{P}$
	\end{algorithmic}
\end{algorithm}

In this section, we will first describe the framework of our proposed method. Then, the viewpoint detector, viewport descriptor, viewport quality aggregator in local branch and the global quality estimator in global branch are introduced in details. Finally, we present the training and testing protocols adopted in VGCN.

\subsection{Overview}

We illustrate the pipeline of VGCN in Fig. \ref{fig:fig2}, it is composed of the viewpoint detector, viewport descriptor, viewport quality aggregator, global quality estimator, and quality regressor inspired by the viewing process of omnidirectional images. In the local branch, we first design a viewpoint detector to sample viewports that are appealing to observers. Based on the HVS that humans are more sensitive to structural information, speeded up robust features (SURF) local feature detector \cite{surf} and 2D Gaussian filter are adopted to generate the heatmap, the viewpoints are selected according to specific roles. Then, we leverage the pre-trained ResNet-18 architecture as the viewport descriptor to extract effective features representing the visual information inside the viewport. After that, the viewport quality aggregator is utilized to model the mutual dependency of different viewports in a single omnidirectional image.

In the global branch, the entire distorted image is used without viewport sampling. Since the distorted 360-degree image covering the spherical scenery is easily accessible in our framework, the pseudo reconstruction step in Fig. \ref{fig:fig1} is omitted for simplicity. We apply a deep bilinear CNN (DB-CNN) proposed in \cite{DBCNN} as the global quality estimator for measuring synthetic and authentic distortions. Finally, the qualities predicted by local and global branches are fused in the quality regressor to acquire perceptual omnidirectional image quality.

\subsection{Local Branch}
We develop the local branch to simulate the viewport information interaction and aggregation process when subjects browsing the spherical scenery. It includes the viewpoint detector, viewport descriptor, and viewport quality aggregator. We will introduce them in details as follows:
\subsubsection{\textbf{Viewpoint Detector}}
It aims to select viewports attractive to observers. Since structural information is usually involved in salient objects and appealing to observers \cite{ssim}, the viewpoint detector adopted in our framework is based on the SURF local feature detection \cite{surf} to select keypoints. At first, we automatically downsample the omnidirectional image \cite{ssim} to avoid the influence of tiny textures. Then, padding in the left and right sides is applied to keep the consistency of omnidirectional content during keypoint detection. The distorted image is denoted as $\bm{I_d}$ and SURF operation is denoted as $Surf(\cdot)$. The selected keypoints are given as follows:
\begin{equation}\label{1}
\bm{S} = Surf(Padding(\bm{I_d})),
\end{equation}
where $\bm{S}$ represents the set of keypoints and $Padding(\cdot)$ is the operation before SURF detection. We annotate each keypoint in the empty omnidirectional image and obtain the keypoint map $\bm{I_s}$. Afterwards, $\bm{I_s}$ is convoluted with a 2D Gaussian Filter $\bm{G}$ \cite{dhp} to acquire the heatmap $\bm{I_h}$:
\begin{equation}\label{2}
\bm{I_h} = \bm{I_s} \otimes \bm{G},
\end{equation}
where $\otimes$ represents the convolution operation. Note that $\bm{I'_s}$ and $\bm{I'_h}$ exhibited in Fig. \ref{fig:fig2} remove the padding areas for better visualization. Also, the padding areas are removed before viewpoint selection to avoid repetitive points. In our pipeline, $N$ viewpoints are selected according to Algorithm \ref{Vselection}. $N$ is set as 20 according to \cite{VCNN} and the angular distance threshold $d_{th}$ is set as $30^{\circ}$ since the selected viewport should spread over the sphere instead of focusing on a particular region. Then, given the distorted omnidirectional image $\bm{I_d}$ and selected viewpoints $\bm{P}$ as central points, $N$ viewports $\{\bm{V_i}\}_{i=1}^N$ are sampled and fed into the viewport descriptor.

\subsubsection{\textbf{Viewport Descriptor}}
To reduce the computational complexity and memory cost, the omnidirectional image is first downsampled to the resolution $512\times1024$ as done in \cite{VCNN,mc360iqa2}. Then, we obtain $N$ viewports covering the $90^{\circ}$ FoV in the size of $256\times256$. The ResNet-18 architecture is adopted as viewport descriptor because it is proved useful in various computer vision tasks \cite{resnet} e.g., classification, recognition, segmentation, etc.

The detailed structure of viewport descriptor is listed in Table \ref{table1}. $conv1$, $conv2\_x$, $conv3\_x$, $conv4\_x$ and $conv5\_x$ in Table \ref{table1} are the same layers in ResNet-18 \cite{resnet}. Since the model takes input with the size $256\times256$, the output of $conv5\_x$ is a $8\times8$ feature map. Thus, we add a max-pooling layer, and the viewport descriptor gives a 512-dimensional vector for feature representation.

\begin{table}[ht]
	\begin{center}
		\captionsetup{justification=centering}
		\caption{\textsc{Detailed Configurations of the Viewport Descriptor.}}
		\label{table1}
		\begin{threeparttable}
			\begin{tabular}{@{}c|cc@{}}
				\toprule
				Layer name                & Output size         & Layer                  \\ \midrule
				conv1                     & $128\times128$      & $7\times7,\,64,\,2$ \\ \midrule
				\multirow{2}{*}{conv2\_x} & \multirow{2}{*}{$64\times64$} & $3\times3$ max-pool, stride 2 \\
				& & $\left[ \begin{array}{l} 3 \times 3,\,64,\,1\\3 \times 3,\,64,\,1\end{array} \right]\times 2$  \\ \midrule
				conv3\_x  				  & $32\times32$  
				& $\left[ \begin{array}{l} 3 \times 3,\,128,\,1\\3 \times 3,\,128,\,1\end{array} \right]\times 2$  \\ \midrule
				conv4\_x                  & $16\times16$                  
				& $\left[ \begin{array}{l} 3 \times 3,\,256,\,1\\3 \times 3,\,256,\,1\end{array} \right]\times 2$  \\ \midrule
				conv5\_x                  & $8\times8$                  
				& $\left[ \begin{array}{l} 3 \times 3,\,512,\,1\\3 \times 3,\,512,\,1\end{array} \right]\times 2$  \\ \midrule
				& $1\times1$              & $8\times8$ max-pool           \\ \bottomrule
			\end{tabular}
			\begin{tablenotes}
				\footnotesize
				\item[1] Convolutional layer: kernel size, channel, stride
			\end{tablenotes}
		\end{threeparttable}
	\end{center}
\end{table}

\subsubsection{\textbf{Viewport Quality Aggregator}}

\begin{figure}[htbp]
	\centering
	\subfigure[connected]{
		\includegraphics[width=4cm]{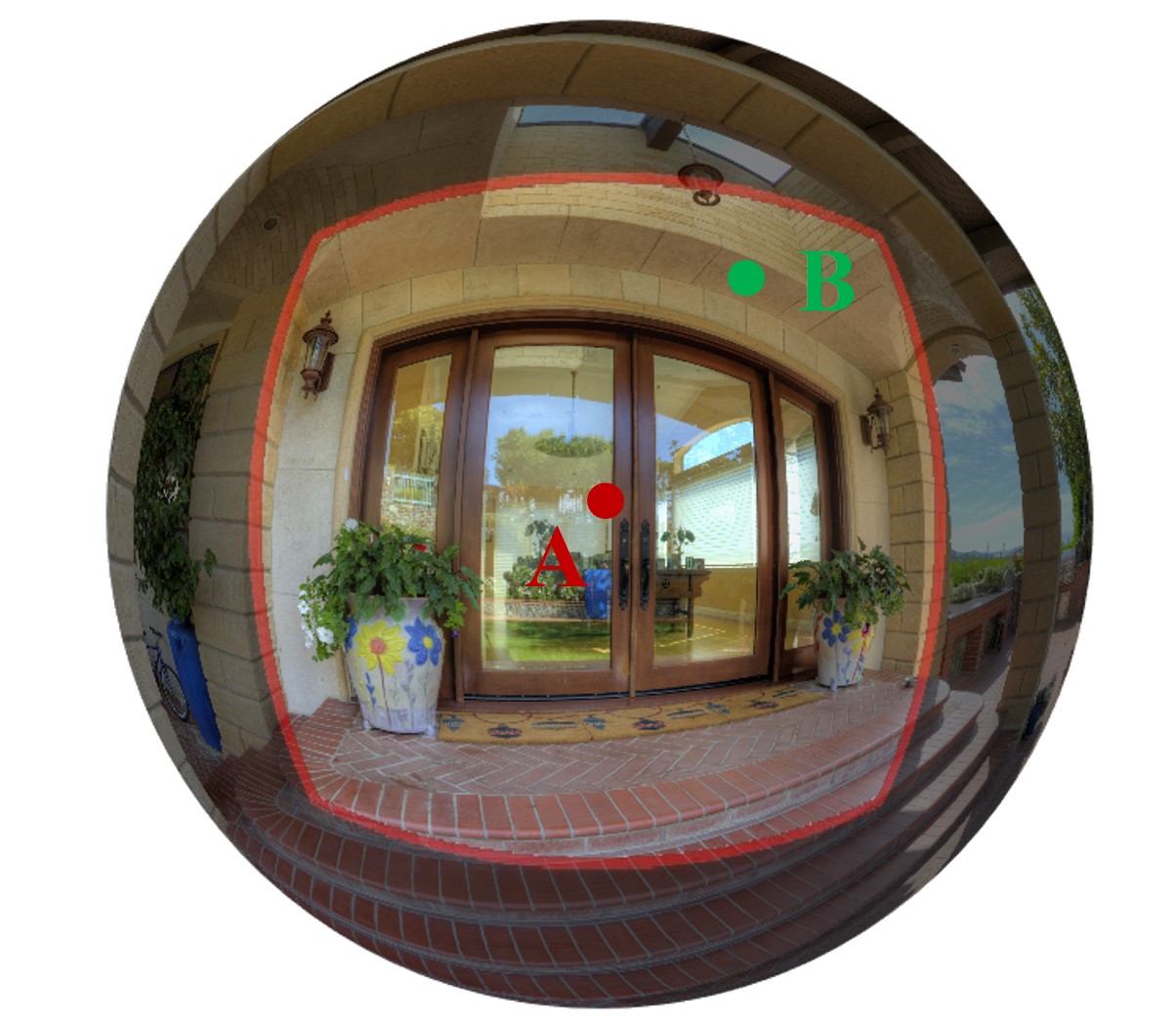}
	}
	\subfigure[separated]{
		\includegraphics[width=4cm]{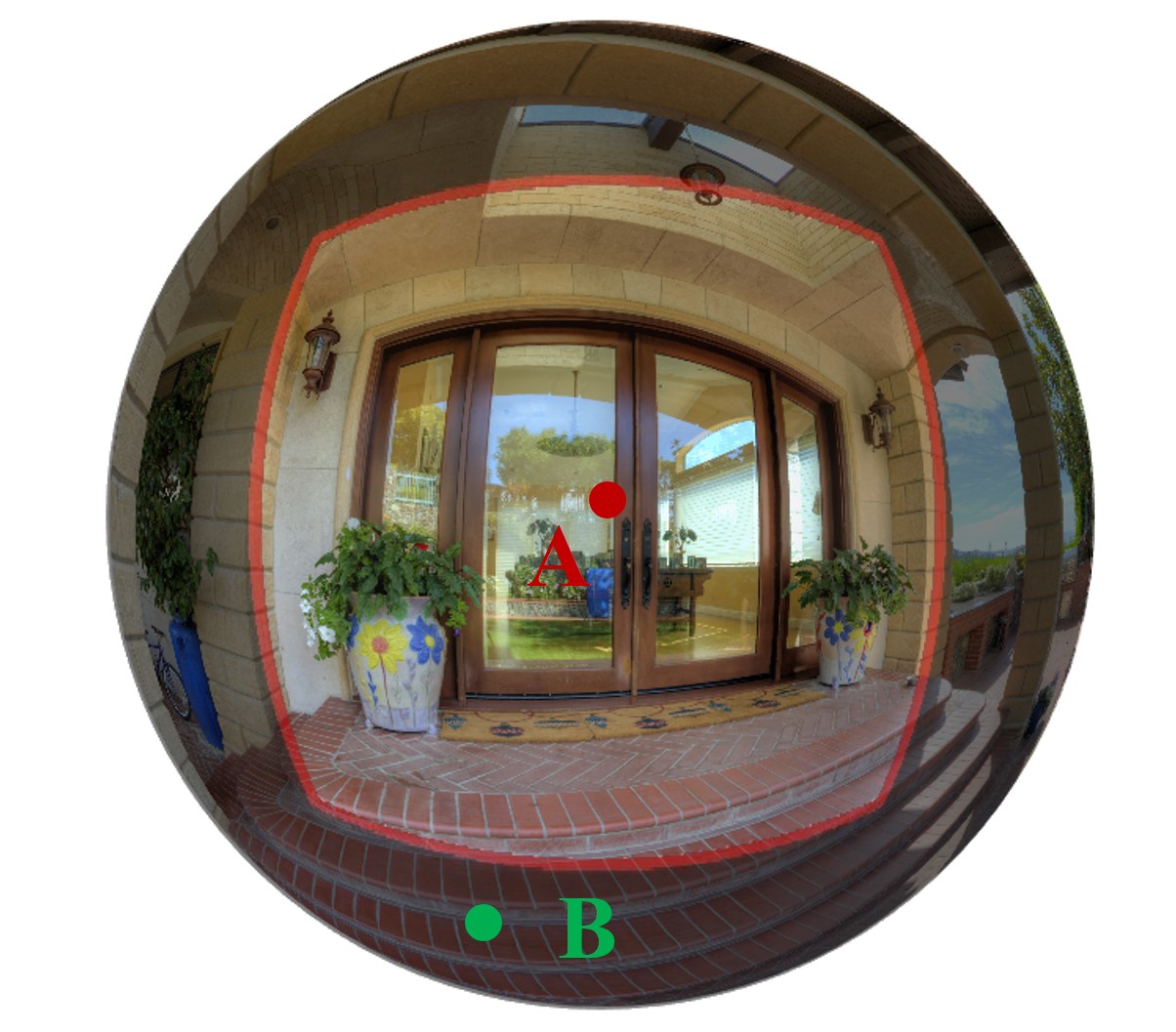}
	}
	\caption{Visual examples of spatial relations. (a) Viewport A and viewport B are connected. (b) Viewport A and viewport B are separated.}
	\label{fig:fig3}
\end{figure}

To model the mutual dependency of different viewports, we build a spatial viewport graph. The graph nodes are defined with $N$ selected viewports and we connect pairs of viewports with different edges. To be specific, as shown in Fig. \ref{fig:fig3}, if the central point of viewport B is within the FoV of viewport A, then viewport A and B are connected, otherwise, they are separated.

\begin{table}[ht]
	\begin{center}
		\captionsetup{justification=centering}
		\caption{\textsc{Detailed Configurations of Tailored S-CNN and Tailored VGG-16.}}
		\label{table2}
		\begin{threeparttable}
			\scalebox{0.88}{
				\begin{tabular}{@{}cc|cc@{}}
					\toprule
					Network                          & Layer name                     & Output size                     & Layer                         \\ \midrule
					\multirow{9}{*}{Tailored S-CNN}  & \multirow{2}{*}{s\_conv1\_x}   & \multirow{2}{*}{$256\times512$} & $3\times3, 48,\,1$                \\
					&                                &                                 & $3\times3, 48,\,2$      \\ \cmidrule(l){2-4} 
					& \multirow{4}{*}{s\_conv2\_x}   & \multirow{4}{*}{$64\times128$}  & $3\times3,\,64,\,1$ \\
					&                                &                                 & $3\times3,\,64,\,2$ \\
					&                                &                                 & $3\times3,\,64,\,1$ \\ 
					&                                &                                 & $3\times3,\,64,\,2$  \\ \cmidrule(l){2-4} 
					& \multirow{3}{*}{s\_conv3\_x}   & \multirow{3}{*}{$32\times64$}   & $3\times3,\,128,\,1$ \\
					&                                &                                 & $3\times3,\,128,\,1$ \\
					&                                &                                 & $3\times3,\,128,\,2$ \\ \midrule
					\multirow{17}{*}{Tailored VGG-16} & \multirow{3}{*}{vgg\_conv1\_x} & \multirow{3}{*}{$256\times512$} & $3\times3,\,64,\,1$   \\
					&                                &                                 &$3\times3,\,64,\,1$  \\
					&                                &                                 & $2\times2$ max-pool, stride 2 \\ \cmidrule(l){2-4} 
					& \multirow{3}{*}{vgg\_conv2\_x} & \multirow{3}{*}{$128\times256$} & $3\times3,\,128,\,1$ \\
					&                                &                                 & $3\times3,\,128,\,1$ \\
					&                                &                                 & $2\times2$ max-pool, stride 2 \\ \cmidrule(l){2-4} 
					& \multirow{4}{*}{vgg\_conv3\_x} & \multirow{4}{*}{$64\times128$}  & $3\times3,\,256,\,1$\\
					&                                &                                 & $3\times3,\,256,\,1$\\
					&                                &                                 & $3\times3,\,256,\,1$\\
					&                                &                                 & $2\times2$ max-pool, stride 2 \\ \cmidrule(l){2-4} 
					& \multirow{4}{*}{vgg\_conv4\_x} & \multirow{4}{*}{$32\times64$}   & $3\times3,\,512,\,1$\\
					&                                &                                 & $3\times3,\,512,\,1$\\
					&                                &                                 & $3\times3,\,512,\,1$\\
					&                                &                                 & $2\times2$ max-pool, stride 2 \\ \cmidrule(l){2-4}
					& \multirow{3}{*}{vgg\_conv5\_x} & \multirow{3}{*}{$32\times64$}   & $3\times3,\,512,\,1$\\
					&                                &                                 & $3\times3,\,512,\,1$\\
					&                                &                                 & $3\times3,\,512,\,1$\\ \bottomrule
			\end{tabular}}
			\begin{tablenotes}
				\footnotesize
				\item[1] Convolutional layer: kernel size, channel, stride.
				\item[2] ReLU layers after convolutional layers are ignored.
			\end{tablenotes}
		\end{threeparttable}
	\end{center}
\end{table}

Formally, we denote the feature representation of $N$ viewports as $\bm{X}=\{\bm{x_1},\bm{x_2},\cdots,\bm{x_N}\}$ and each viewport feature $\bm{x_i}$ is a $512$-dimensional vector corresponding to the viewport descriptor. Then, the affinity between every two viewports can be represented as:
\begin{equation}\label{3}
\bm{A}(\bm{x_i},\bm{x_j}) = \left\{ \begin{array}{l}
1,\,if\,AngularDist[({\phi _i},{\theta _i}),({\phi _j},{\theta _j})] \le {45^ \circ }\\
0,\,otherwise
\end{array} \right.,
\end{equation}
where $\bm{A}$ is the affinity matrix and $AugularDist(\cdot)$ computes the angular distance between two viewpoints on the sphere. $({\phi _i},{\theta _i})$ and $({\phi _j},{\theta _j})$ denote the longitudes and latitudes of viewpoint $i$ and $j$. Since the viewport size adopted in our framework is ${90^\circ}$, we set the angular distance threshold in Eq. \ref{3} as half of the viewport size equaling ${45^\circ}$. Afterwards, normalization \cite{kipf2016semi,yan2018spatial} is performed as follows: 
\begin{equation}\label{4}
\bm{\hat{A}} = {\bm{D}^{ - \frac{1}{2}}}\bm{A}{\bm{D}^{ - \frac{1}{2}}},
\end{equation}
where $\bm{\hat{A}}$ is the adjacency matrix of the undirected spatial viewport graph after normalization. $\bm{D}$ represents the diagonal matrix and $\bm{D}_{ii}=\sum_j\bm{A}_{ij}$. Then, the graph convolution network is implemented with the following layer-wise propagation rule:
\begin{equation}\label{5}
{\bm{H^{(l + 1)}}} = \sigma (B{N_{\gamma ,\beta }}(\bm{{\hat A}{H^{(l)}}{W^{(l)}}})),
\end{equation}
where $\sigma(\cdot)$ represents the Softplus activation function \cite{glorot2011deep} and $\sigma(x)=log(1+e^x)$. $BN_{\gamma ,\beta }(\cdot)$ is the batch normalizing transform and $\gamma ,\beta$ are the parameters to be learned. $\bm{W^{(l)}}$ denotes the layer-specific trainable weight matrix and $\bm{H^{(l)}}$ denotes the matrix after activations in the $l^{th}$ layer, $\bm{H^{(0)}}=\bm{X}$. Based on the graph convolutions, we can pass the message of different viewports inside the graph \cite{wang2018videos} and update features of each viewport node. Five graph convolution layers are adopted and the viewport feature dimensions after each graph convolution layer are $[256, 128, 64, 32, 1]$. Finally, an average pooling layer is applied to aggregate the viewport features and obtain the local quality aggregation $Q_L$.

\subsection{Global Branch and Quality Regressor}
The global branch corresponds to the global quality estimation step in Fig. \ref{fig:fig1}. When observers browse the 360-degree content, it is difficult for them to view the entire scenery at one time, thus they need to `reconstruct' the spherical scenery in the hallucination based on the viewed viewports, and acquire the global quality. Strictly speaking, we need to reconstruct the omnidirectional image according to the viewports sampled in the local branch in our framework. However, since the distorted omnidirectional image covering the whole scene is easily accessible, we remove the pseudo reconstruction step for simplicity and directly takes the omnidirectional image in ERP format as input.

\subsubsection{\textbf{Global Quality Estimator}}
The DB-CNN proposed in \cite{DBCNN} is utilized as the global quality estimator to measure degradations of image quality. It is composed of two streams, namely tailored S-CNN for measuring synthetic distortions and tailored VGG-16 \cite{vgg} for measuring authentic distortions. According to \cite{DBCNN}, synthetic distortions are simulated via computers while authentic distortions are introduced during acquisition. Before being fed into DB-CNN, the omnidirectional image is resized to $512\times1024$ corresponding to the viewport size in local branch. The detailed structures of tailored S-CNN and tailored VGG-16 are depicted in Table \ref{table2}. 

As we can see from Table \ref{table2}, the output of tailored S-CNN $\bm{Y_1}$ ($h\times w\times d_1$) and tailored VGG-16 $\bm{Y_2}$ ($h\times w\times d_2$) have the same shape but different channels. Thus, bilinear pooling \cite{DBCNN} is adopted to combine features as follows:
\begin{equation}\label{6}
\bm{B} = \bm{Y_1}^T\bm{Y_2},
\end{equation}
where $\bm{B}$ denotes the combined feature with size $d_1\times d_2$. Afterwards, we apply bilinear representation \cite{DBCNN,pennec2006riemannian} for mapping $\bm{B}$ to $\bm{\tilde {B}}$ in the Euclidean space:
\begin{equation}\label{7}
\bm{\tilde {B}} = \frac{{sign(\bm{B}) \odot \sqrt {\left| \bm{B} \right|} }}{{{{\left\| {sign(\bm{B}) \odot \sqrt {\left| \bm{B} \right|} } \right\|}_2}}},
\end{equation}
where $sign(\cdot)$ denotes the sign function and $\odot$ represents the element-wise product. Then, $\bm{\tilde {B}}$ is regressed onto the global quality $Q_G$ using a fully connected layer.

\subsubsection{\textbf{Quality Regressor}}
Finally, after getting the local quality aggregation $Q_L$ and global quality $Q_G$, we utilize a fully connected layer to automatically assign weight for $Q_L$ and $Q_G$ for obtaining the perceptual quality as shown in Fig. \ref{fig:fig2}.
  
\subsection{Training and Testing}
A large amount of labeled training data promotes the development of deep learning technologies. However, existing OIQA databases \cite{OIQADatabse,CVIQDDatabase} only provide limited omnidirectional images for training networks. Thus, we utilize the pre-trained weight of ResNet-18, VGG-16 on ImageNet \cite{imagenet} and pre-trained weight of S-CNN on Waterloo Exploration Database \cite{exploration} and PASCAL VOC 2012 \cite{pascal} for initialization. Moreover, 2D IQA database \cite{live} is leveraged in our experiment for pre-training the viewport descriptor.

In blind omnidirectional image quality assessment, it is difficult to accurately predict the subjective MOS value \cite{MEON}. Therefore, the training stage is divided into three steps: 1) pre-training the viewport descriptor on LIVE IQA Database \cite{live}; 2) pre-training the local and global branches; 3) jointly optimizing the entire network VGCN.

At first, we adopt the ResNet-18 architecture with pre-trained weight on ImageNet to predict the 2D image quality, the loss function $l_1$ is denoted as:
\begin{equation}\label{8}
{l_1} = \frac{1}{K}\sum\nolimits_k {{{||q_{2d}^{(k)} - \widehat q_{2d}^{(k)}||}_2}},
\end{equation}
\begin{equation}\label{9}
\widehat q_{2d}^{(k)} = {f_{1}}(\bm{I^{(k)}_{2d}};\bm{w_{1}}),
\end{equation}
where $K$ is the batch size, $q_{2d}^{(k)}$ and $\widehat q_{2d}^{(k)}$ represent the subjective MOS label and objective predicted score of the $k$-th input 2D image $\bm{I^{(k)}_{2d}}$ in the mini-batch. Note that 2D images are randomly cropped to the size $256\times256$ and then taken as input for ResNet-18. The weight $\bm{w_{1}}$ for ResNet-18 $f_{1}$ is updated by minimizing $l_1$ as follows:
\begin{equation}\label{10}
\bm{w_{1}^{'}} = \arg \mathop {\min }\limits_{\bm{w_1}} {l_1},
\end{equation}

Secondly, the local and global branches are trained respectively. The viewport descriptor in local branch is initialized with $\bm{w_{1}^{'}}$. The tailored S-CNN and tailored VGG-16 in global branch are initialized with the pre-trained weight on Waterloo Exploration Database, PASCAL VOC 2012 and ImageNet as described in \cite{DBCNN}. The optimization process for local and global branches are given as:
\begin{equation}\label{11}
{l_{2L}} = \frac{1}{K}\sum\nolimits_k {{{||q^{(k)} - \widehat q_{L}^{(k)}||}_2}},
\end{equation}
\begin{equation}\label{12}
{l_{2G}} = \frac{1}{K}\sum\nolimits_k {{{||q^{(k)} - \widehat q_{G}^{(k)}||}_2}},
\end{equation}
\begin{equation}\label{13}
\widehat q_{L}^{(k)} = {f_{L}}(\bm{I^{(k)}_{d}};\bm{w_{L}}),\ 
\widehat q_{G}^{(k)} = {f_{G}}(\bm{I^{(k)}_{d}};\bm{w_{G}}),
\end{equation}
\begin{equation}\label{14}
\bm{w_{L}^{'}} = \arg \mathop {\min }\limits_{\bm{w_L}} {l_{2L}},\ 
\bm{w_{G}^{'}} = \arg \mathop {\min }\limits_{\bm{w_G}} {l_{2G}},
\end{equation}
where $\bm{I^{(k)}_{d}}$, $q^{(k)}$ represent the $k$-th input distorted omnidirectional image with its subjective score in a mini-batch. $\widehat q_{L}^{(k)}$, $\widehat q_{G}^{(k)}$ are the predicted scores of local and global branches. $l_{2L}$ refers to the loss function for updating the weight $\bm{w_L}$ of local branch $f_L$, and $l_{2G}$ denotes the loss function to update $\bm{w_G}$ of global branch $f_G$.

Finally, we jointly optimize the local, global branches and quality regressor using distorted 360-degree images. Considering the purpose of VGCN is to predict the perceptual quality of input omnidirectional images, we again adopt $l_2$-norm as the loss function $l_3$ to update VGCN parameters as follows:
\begin{equation}\label{15}
l_3 = \frac{1}{K}\sum\nolimits_k {{{||q^{(k)} - \widehat q^{(k)}||}_2}},
\end{equation}
\begin{equation}\label{16}
\widehat q^{(k)} = f(\bm{I^{(k)}_{d}};\bm{w_L},\bm{w_G},\bm{w_R}),
\end{equation}
\begin{equation}\label{17}
\bm{w_{L}^{*}},\bm{w_{G}^{*}},\bm{w_{R}^{*}} = \arg \mathop {\min }\limits_{\bm{w_L},\bm{w_G},\bm{w_R}} {l_3},
\end{equation}
where $\widehat q^{(k)}$ is the final predicted quality. $f$ denotes the VGCN network, the weight $\bm{w_R}$ for quality regressor is trained from scratch while $(\bm{w_L},\bm{w_G})$ are initialized with pre-trained weight $(\bm{w_{L}^{'}},\bm{w_{G}^{'}})$.

During testing, the distorted omnidirectional image is directly fed into VGCN, then the network returns the predicted quality of the given image.

\section{Experimental Results and Analysis}
In this section, the databases and performance measures used in our experiment are introduced at first. Then, we conduct the performance comparison of VGCN with other metrics on individual and cross databases. Finally, the effectiveness of each component in VGCN is verified via ablation study. 

\subsection{Databases and Performance Measures}
\subsubsection{\textbf{Database}}
We utilize two benchmark omnidirectional image quality assessment databases in this experiment, namely OIQA Database \cite{OIQADatabse} and CVIQD Database \cite{CVIQDDatabase}.

\textbf{OIQA Database \cite{OIQADatabse}:} This database includes 320 distorted omnidirectional images deriving from 16 reference images with 4 distortion types and 5 distortion levels. Specifically, the degradations are JPEG compression (JPEG), JPEG2000 compression (JP2K), Gaussian blur (BLUR) and Gaussian white noise (WN). Subjective ratings of MOS are given in the range [1, 10], where higher score means better visual quality. 

\textbf{CVIQD Database \cite{CVIQDDatabase}:} It is the largest omnidirectional image quality assessment database containing 528 compressed images originating from 16 reference images. Three commonly used coding technologies are involved in this database, namely JPEG, H.264/AVC and H.265/HEVC. The MOS values are normalized and rescaled to the range [0, 100].

\subsubsection{\textbf{Performance Measures}}
Both standard measures \cite{PerformanceMeasure} and Krasula methodology \cite{krasula2016accuracy} are utilized in our experiment to evaluate the performance of different metrics.

\textbf{Standard measures \cite{PerformanceMeasure}:} Three standard measures are adopted, including Spearman’s rank order correlation coefficient (SROCC), Pearson’s linear correlation coefficient (PLCC) and root mean squared error (RMSE). SROCC measures the prediction monotonicity, while PLCC and RMSE measure the prediction accuracy. For SROCC and PLCC, the higher the better. In contrast, the lower RMSE means the smaller distance between MOS value and predicted score. Before computing PLCC and RMSE, to maximize the correlation between subjective rating and objective score, we apply a five-parameter logistic function:
\begin{equation}\label{18}
y = {\beta _1}(\frac{1}{2} - \frac{1}{{1 + \exp ({\beta _2}(x - {\beta _3}))}}) + {\beta _4}x + {\beta _5},
\end{equation}
where $x$ refers to the predicted quality of objective metrics and $y$ denotes the mapped score. $\beta1$ to $\beta_5$ are five parameters to fit the logistic function. 

\textbf{Krasula methodology \cite{krasula2016accuracy}:} The commonly used PLCC and SROCC measures remain some issues \cite{aldahdooh2018improved}, to overcome the drawbacks of traditional performance measures, Krasula methodology takes into account the impact of mapping functions and the statistical significance of the subjective scores. It tests the model reliability by checking whether it can differentiate different/similar pairs as well as better/worse pairs. Thus, to apply Krasula methodology, we select pairs of 360-degree images to calculate the area under the ROC curve of the \textit{Different vs. Similar} categories (AUC-DS), area under the ROC curve of the \textit{Better vs. Worse} categories (AUC-BW) and percentage of correct classification ($C_0$) as described in \cite{krasula2016accuracy}. A good IQA metric is able to achieve AUC-DS, AUC-BW and CC close to 1.

\subsection{Performance Evaluation}

\begin{table*}[ht]
	\renewcommand\arraystretch{1.5}
	\begin{center}
		\captionsetup{justification=centering}
		\caption{\textsc{Performance Comparison on OIQA Database. VGCN (local) Denotes the Local Branch in Our Proposed Model. The Best Performing FR and NR Metrics are Highlighted in Bold.}}
		\label{table3}
		\scalebox{0.82}{
			\begin{tabular}{@{}cc|ccc|ccc|ccc|ccc|ccc@{}}
				\toprule
				&              & \multicolumn{3}{c|}{JPEG}                           & \multicolumn{3}{c|}{JP2K}                           & \multicolumn{3}{c|}{WN}                             & \multicolumn{3}{c|}{BLUR}                           & \multicolumn{3}{c}{ALL}                             \\ \midrule
				&              & PLCC            & SROCC           & RMSE            & PLCC            & SROCC           & RMSE            & PLCC            & SROCC           & RMSE            & PLCC            & SROCC           & RMSE            & PLCC            & SROCC           & RMSE            \\
				\multirow{8}{*}{FR} & PSNR         & 0.6941          & 0.7060          & 1.6141          & 0.8632          & 0.7821          & 1.1316          & 0.9547          & 0.9500          & 0.5370          & 0.9282          & 0.7417          & 0.8299          & 0.5812          & 0.5226          & 1.7005          \\
				& S-PSNR \cite{SPSNR}& 0.6911          & 0.6148          & 1.6205          & 0.9205          & 0.7250          & 0.8757          & 0.9503          & 0.9357          & 0.5620          & 0.8282          & 0.7525          & 1.0910          & 0.5997          & 0.5399          & 1.6721          \\
				& WS-PSNR \cite{WSPSNR}& 0.7133          & 0.6792          & 1.5713          & 0.9344          & 0.7500          & 0.9128          & 0.9626          & 0.9500          & 0.4890          & 0.8190          & 0.7668          & 1.1172          & 0.5819          & 0.5263          & 1.6994          \\
				& CPP-PSNR \cite{CPPPSNR}& 0.6153          & 0.5362          & 1.7693          & 0.8971          & 0.7250          & 0.9904          & 0.9276          & 0.9143          & 0.6739          & 0.7969          & 0.7185          & 1.1728          & 0.5683          & 0.5149          & 1.7193          \\
				& SSIM \cite{ssim}& 0.9077          & \textbf{0.9008} & 0.9406          & 0.9783          & 0.9679          & 0.4643          & 0.8828          & 0.8607          & 0.8474          & 0.9926          & 0.9777          & 0.2358          & 0.8718          & 0.8588          & 1.0238          \\
				& MS-SSIM \cite{msssim}& \textbf{0.9102} & 0.8937          & \textbf{0.9288} & 0.9492          & 0.9250          & 0.7052          & 0.9691          & 0.9571          & 0.4452          & 0.9251          & 0.8990          & 0.7374          & 0.7710          & 0.7379          & 1.3308          \\
				& FSIM \cite{fsim}& 0.8938          & 0.8490          & 1.0057          & 0.9699          & 0.9643          & 0.5454          & 0.9170          & 0.8893          & 0.7197          & \textbf{0.9914} & \textbf{0.9902} & \textbf{0.2544} & 0.9014          & 0.8938          & 0.9047          \\
				& DeepQA \cite{DeepQA}& 0.8301          & 0.8150          & 1.2506          & \textbf{0.9905} & \textbf{0.9893} & \textbf{0.3082} & \textbf{0.9709} & \textbf{0.9857} & \textbf{0.4317} & 0.9623          & 0.9473          & 0.5283          & \textbf{0.9044} & \textbf{0.8973} & \textbf{0.8914} \\ \midrule
				\multirow{6}{*}{NR} & BRISQUE \cite{BRISQUE}& 0.9160          & 0.9392          & 0.8992          & 0.7397          & 0.6750          & 1.5082          & \textbf{0.9818}          & 0.9750          & \textbf{0.3427}          & 0.8663          & 0.8508          & 0.9697          & 0.8424          & 0.8331          & 1.1261          \\
				& BMPRI \cite{BMPRI}& 0.9361          & 0.8954          & 0.7886          & 0.8322          & 0.8214          & 1.2428          & 0.9673          & 0.9821          & 0.4572          & 0.5199          & 0.3807          & 1.6584          & 0.6503          & 0.6238          & 1.5874          \\
				& DB-CNN \cite{DBCNN}& 0.8413          & 0.7346          & 1.2118          & 0.9755          & \textbf{0.9607} & 0.4935          & 0.9772 & \textbf{0.9786} & 0.3832 & 0.9536          & 0.8865          & 0.5875          & 0.8852          & 0.8653          & 0.9717          \\
				& MC360IQA \cite{mc360iqa2}& 0.9459          & 0.9008          & 0.7272          & 0.9165          & 0.9036          & 0.8966          & 0.9718          & 0.9464          & 0.4251          & 0.9526          & 0.9580          & 0.5907          & 0.9267          & 0.9139          & 0.7854         \\
				& VGCN (local) & 0.9508          & 0.8972          & 0.6949          & \textbf{0.9793} & 0.9439          & \textbf{0.4541} & 0.9682          & 0.9714          & 0.4515          & 0.9838          & \textbf{0.9759} & 0.3479          & 0.9529          & 0.9444          & 0.6340          \\
				& VGCN         & \textbf{0.9540} & \textbf{0.9294} & \textbf{0.6720} & 0.9771          & 0.9464          & 0.4772          & 0.9811          & 0.9750          & 0.3493          & \textbf{0.9852} & 0.9651          & \textbf{0.3327} & \textbf{0.9584} & \textbf{0.9515} & \textbf{0.5967} \\ \bottomrule
		\end{tabular}}
	\end{center}
\end{table*}

\begin{figure*}[h]
	\centering
	\subfigure[PSNR]{
		\includegraphics[height=2.4cm]{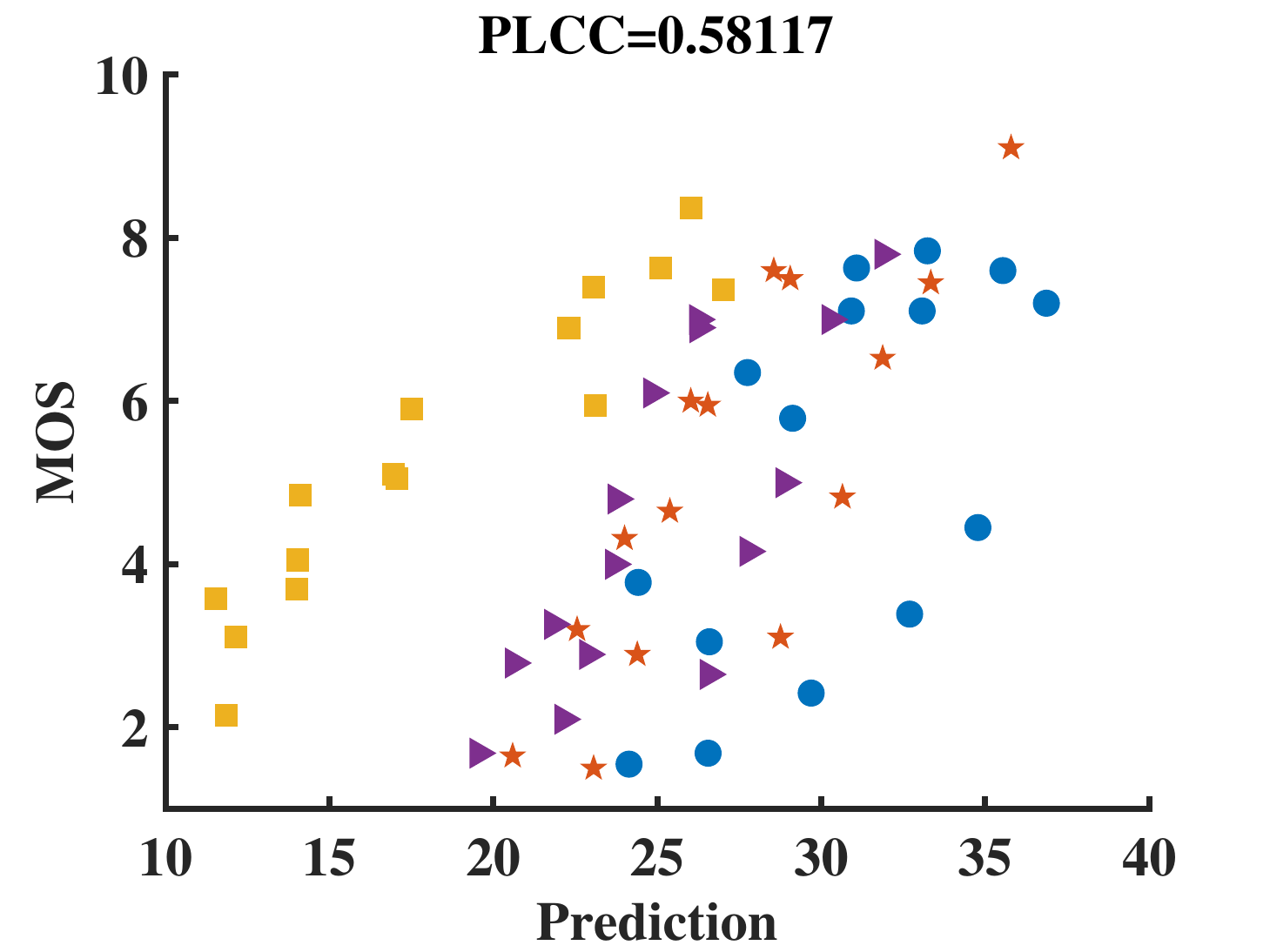}
	}
	\subfigure[S-PSNR]{
		\includegraphics[height=2.4cm]{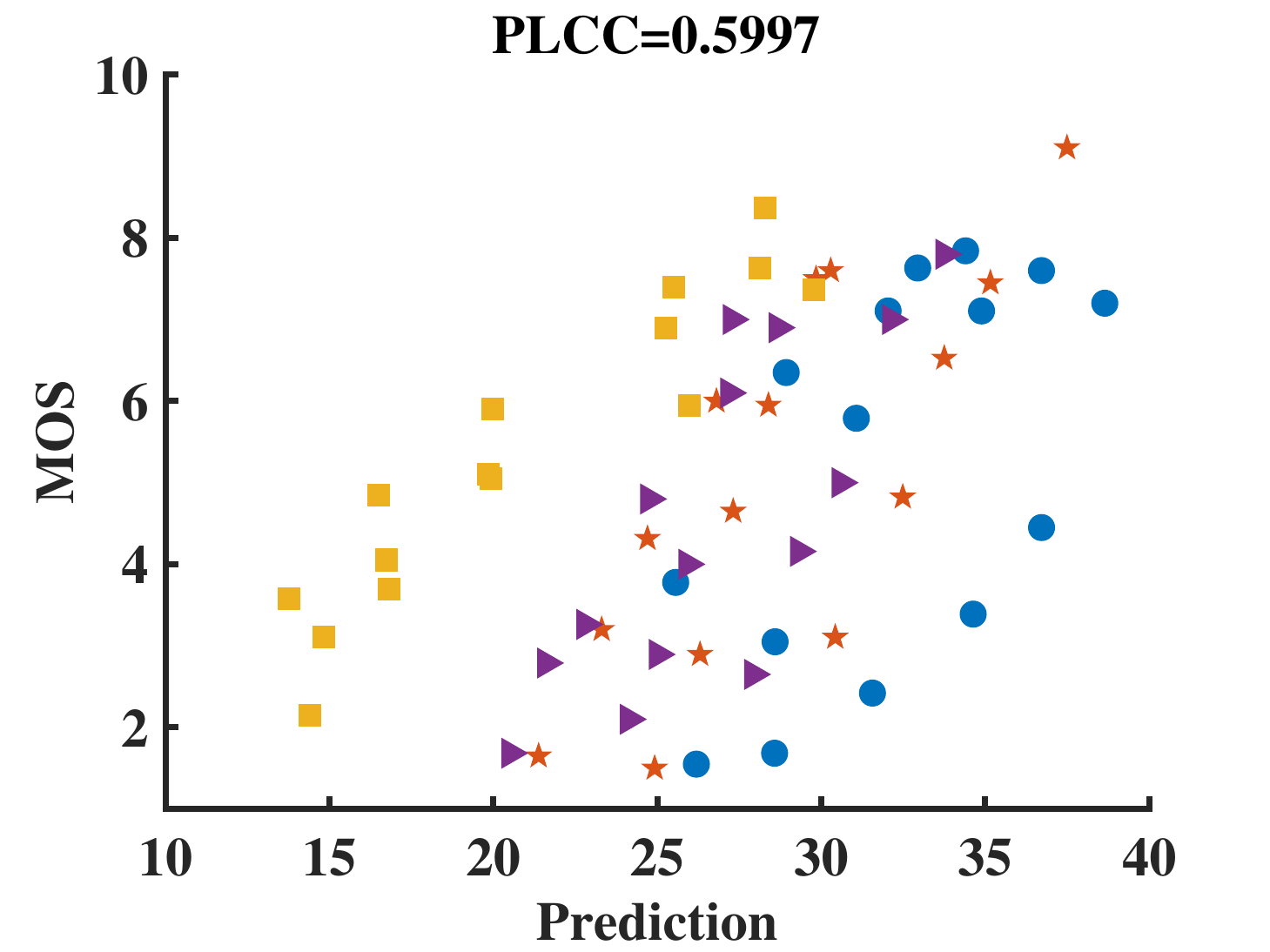}
	}
	\subfigure[WS-PSNR]{
		\includegraphics[height=2.4cm]{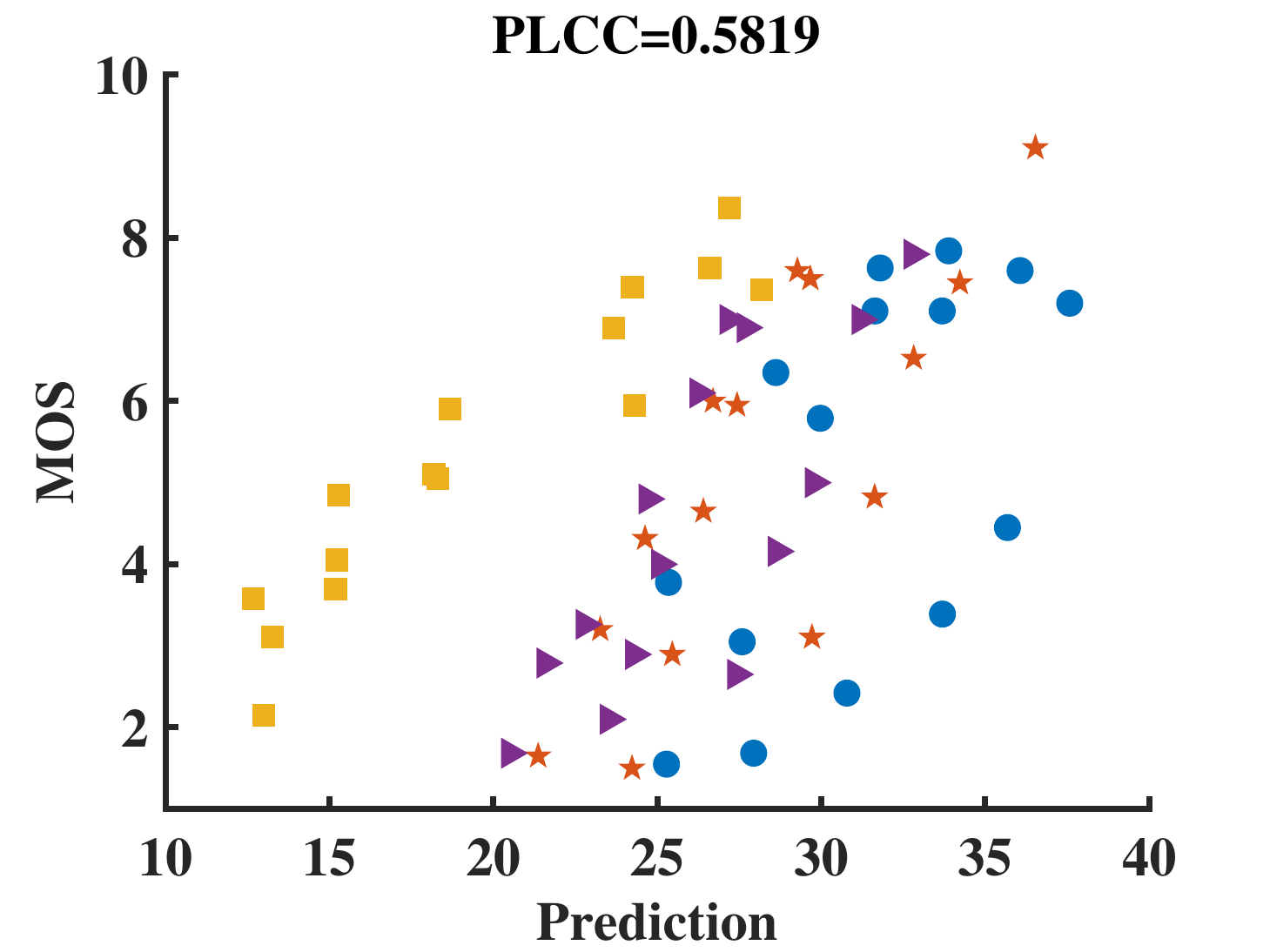}
	}
	\subfigure[CPP-PSNR]{
		\includegraphics[height=2.4cm]{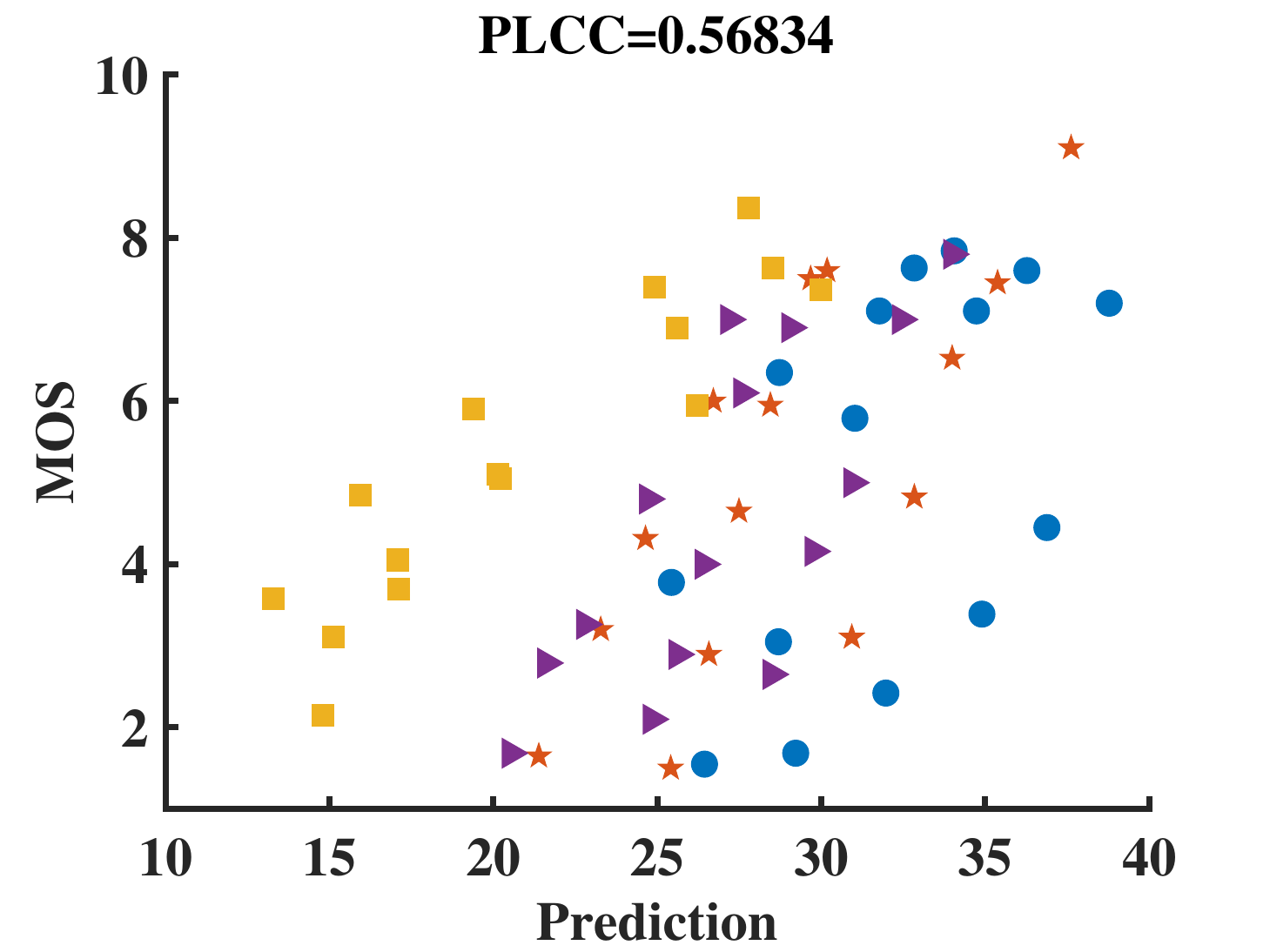}
	}
	\subfigure[SSIM]{
		\includegraphics[height=2.4cm]{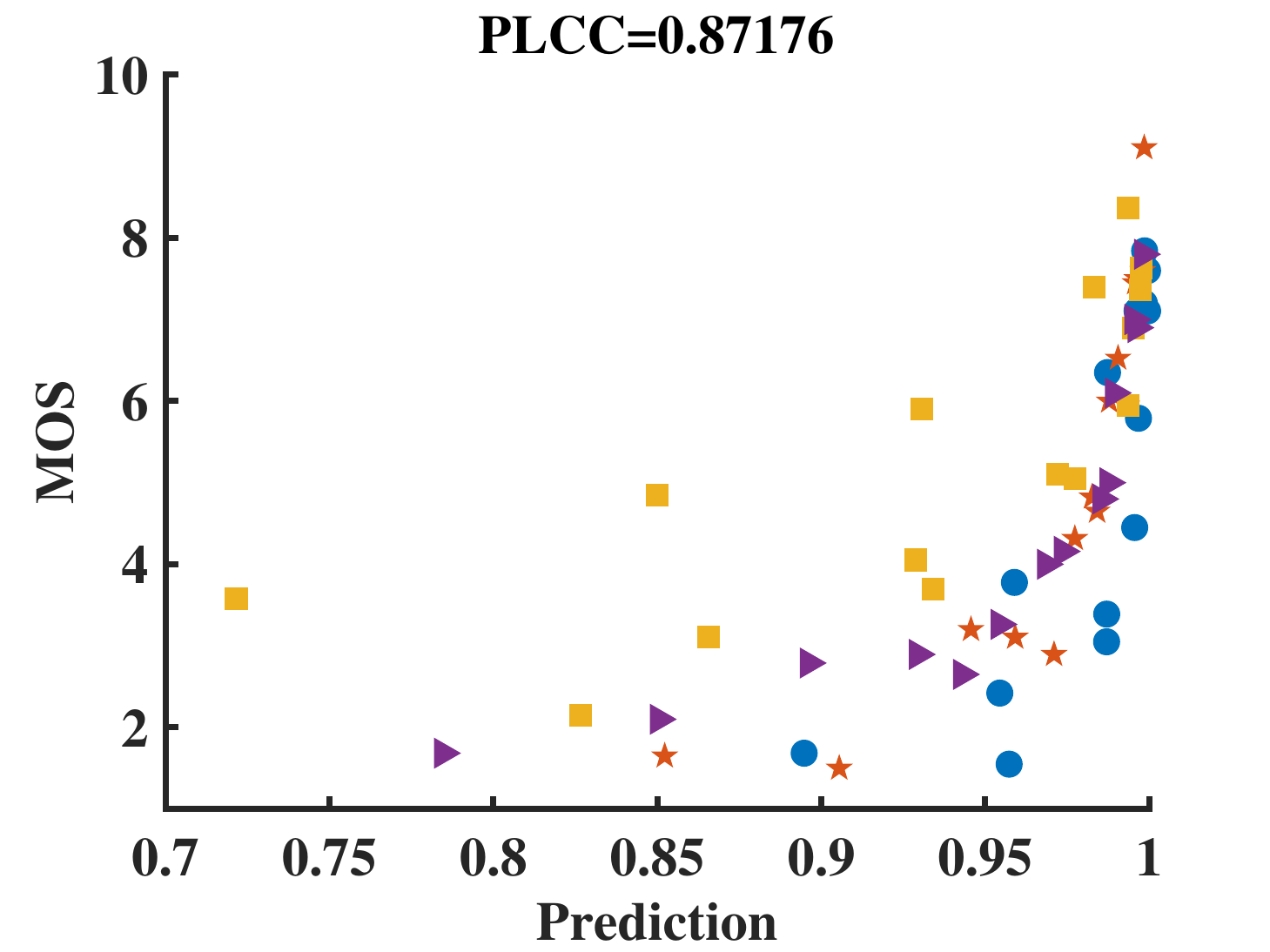}
	}
	\subfigure[MS-SSIM]{
		\includegraphics[height=2.4cm]{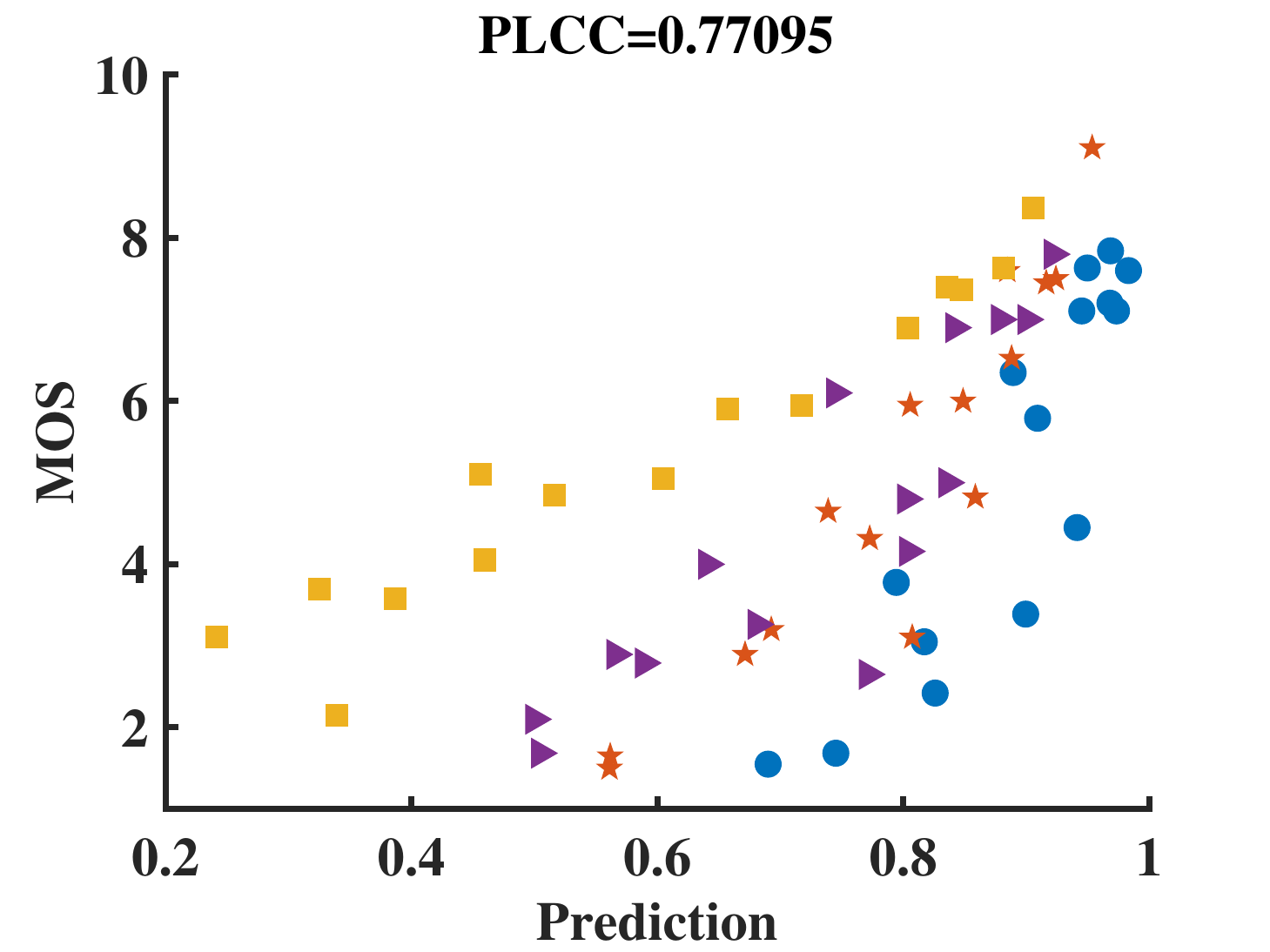}
	}
	\subfigure[FSIM]{
		\includegraphics[height=2.4cm]{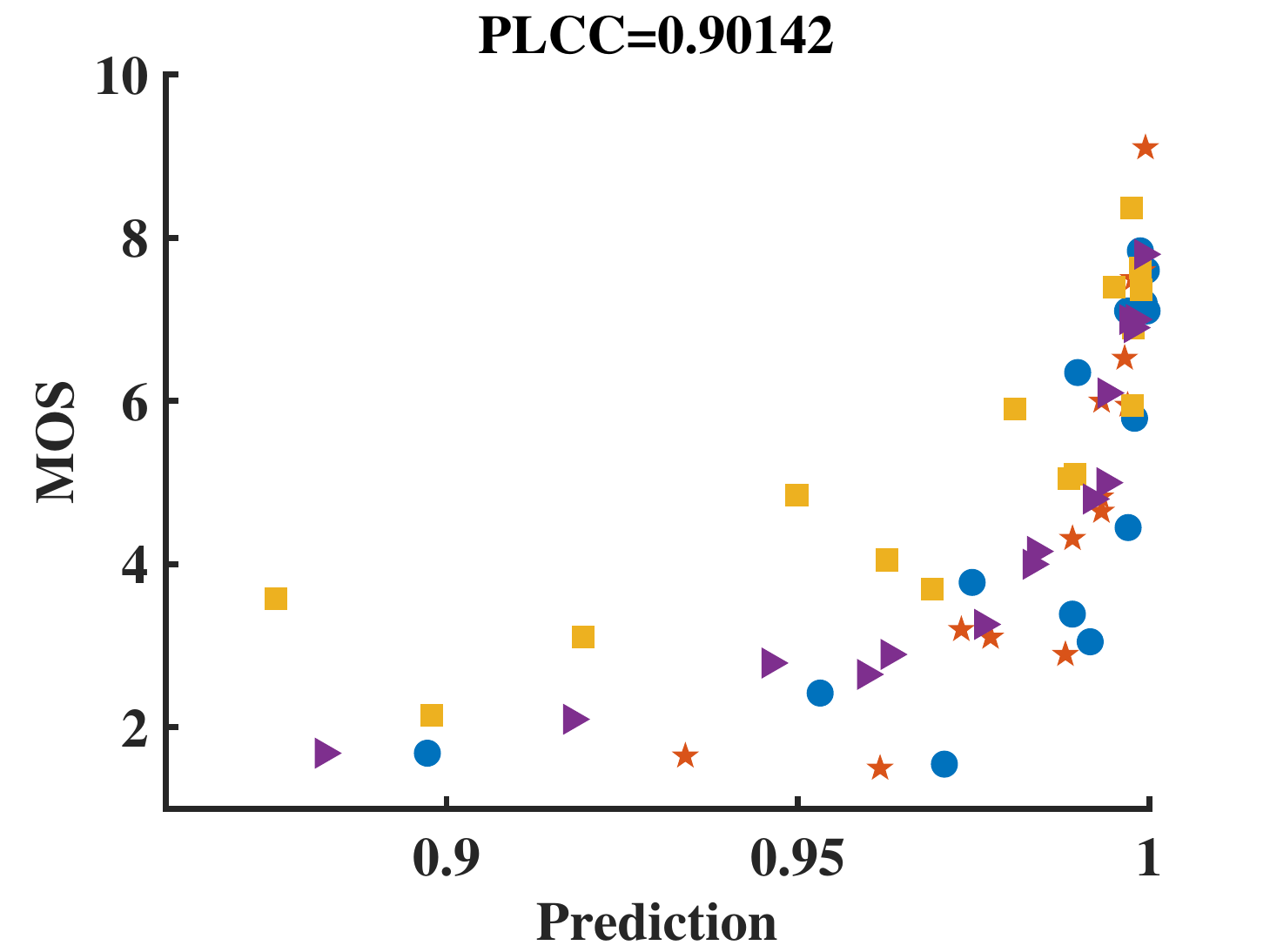}
	}
	\subfigure[DeepQA]{
		\includegraphics[height=2.4cm]{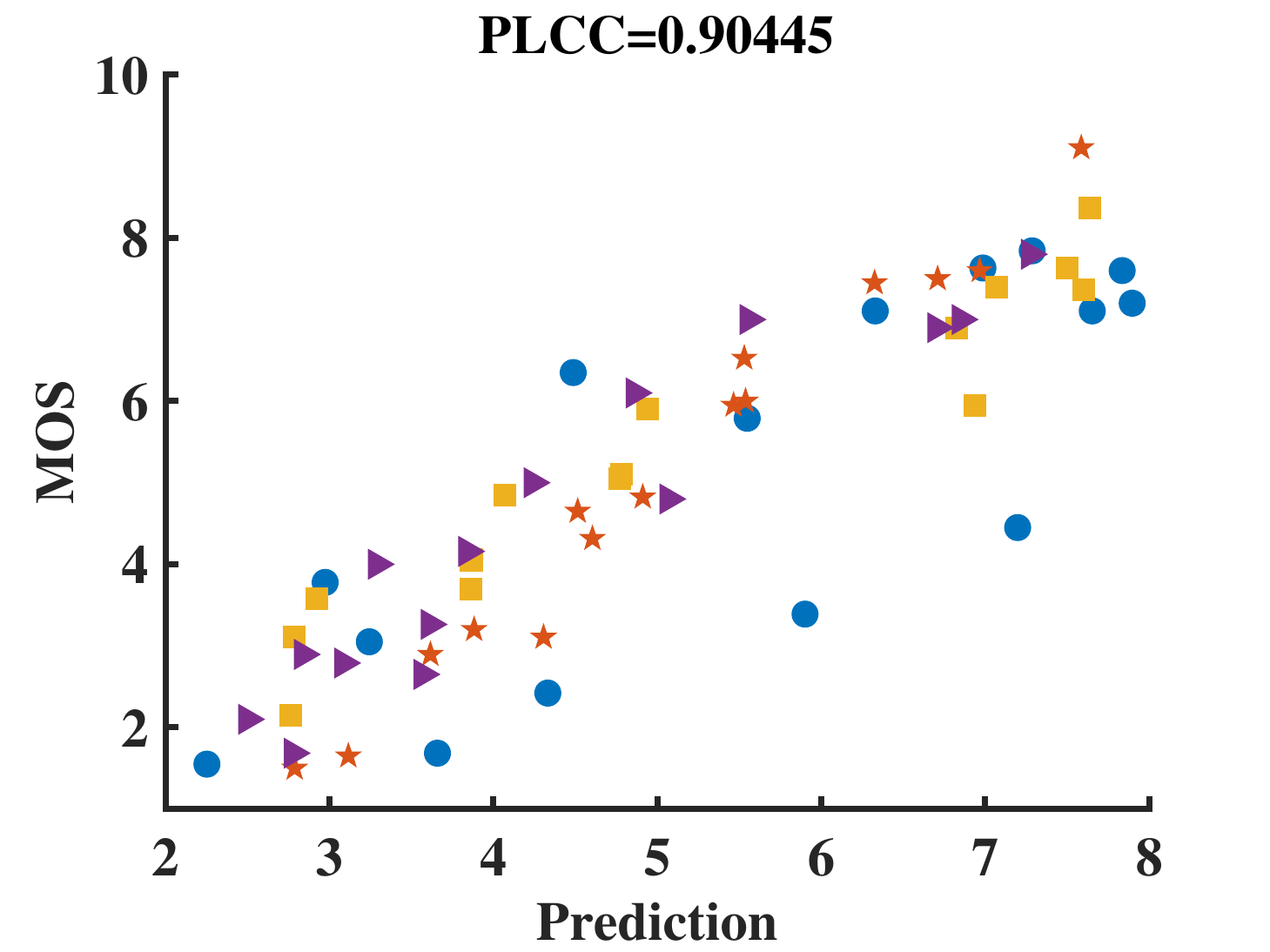}
	}
	\subfigure[BRISQUE]{
		\includegraphics[height=2.4cm]{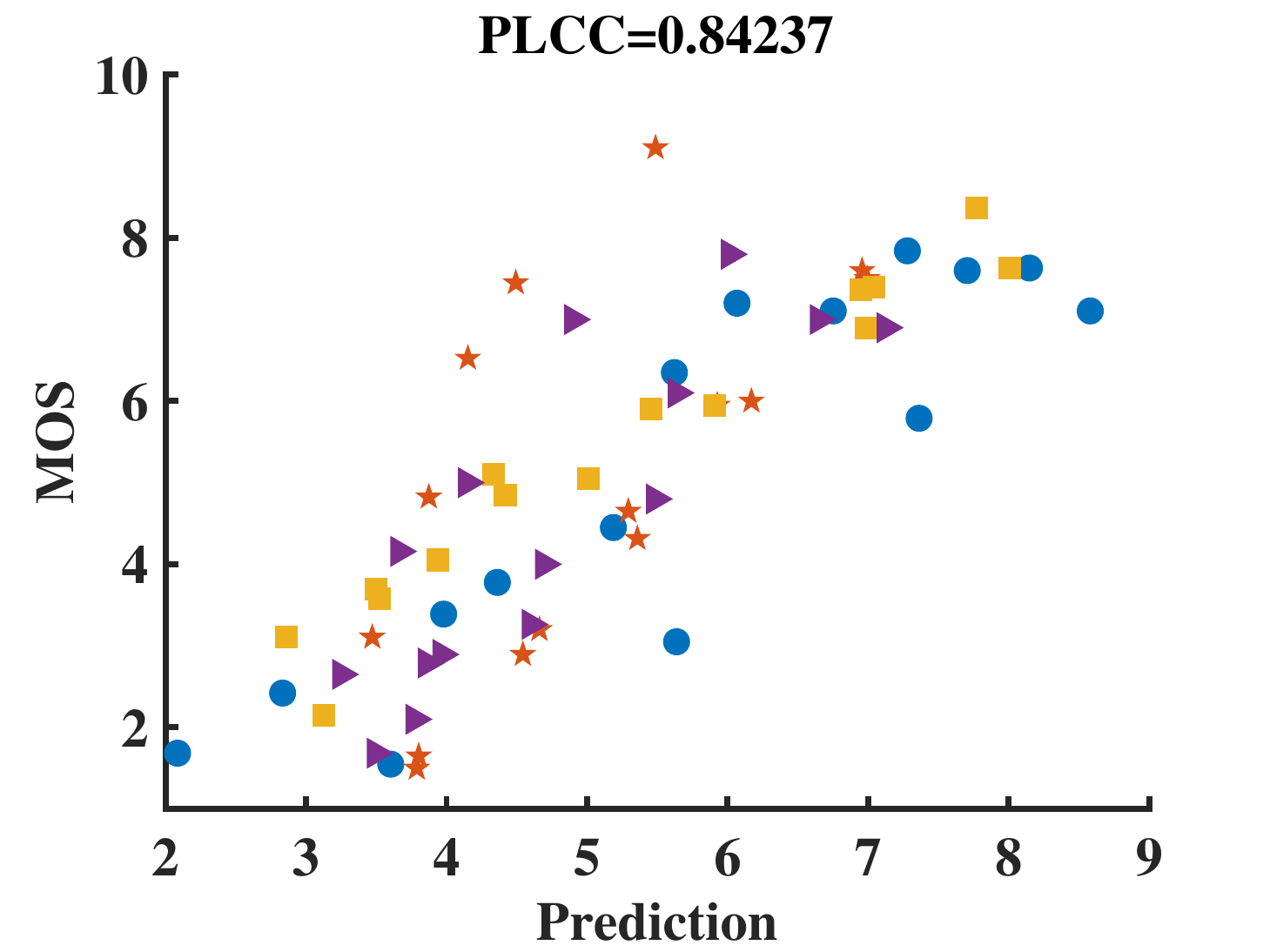}
	}
	\subfigure[BMPRI]{
		\includegraphics[height=2.4cm]{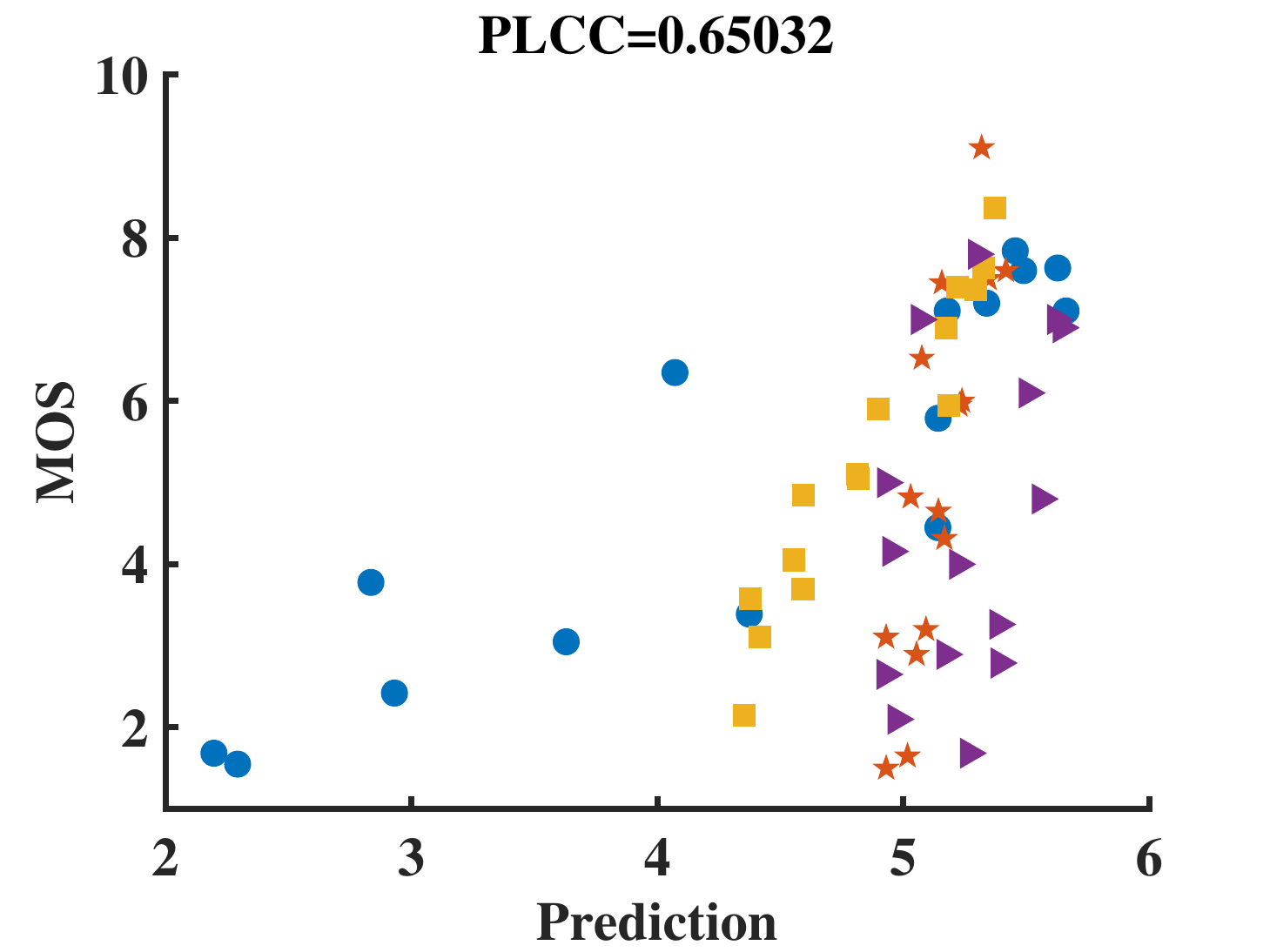}
	}
	\subfigure[DB-CNN]{
		\includegraphics[height=2.4cm]{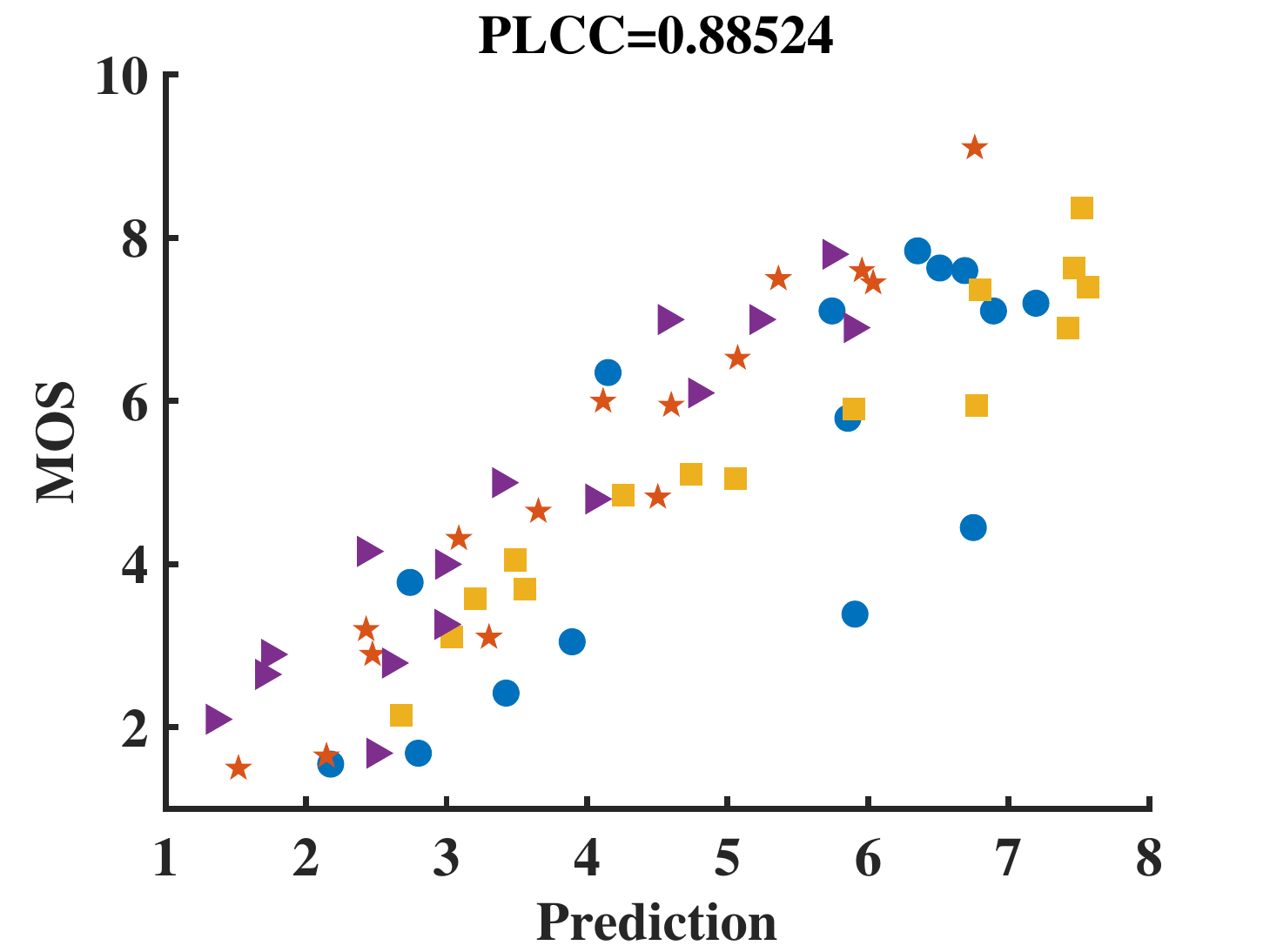}
	}
	\subfigure[MC360IQA]{
		\includegraphics[height=2.4cm]{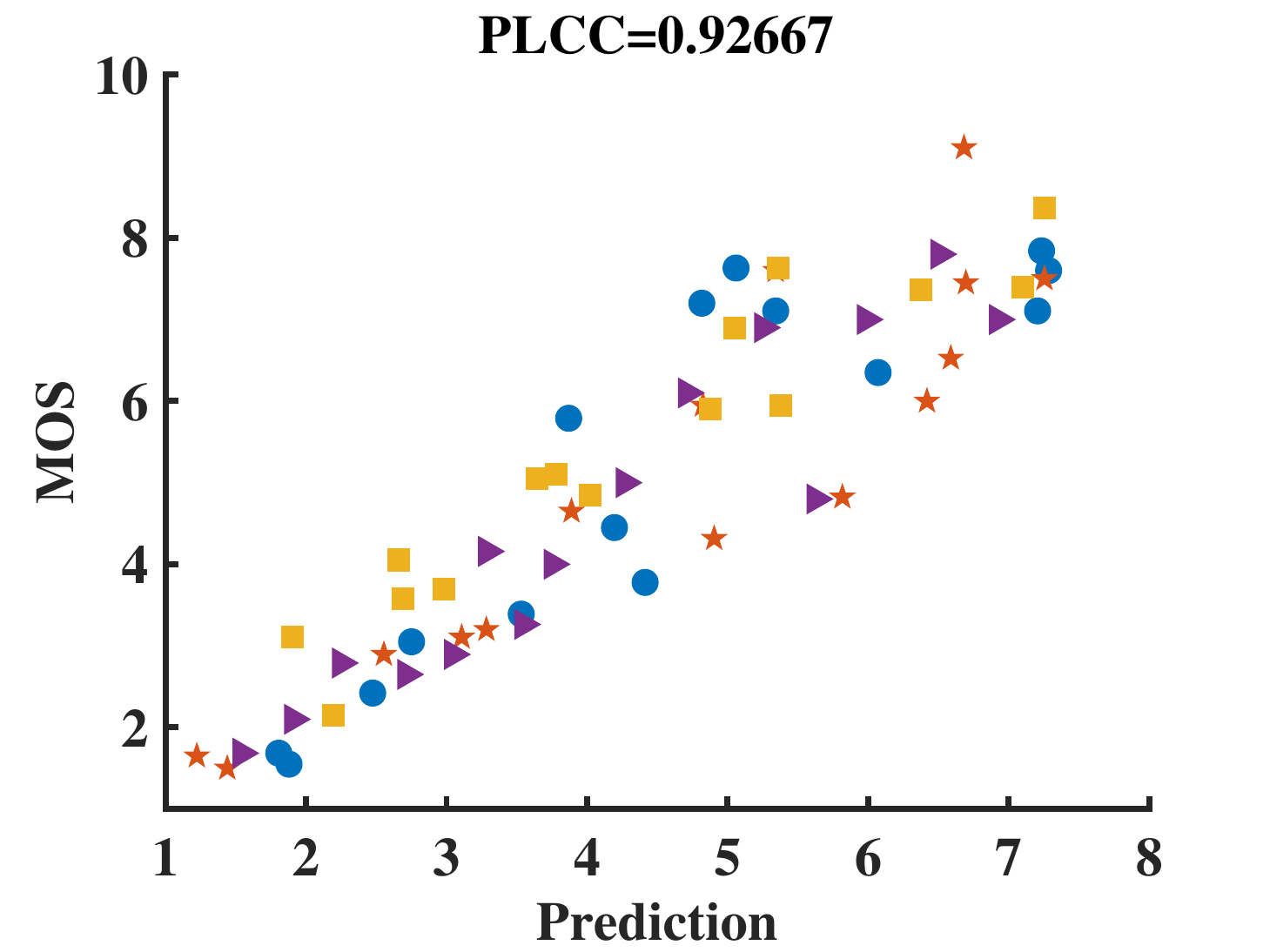}
	}
	\subfigure[VGCN (local)]{
		\includegraphics[height=2.4cm]{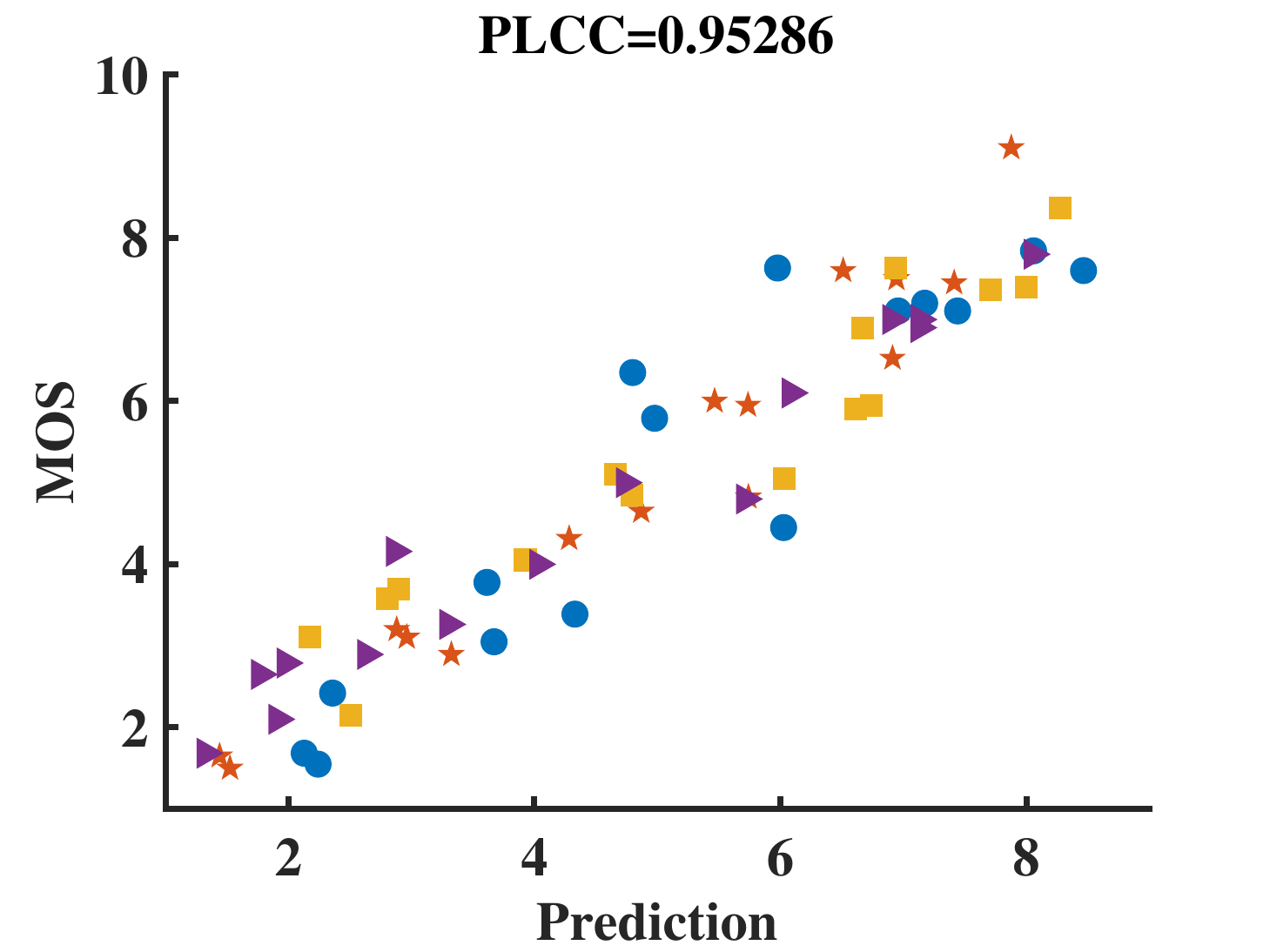}
	}
	\subfigure[VGCN]{
		\includegraphics[height=2.4cm]{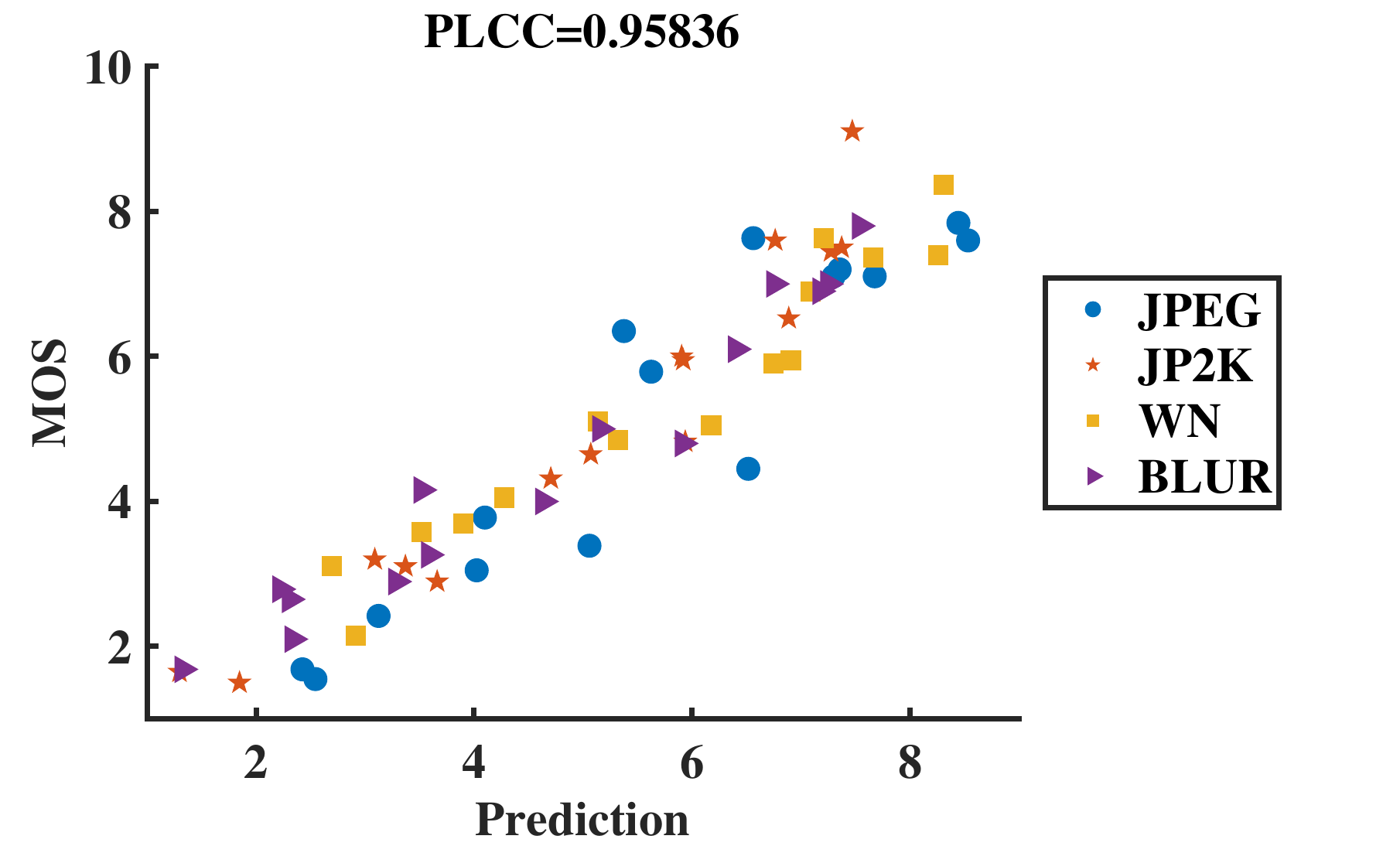}
	}
	\caption{Scatter plots of MOS values against predictions by IQA metrics for individual distortion type on the testing set of OIQA Database.}
	\label{fig:fig4}
\end{figure*}

We split the database into the training and testing set following the standard method mentioned in \cite{VQA-ODV,VCNN}, which means in the entire database, distorted images corresponding to 3 reference images are randomly selected as testing set and the remaining impaired images of 13 reference images are regarded as the training set. The input omnidirectional is first downsampled to the size $512\times1024$ as described in Section II. Adaptive moment estimation optimizer (Adam) is employed in three training steps but set as different learning rates. In step I, we achieve the pre-trained weight $\bm{w_1^{'}}$ at 60 epochs with an initial learning rate $10^{-4}$. In step II, the weight $\bm{w_L^{'}}$ for local branch is obtained after 200 epochs for OIQA database and 30 epochs for CVIQD database, the initial learning rate is set as $10^{-3}$ and scaled by 0.25 every 40 epochs. Note that learning rate for viewport descriptor is fixed at $10^{-6}$. In the global branch, the pre-trained weight of tailored S-CNN and VGG-16 are introduced and fixed before training, then we initialize the learning rate with $10^{-2}$ and lower it by a factor of 10 at 400 epochs. The pre-trained global branch is obtained after 1000 epochs. In step III, we fix the local and global branches with a learning rate $10^{-6}$, and jointly optimize VGCN with an initial learning rate $10^{-2}$. The final model is obtained after 20 epochs. The mini-batch size is set to 16 during pre-training, and 8 in joint optimization. The VGCN is implemented with Pytorch and will be publicly available online.

\begin{table*}[ht]
	\renewcommand\arraystretch{1.5}
	\begin{center}
		\captionsetup{justification=centering}
		\caption{\textsc{Performance Comparison on CVIQD Database. VGCN (local) Denotes the Local Branch in Our Proposed Model. The Best Performing FR and NR Metrics are Highlighted in Bold.}}
		\label{table4}
		\scalebox{0.82}{
			\begin{tabular}{@{}cc|ccc|ccc|ccc|ccc@{}}
				\toprule
				&              & \multicolumn{3}{c|}{JPEG}                            & \multicolumn{3}{c|}{AVC}                             & \multicolumn{3}{c|}{HEVC}                            & \multicolumn{3}{c}{ALL}                             \\ \midrule
				&              & PLCC            & SROCC           & RMSE            & PLCC            & SROCC           & RMSE            & PLCC            & SROCC           & RMSE            & PLCC            & SROCC           & RMSE            \\
				\multirow{8}{*}{FR} & PSNR         & 0.8682          & 0.6982          & 8.0429          & 0.6141          & 0.5802          & 10.5520         & 0.5982          & 0.5762          & 9.4697          & 0.7008          & 0.6239          & 9.9599          \\
				& S-PSNR \cite{SPSNR}      & 0.8661          & 0.7172          & 8.1008          & 0.6307          & 0.6039          & 10.3760         & 0.6514          & 0.6150          & 8.9585          & 0.7083          & 0.6449          & 9.8564          \\
				& WS-PSNR \cite{WSPSNR}     & 0.8572          & 0.6848          & 8.3465          & 0.5702          & 0.5521          & 10.9841         & 0.5884          & 0.5642          & 9.5473          & 0.6729          & 0.6107          & 10.3283         \\
				& CPP-PSNR \cite{CPPCNR}    & 0.8585          & 0.7059          & 8.3109          & 0.6137          & 0.5872          & 10.5615         & 0.6160          & 0.5689          & 9.3009          & 0.6871          & 0.6265          & 10.1448         \\
				& SSIM \cite{ssim}        & 0.9822          & 0.9582          & 3.0468          & 0.9303          & 0.9174          & 4.9029          & 0.9436          & 0.9452          & 3.9097          & 0.9002          & 0.8842          & 6.0793          \\
				& MS-SSIM \cite{msssim}     & 0.9636          & 0.9047          & 4.3355          & 0.7960          & 0.7650          & 8.0924          & 0.8072          & 0.8011          & 6.9693          & 0.8521          & 0.8222          & 7.3072          \\
				& FSIM \cite{fsim}        & \textbf{0.9839} & \textbf{0.9639} & \textbf{2.8928} & \textbf{0.9534} & \textbf{0.9439} & \textbf{4.0327} & \textbf{0.9617} & \textbf{0.9532} & \textbf{3.2385} & 0.9340          & 0.9152          & 4.9864          \\
				& DeepQA \cite{DeepQA}      & 0.9526          & 0.9001          & 4.9290          & 0.9477          & 0.9375          & 4.2683          & 0.9221          & 0.9288          & 4.5694          & \textbf{0.9375} & \textbf{0.9292} & \textbf{4.8574} \\ \midrule
				\multirow{6}{*}{NR} & BRISQUE \cite{BRISQUE}     & 0.9464          & 0.9031          & 5.2442          & 0.7745          & 0.7714          & 8.4573          & 0.7548          & 0.7644          & 7.7455          & 0.8376          & 0.8180          & 7.6271          \\
				& BMPRI \cite{BMPRI}       & 0.9874          & 0.9562          & 2.5597          & 0.7161          & 0.6731          & 9.3318          & 0.6154          & 0.6715          & 9.3071          & 0.7919          & 0.7470          & 8.5258          \\
				& DB-CNN \cite{DBCNN}       & 0.9779          & 0.9576          & 3.3862          & 0.9564          & 0.9545          & 3.9063          & 0.8646          & 0.8693          & 5.9335          & 0.9356          & 0.9308          & 4.9311          \\
				& MC360IQA \cite{mc360iqa2}    & 0.9698          & 0.9693          & 3.9517          & 0.9487          & 0.9569          & 4.2281          & 0.8976          & 0.9104          & 5.2557          & 0.9429          & 0.9428          & 4.6506          \\
				& VGCN (local) & 0.9857          & 0.9666          & 2.7310          & 0.9684          & 0.9622          & 3.3328          & 0.9367          & 0.9422          & 4.1329          & 0.9597          & 0.9539          & 3.9220          \\
				& VGCN         & \textbf{0.9894} & \textbf{0.9759} & \textbf{2.3590} & \textbf{0.9719} & \textbf{0.9659} & \textbf{3.1490} & \textbf{0.9401} & \textbf{0.9432} & \textbf{4.0257} & \textbf{0.9651} & \textbf{0.9639} & \textbf{3.6573} \\ \bottomrule
		\end{tabular}}
	\end{center}
\end{table*}

\begin{figure*}[h]
	\centering
	\subfigure[PSNR]{
		\includegraphics[height=2.4cm]{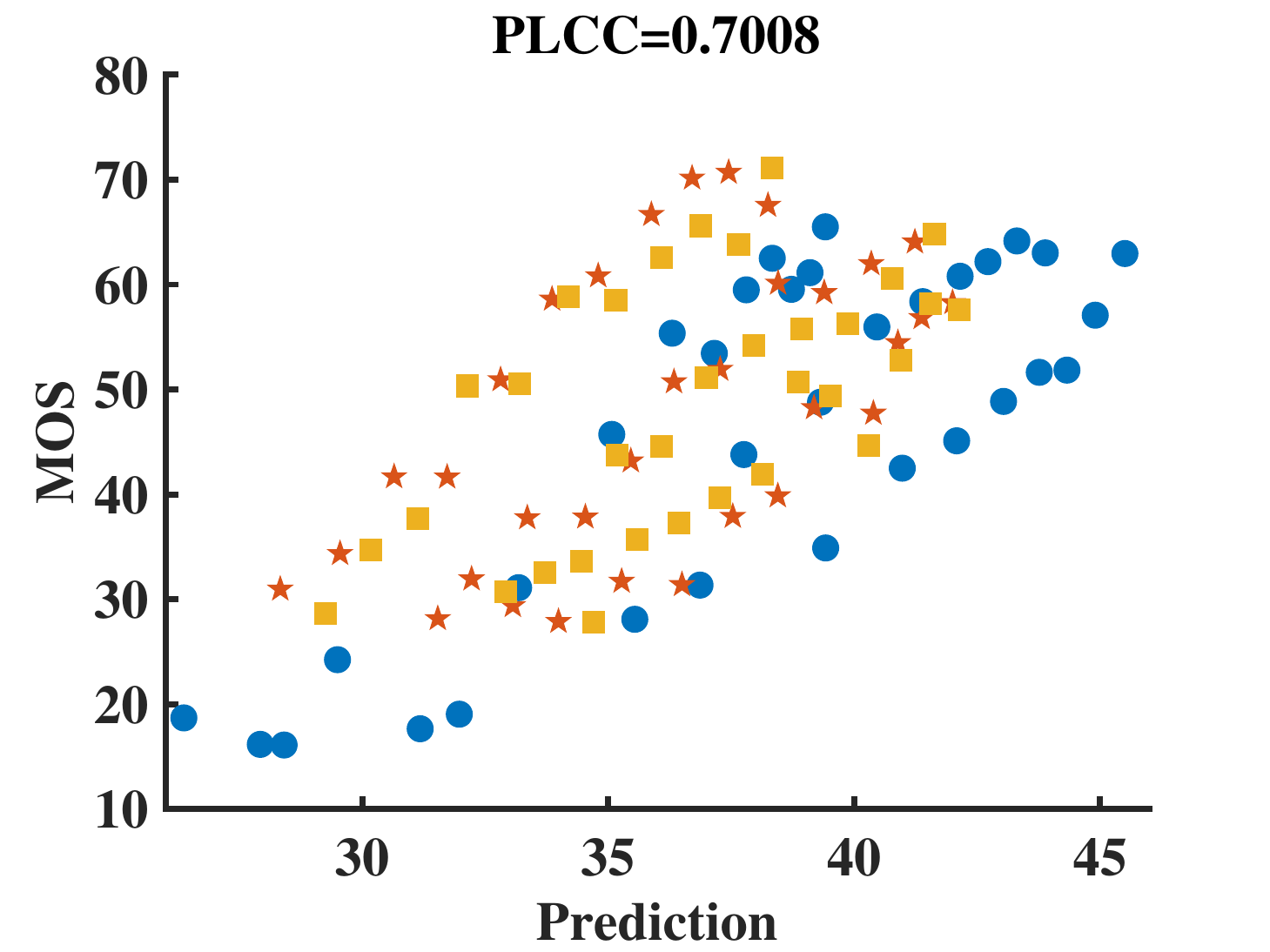}
	}
	\subfigure[S-PSNR]{
		\includegraphics[height=2.4cm]{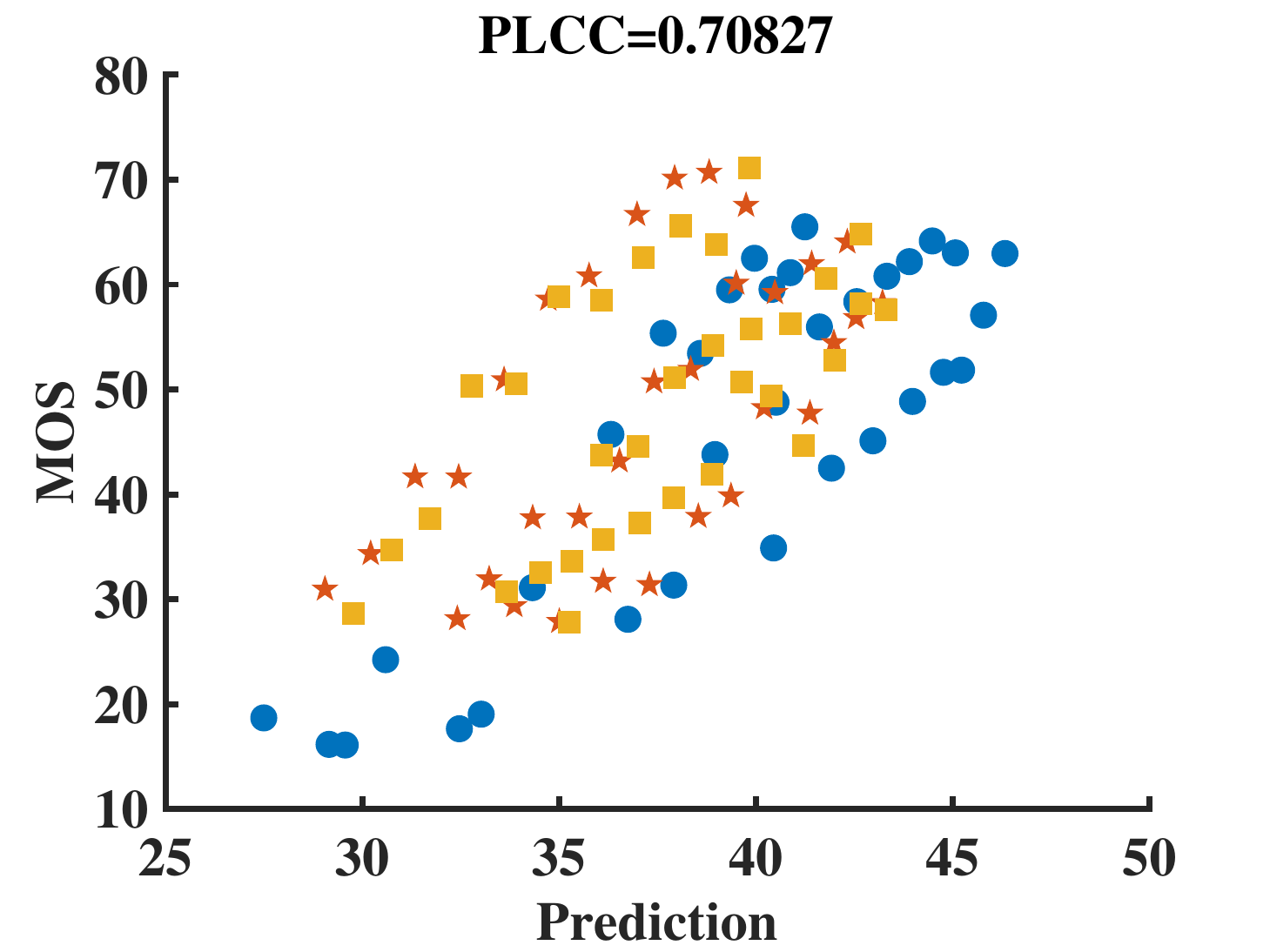}
	}
	\subfigure[WS-PSNR]{
		\includegraphics[height=2.4cm]{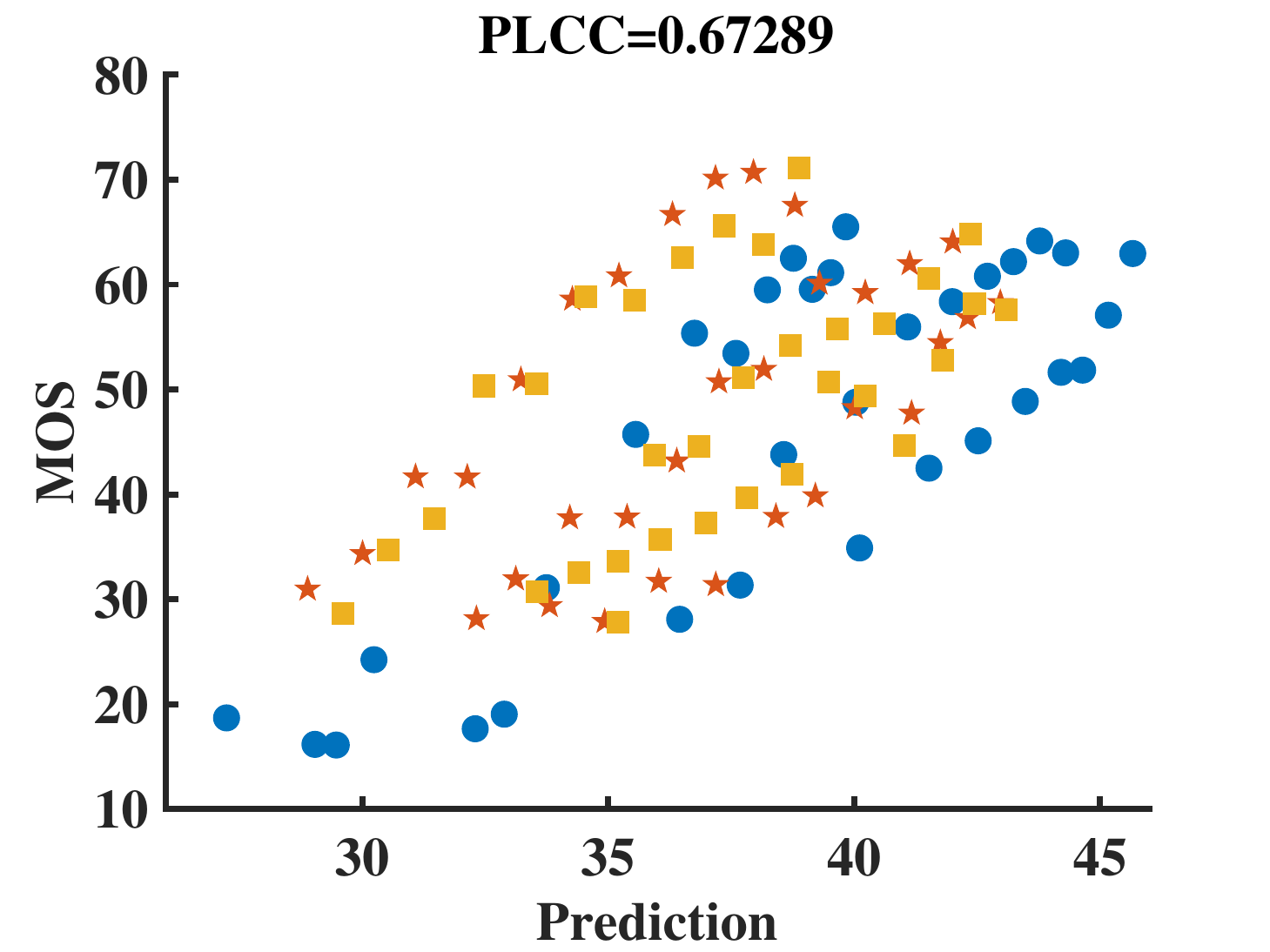}
	}
	\subfigure[CPP-PSNR]{
		\includegraphics[height=2.4cm]{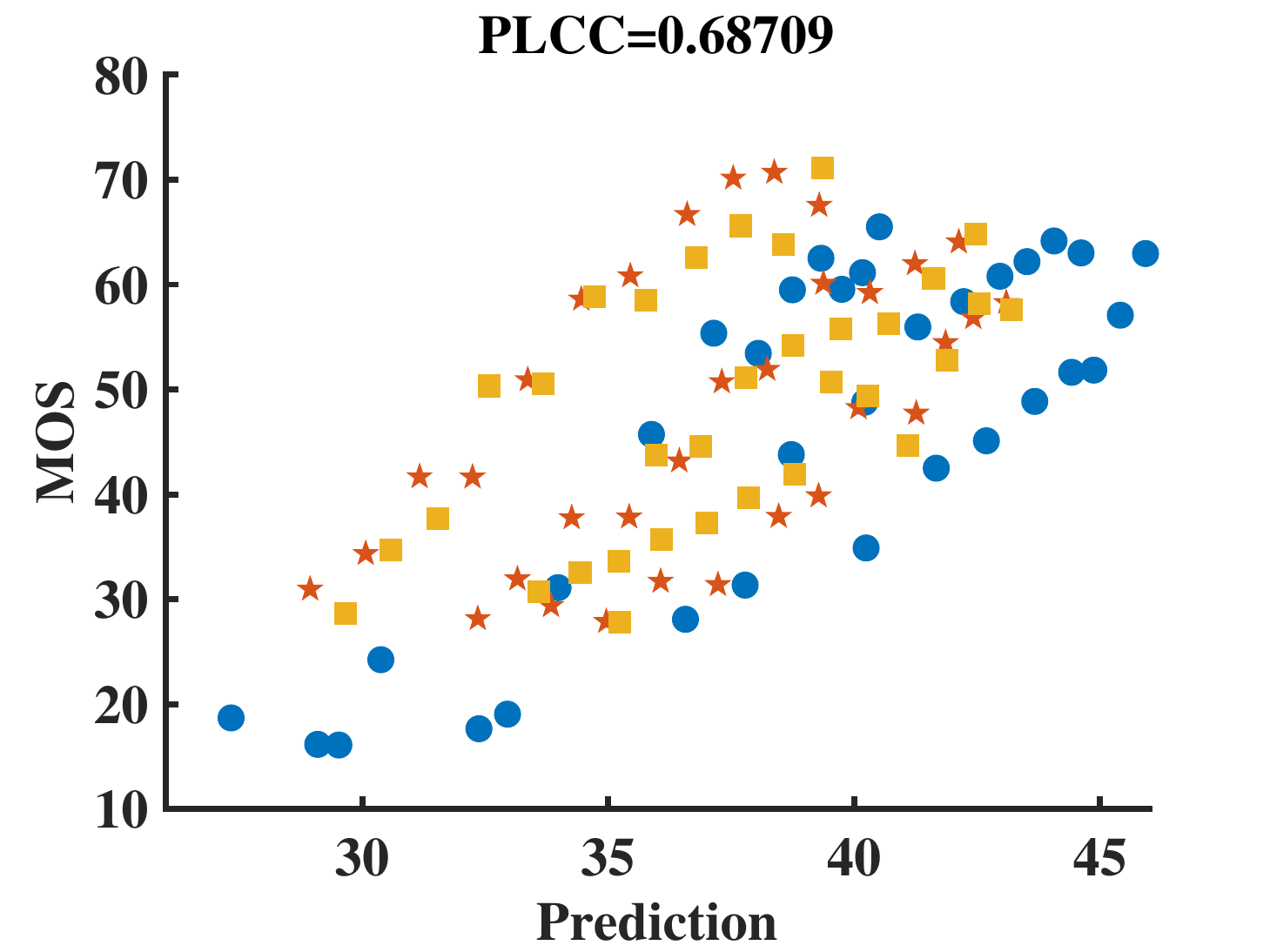}
	}
	\subfigure[SSIM]{
		\includegraphics[height=2.4cm]{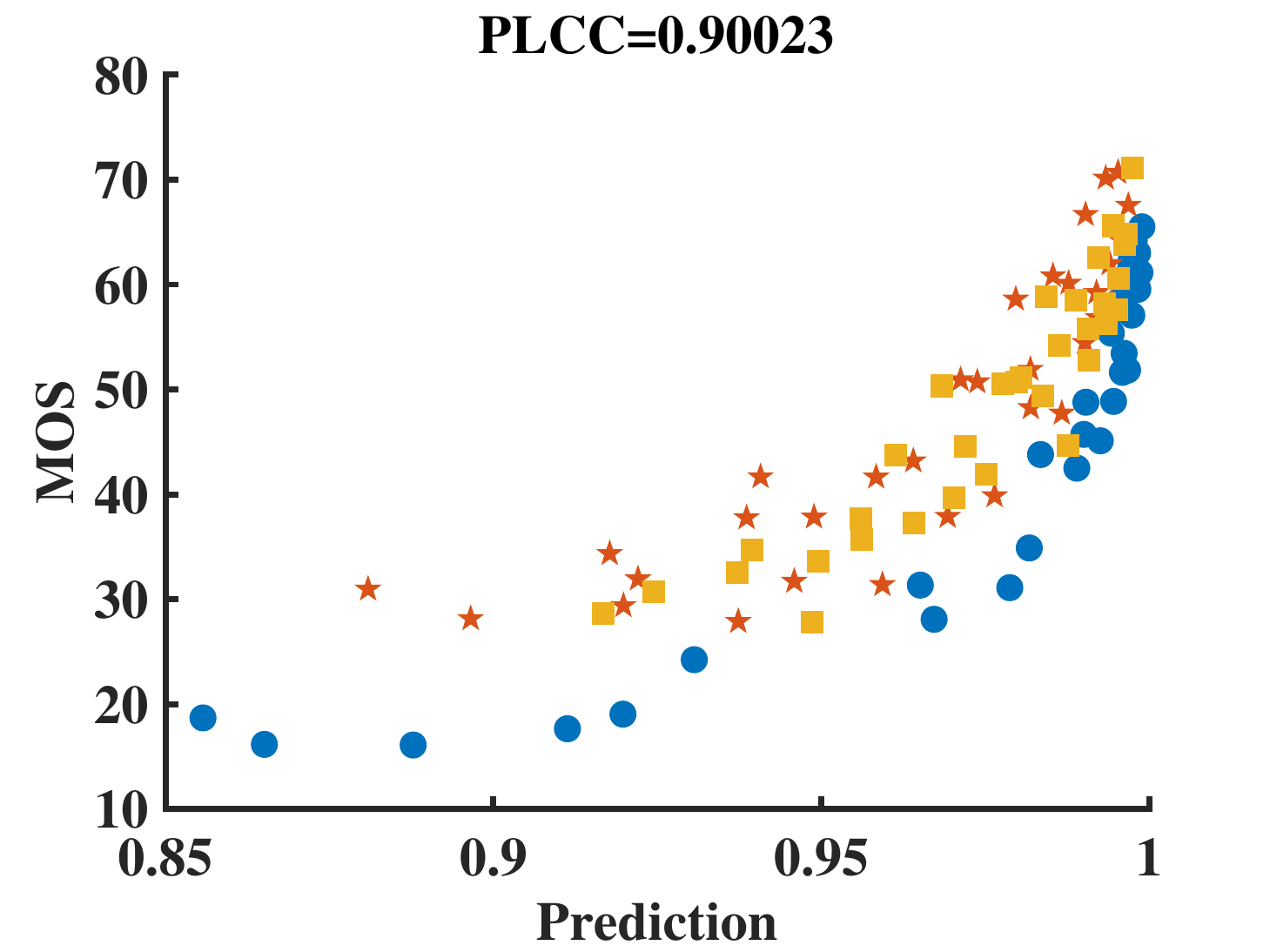}
	}
	\subfigure[MS-SSIM]{
		\includegraphics[height=2.4cm]{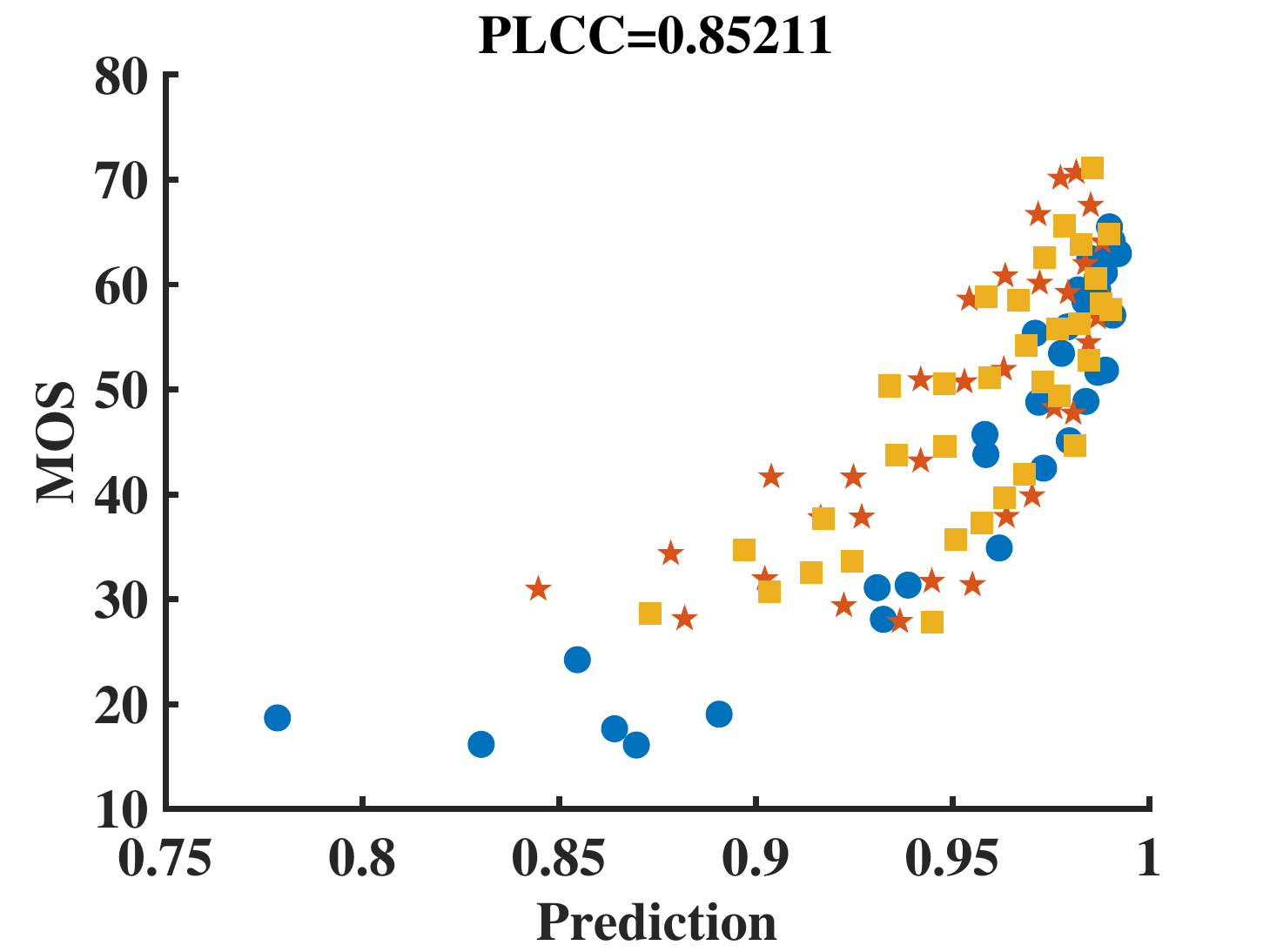}
	}
	\subfigure[FSIM]{
		\includegraphics[height=2.4cm]{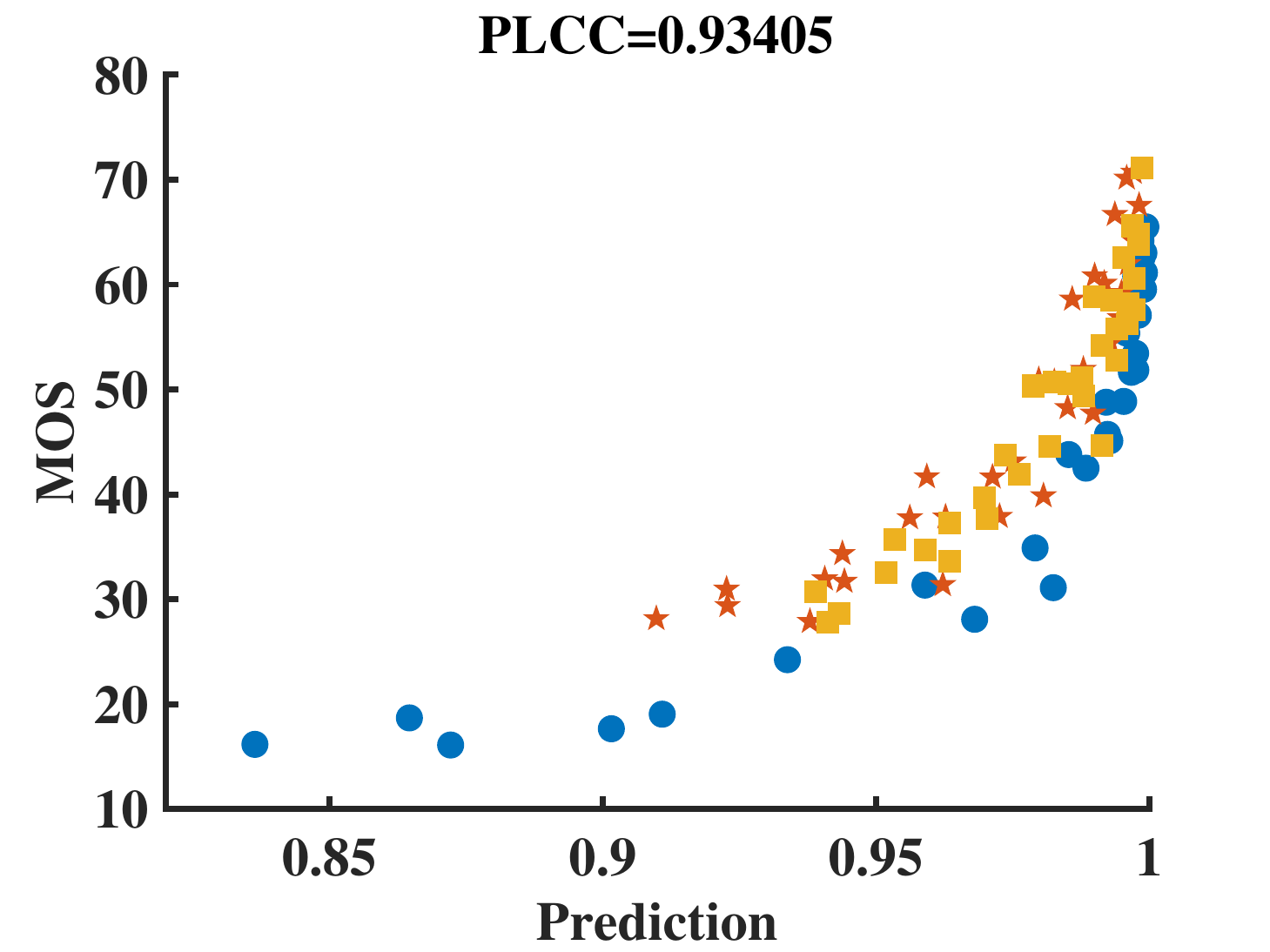}
	}
	\subfigure[DeepQA]{
		\includegraphics[height=2.4cm]{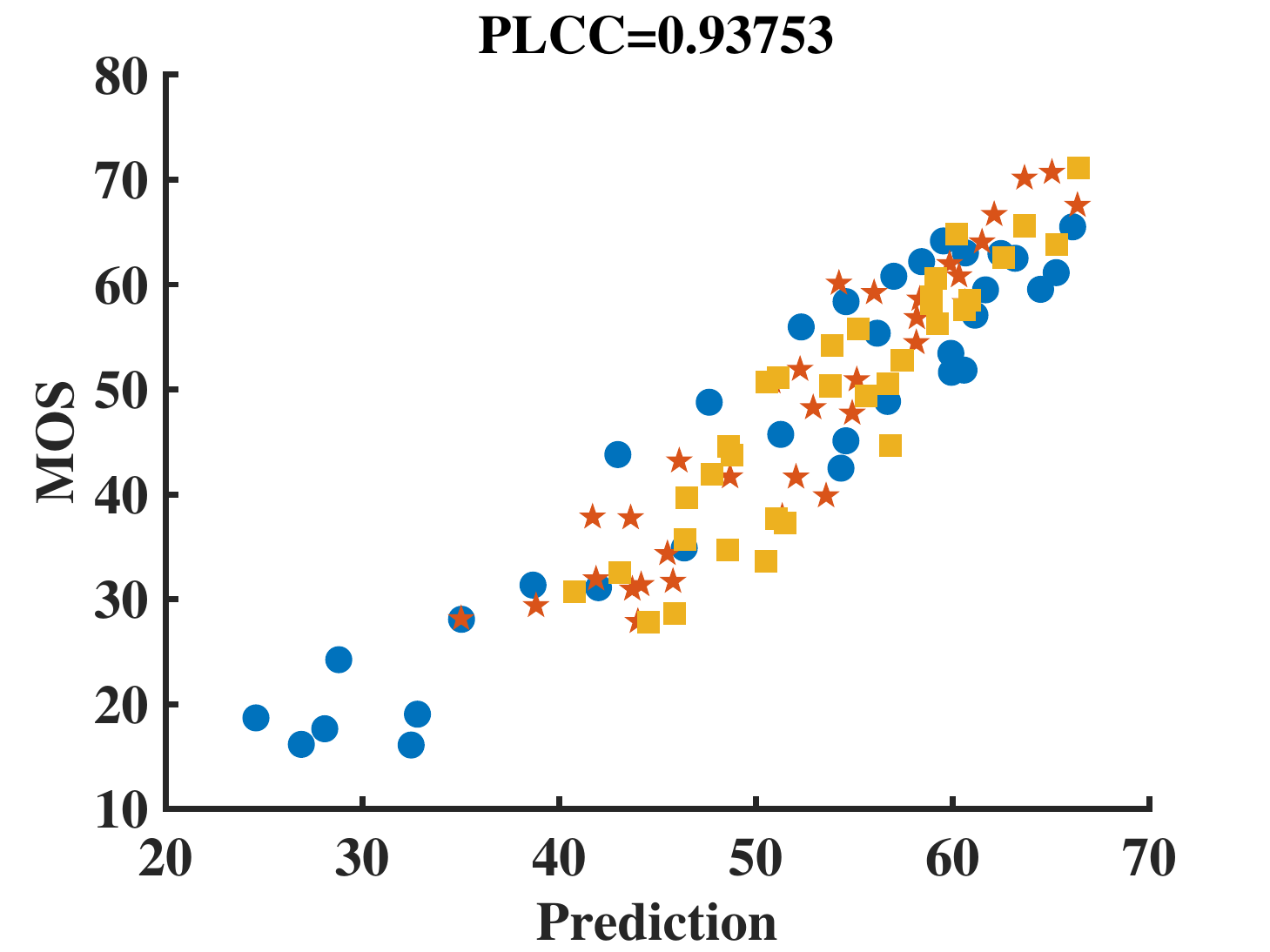}
	}
	\subfigure[BRISQUE]{
		\includegraphics[height=2.4cm]{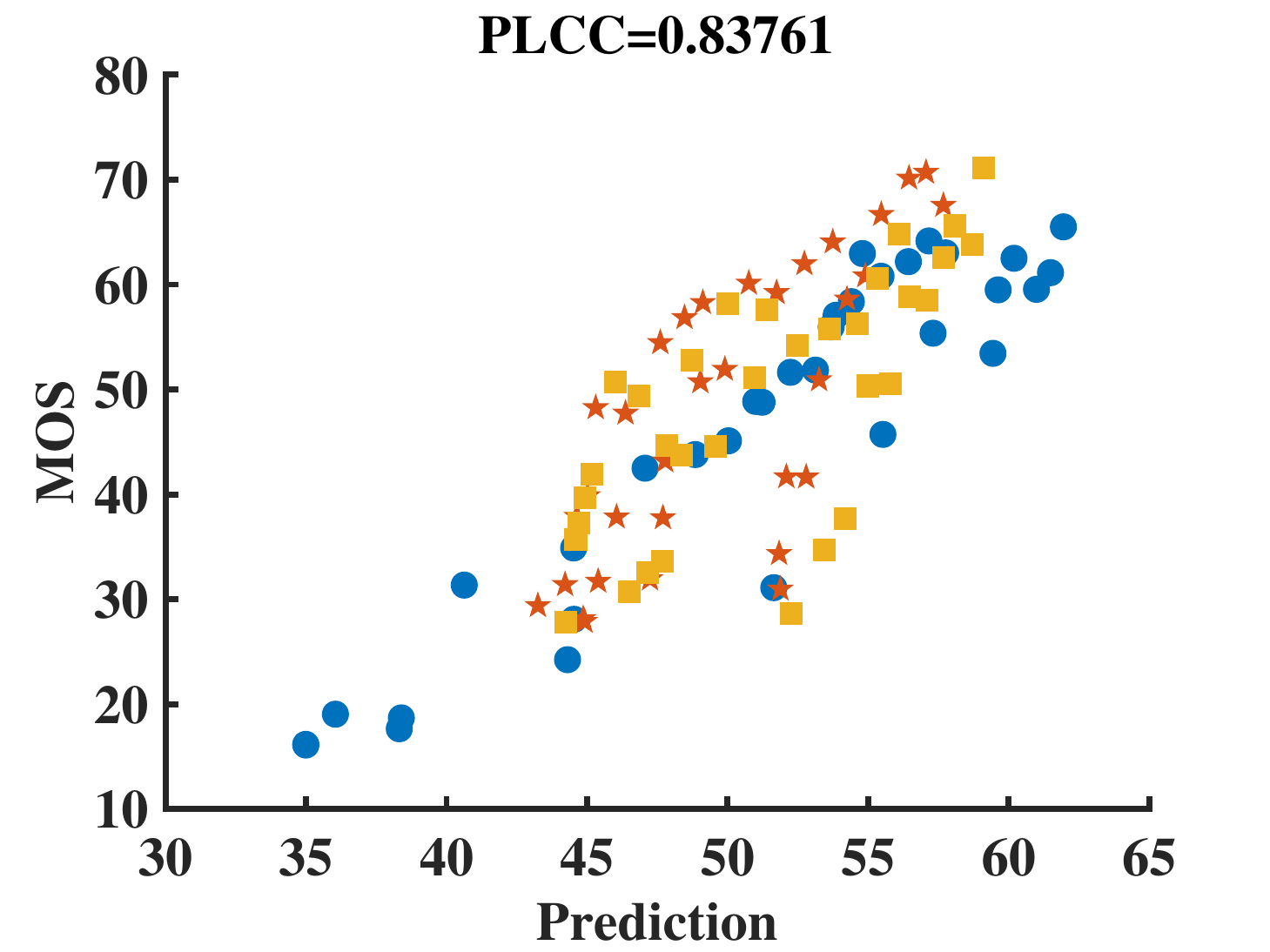}
	}
	\subfigure[BMPRI]{
		\includegraphics[height=2.4cm]{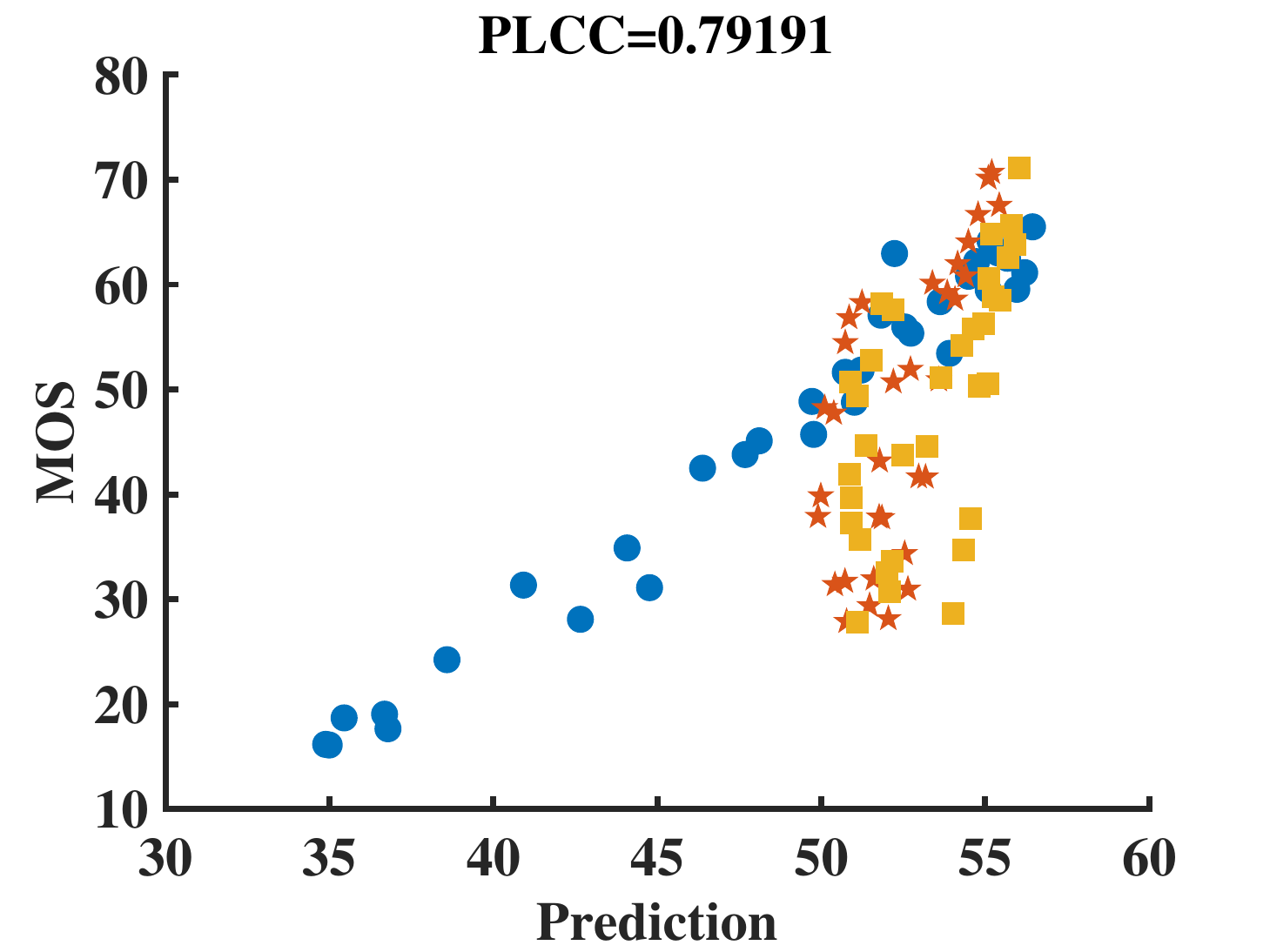}
	}
	\subfigure[DB-CNN]{
		\includegraphics[height=2.4cm]{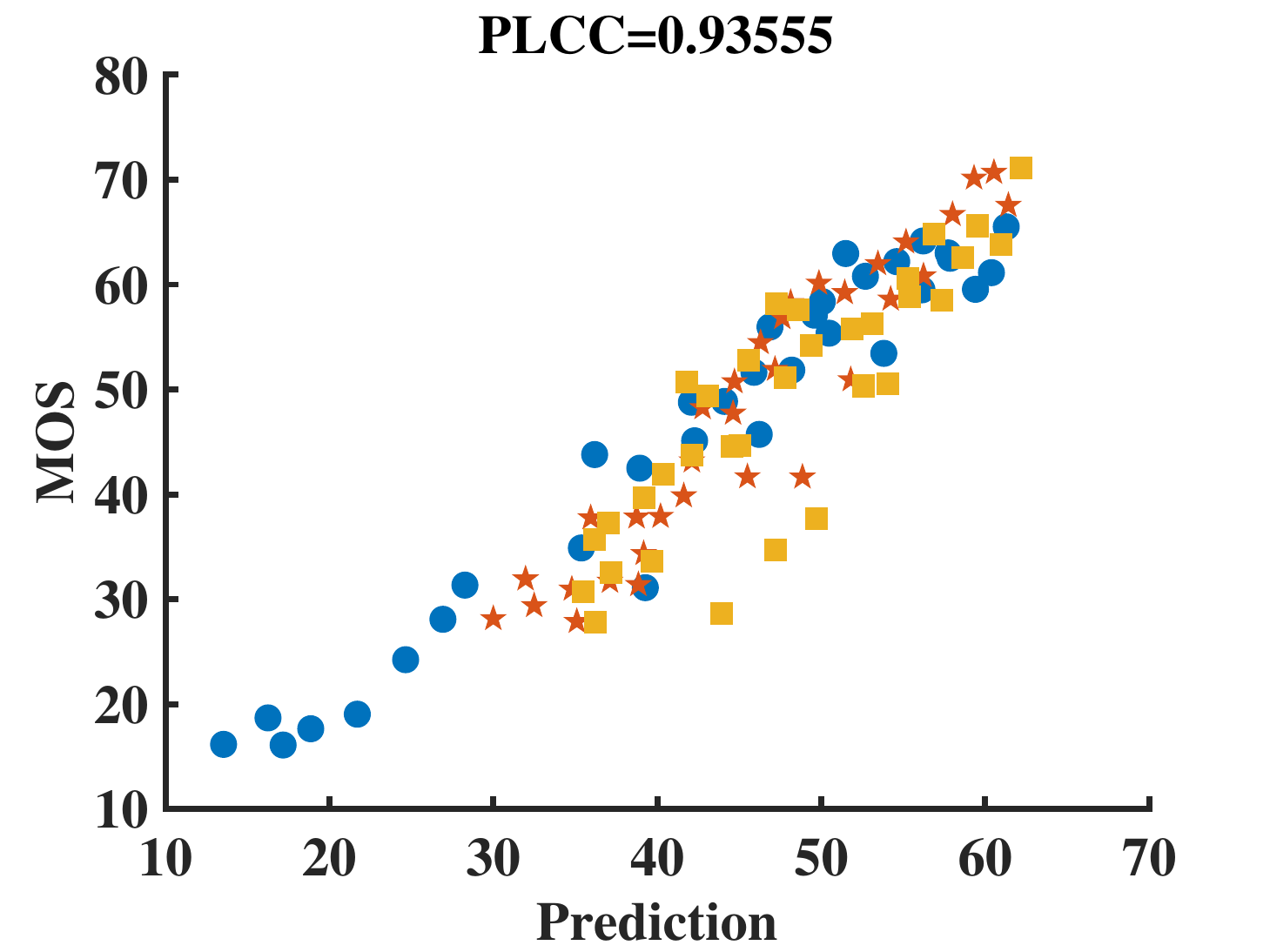}
	}
	\subfigure[MC360IQA]{
		\includegraphics[height=2.4cm]{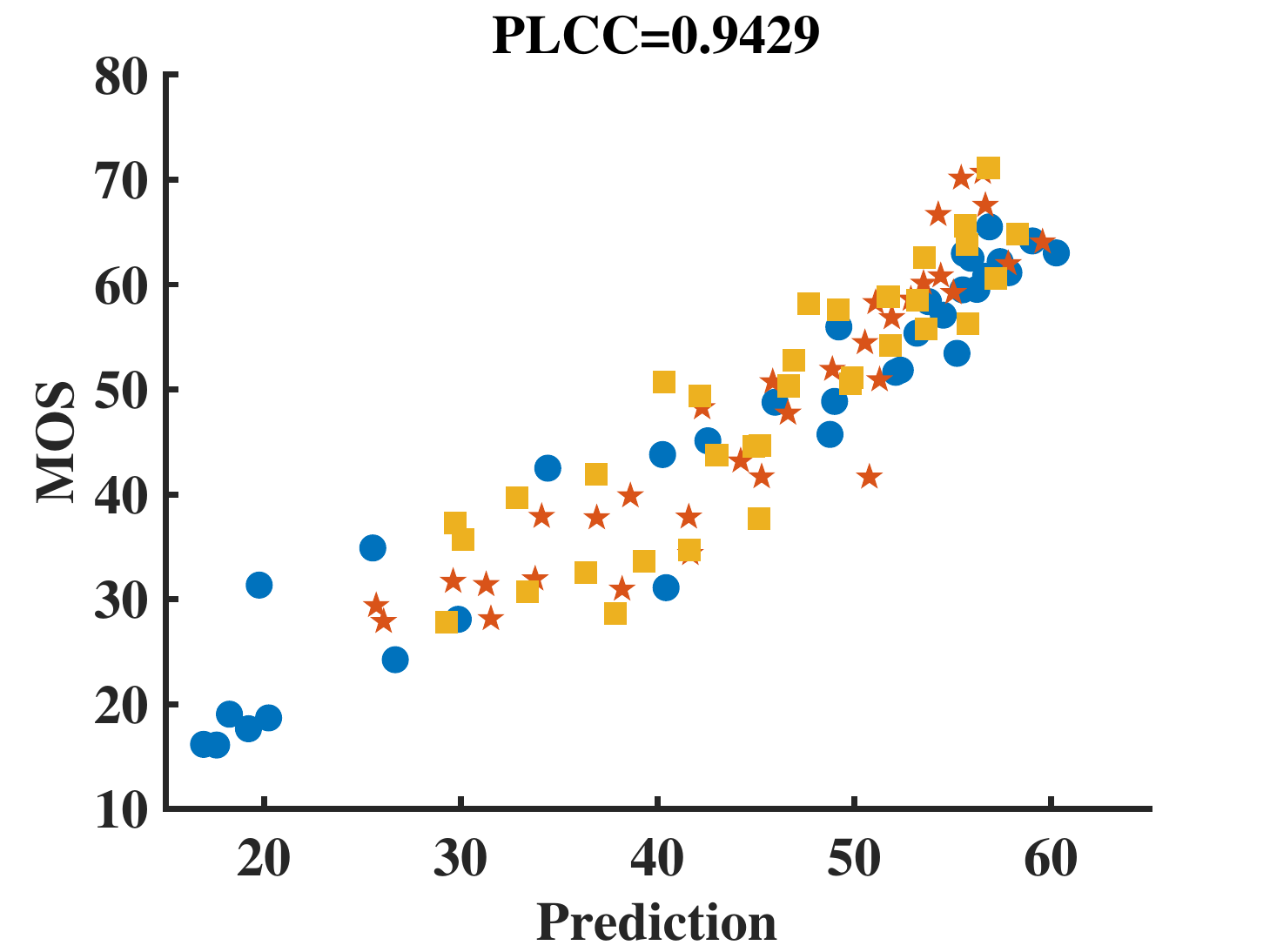}
	}
	\subfigure[VGCN (local)]{
		\includegraphics[height=2.4cm]{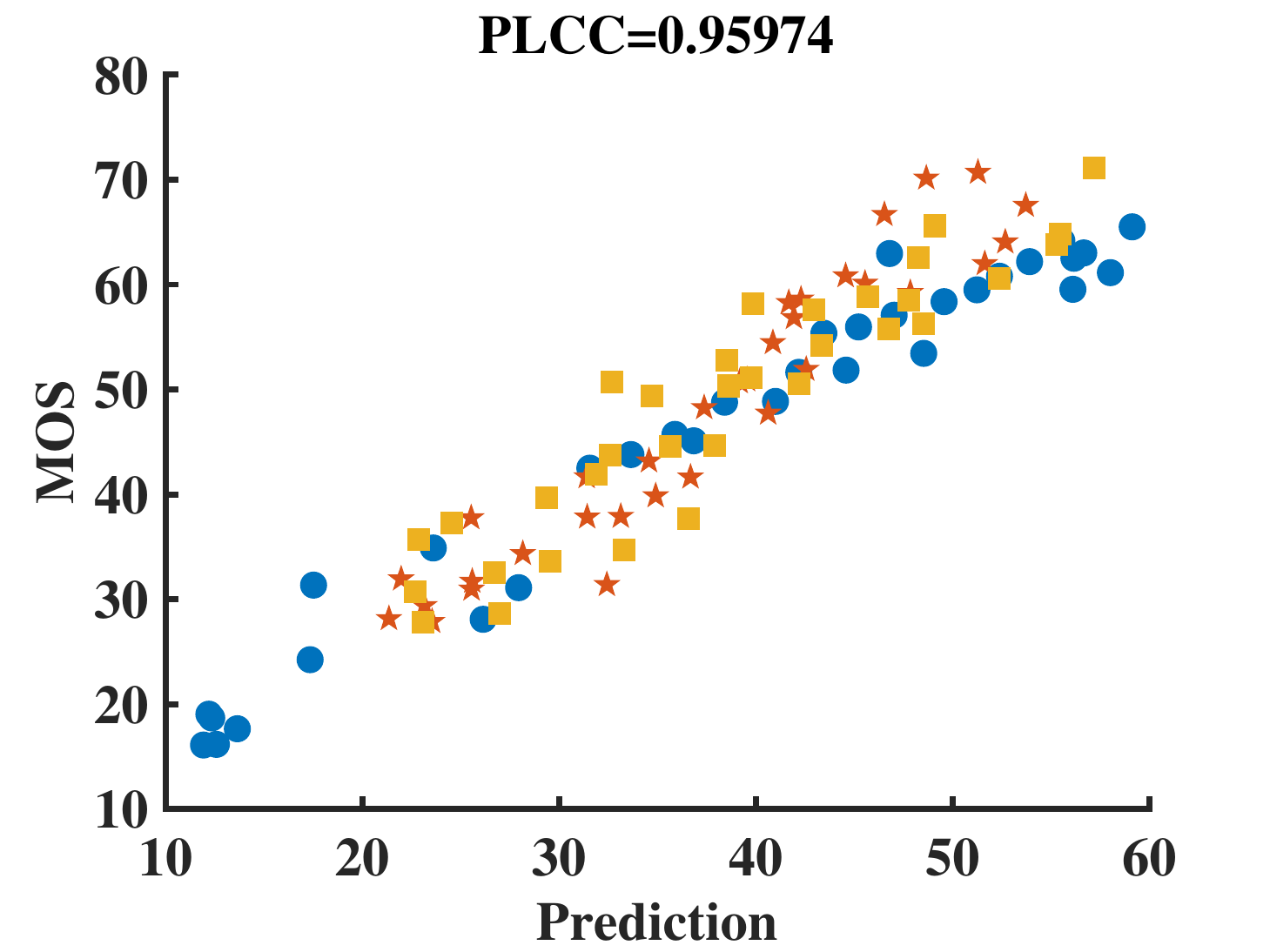}
	}
	\subfigure[VGCN]{
		\includegraphics[height=2.4cm]{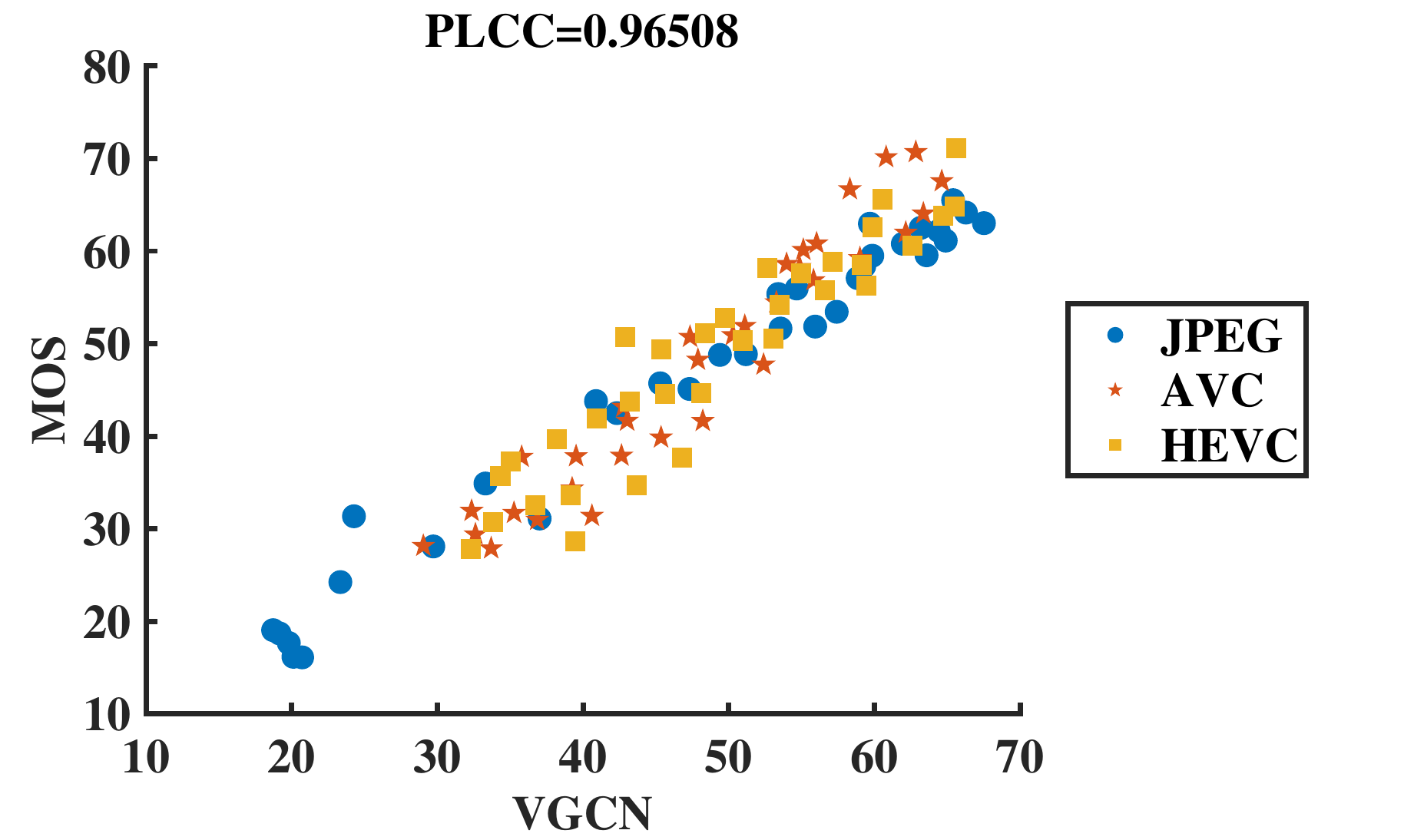}
	}
	\caption{Scatter plots of MOS values against predictions by IQA metrics for individual distortion type on the testing set of CVIQD Database.}
	\label{fig:fig5}
\end{figure*}

We compare our proposed VGCN with other state-of-the-art algorithms including traditional FR IQA metrics PSNR, SSIM \cite{ssim}, MS-SSIM \cite{msssim}, FSIM \cite{fsim}; learning-based FR IQA metric DeepQA \cite{DeepQA}; learning-based NR IQA metrics BRISQUE \cite{BRISQUE}, BMPRI \cite{BMPRI}, DB-CNN \cite{DBCNN}; traditional FR OIQA metrics S-PSNR \cite{SPSNR}, WS-PSNR \cite{WSPSNR}, CPP-PSNR \cite{CPPCNR} and learning-based NR OIQA metric MC360IQA \cite{mc360iqa2}. PSNR and its variants for omnidirectional contents namely S-PSNR, WS-PSNR, CPP-PSNR evaluate the pixel-level fidelity. SSIM assumes that the HVS has an aptitude for extracting structural information and measures structural similarity. MS-SSIM and FSIM are the variants of SSIM, MS-SSIM takes into account the effect of scale and FSIM measures the similarity in the feature domain. DeepQA learns the visual sensitivity maps to weight the error maps of reference and distorted images in existing IQA databases without prior knowledge of the HVS. Natural scene statistic (NSS)-based features in the spatial domain are utilized in BRISQUE to automatically predict the quality with support vector regression. BMPRI applies four types and five levels of distortion to generate multiple pseudo reference images (MPRIs) and describes the similarities between the distorted image and the MPRIs with local binary pattern features. DB-CNN implements the tailored S-CNN and VGG-16 to deal with synthetic and authentic distortions, followed by the bilinear pooling for feature fusion. MC360IQA introduces six parallel sub-networks for assessing the quality of viewport images covering the spherical surface and leverage an image quality regressor to map viewport features onto the perceptual quality. 

For fair comparison, learning-based methods are all re-trained using the same training/testing split scheme adopted in VGCN. The results of performance comparison are listed in Table \ref{table3} and \ref{table4}, and the best performing FR and NR results are highlighted in bold. As we can observe from the tables, PSNR-based OIQA metrics are inferior to traditional SSIM-based IQA metrics. The explanation is that PSNR only evaluates the pixel-level error, but SSIM measures the structural distortion related to the HVS. Besides, the properties of omnidirectional images are not dealt with in 2D IQA metrics, e.g. sphere representation, viewport information. There still exists a gap between 2D IQA and 2D OIQA, resulting in the quality predicted by 2D IQA metrics that cannot correlate well with subjective ratings. Thus, converting the omnidirectional image from ERP format to six viewport images in MC360IQA \cite{mc360iqa2} brings a huge improvement to the prediction accuracy. Motivated by the HVS and viewing process of 360-degree images, our proposed VGCN (local) and VGCN outperform most FR and NR metrics listed in Table \ref{table3} and \ref{table4} owing to 1) the well consideration of viewport information interactions and 2) the incorporation of local and global predictions. Note that we do not list VGCN (global) in the performance comparison because DB-CNN is representative of the global branch.

Then, we illustrate the performance comparison for individual distortion type on the OIQA \cite{OIQADatabse} and CVIQD \cite{CVIQDDatabase} databases. It can be observed from Table \ref{table3} and \ref{table4} that our proposed VGCN achieves competitive performance for most of the distortion types in terms of PLCC, SROCC and RMSE. For compression distortions, e.g. JPEG, AVC, HEVC, VGCN achieves the best performance among NR metrics both on OIQA and CVIQD databases. For other distortions, local branch (local VGCN) or global branch (DB-CNN) is slightly superior to the entire VGCN. Fig. \ref{fig:fig4} and \ref{fig:fig5} exhibit the scatter plots of MOS values versus predictions of IQA models for individual distortion types. For OIQA database, the linear correlation  between subjective score and objective prediction of JPEG compression is generally lower than other three distortions. As observed from Fig. \ref{fig:fig4}, the points denoting JPEG compression deviate from the line distribution. For CVIQD database, as shown in Table \ref{table4} and Fig. \ref{fig:fig5}, it gets more and more difficult to accurately predict the perceived quality of omnidirectional images with the development of coding technologies. The reason may be that the artifacts brought by new codecs are much more complicated than blockiness and blurriness, and the quality ranges of AVC/HEVC are narrower than that of JPEG, making it tough for quality prediction.

\begin{figure}[t]
	\centering
	\subfigure[Different vs. Similar (AUC)]{
		\includegraphics[width=8.5cm]{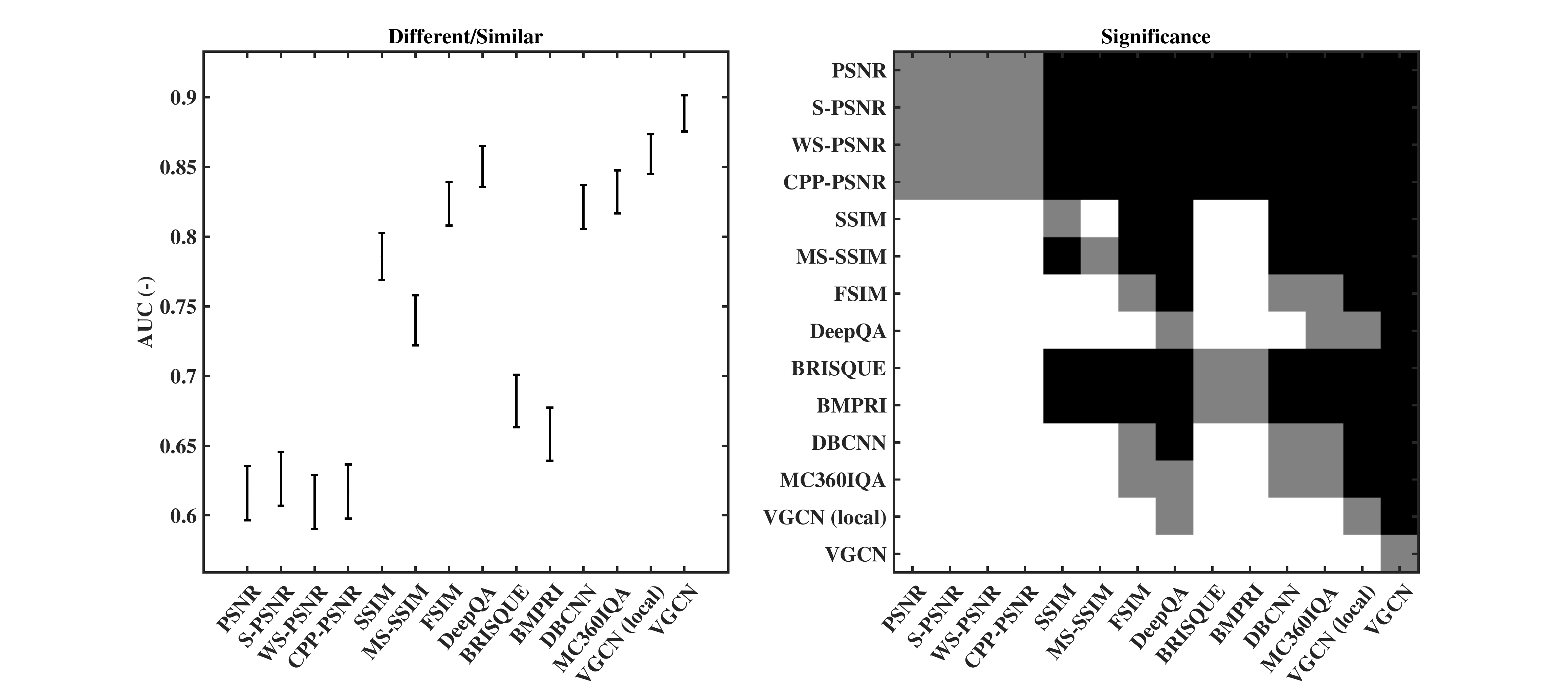}
	}
	\subfigure[Better vs. Worse ($C_0$)]{
		\includegraphics[width=8.5cm]{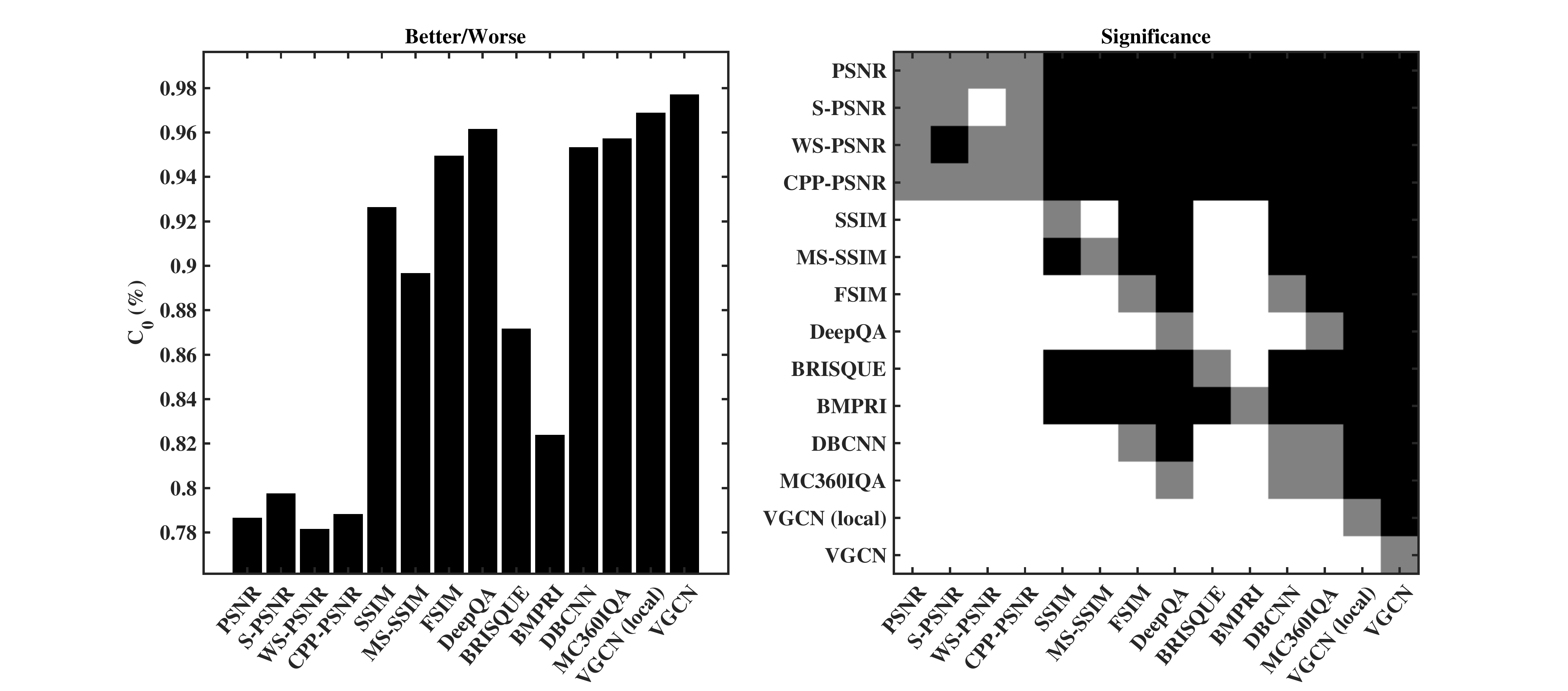}
	}
	\subfigure[Better vs. Worse (AUC)]{
		\includegraphics[width=8.5cm]{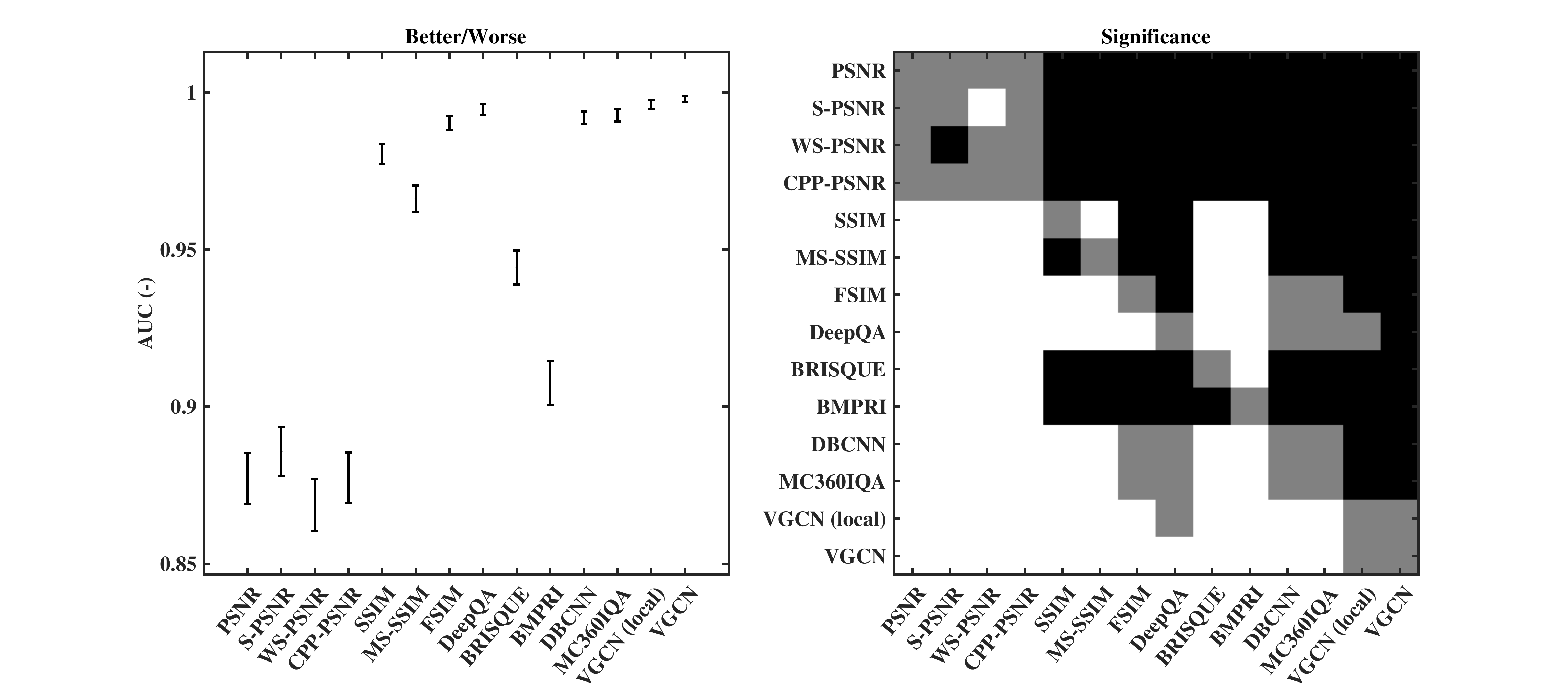}
	}
	\caption{The results and statistical analysis for the testing split of CVIQD database.}
	\label{fig:fig6}
\end{figure}

Finally, we implement the Krasula criteria \cite{krasula2016accuracy} to evaluate the capacity of state-of-the-art metrics for distinguishing different/similar and better/worse pairs. Individual scores are required for significance analysis of the subjective ratings to employ the Krasula methodology, and we only have access to CVIQD database with individual scores, thus the experiments are conducted on the CVIQD database. Fig. \ref{fig:fig6} depicts the AUC of \textit{Different vs. Similar}, \textit{Better vs. Worse} categories and percentage of correct classification. The error bars denote 95\% confidence intervals. Superiority of the proposed VGCN is demonstrated according to AUC-DS, AUC-BW and $C_0$. Besides, the performance of AUC-BW is generally higher than that of AUC-DS, which means distinguishing different/similar pairs remains a challenging problem compared with better/worse pairs. In the significance plots of Fig. \ref{fig:fig6}, white(black) boxes refer to the case that the metric in the row is significantly better(worse) than the metric in the column, and gray boxes denote that the two metrics are statistically indistinguishable. Then, we come to the conclusions that 1) VGCN significantly outperforms other metrics in the first and second analysis (Fig \ref{fig:fig6} (a) and Fig \ref{fig:fig6} (b)), and 2) VGCN and its local branch have competitive performance in the third analysis (Fig. \ref{fig:fig6}(c)) since both of them have AUC-BW sufficiently close to 1.

\subsection{Cross Database Validation}

To verify the generalization ability of the proposed VGCN, cross database validation is conducted. To be specific, we train VGCN with OIQA (CVIQD) database and utilize CVIQD (OIQA) database to test the model. The corresponding results are listed in Table \ref{table5}. Since OIQA database contains more distortions, e.g., compression artifacts, blurriness and noise, while CVIQD database only includes compression degradations, the generalization ability of the network trained on OIQA database is much better than that trained on CVIQD database as shown in Table \ref{table5}. Moreover, the viewport descriptor in VGCN is pre-trained on 2D IQA databases with various distortions, thus the performance of VGCN is superior to other state-of-the-art metrics by a large margin. To conclude, our proposed VGCN demonstrate a strong generalization ability to different databases.

\begin{table}[ht]
	\begin{center}
		\captionsetup{justification=centering}
		\caption{\textsc{Performance Evaluation of Cross Database Test. The Best Performing Results are Highlighted in Bold.}}
		\label{table5}
		\scalebox{0.9}{
		\begin{tabular}{@{}c|ccc|ccc@{}}
			\toprule
			& \multicolumn{3}{c|}{Train OIQA/Test CVIQD}          & \multicolumn{3}{c}{Train CVIQD/Test OIQA}           \\ \midrule
			& PLCC            & SROCC           & RMSE            & PLCC            & SROCC           & RMSE            \\ \midrule
			BRISQUE \cite{BRISQUE}      & 0.7543          & 0.6891          & 9.3805          & 0.6816          & 0.5238          & 1.5471          \\
			BMPRI \cite{BMPRI}       & 0.8007          & 0.7492          & 8.5600          & 0.6483          & 0.5890          & 1.6097          \\
			DBCNN \cite{DBCNN}       & 0.7896          & 0.7684          & 8.7669          & 0.5817          & 0.5299          & 1.7198          \\
			MC360IQA \cite{mc360iqa2}    & 0.8886          & 0.8629          & 6.5526          & 0.4375          & 0.3329          & 1.9012          \\
			VGCN         & \textbf{0.9241} & \textbf{0.9050} & \textbf{5.4616} & \textbf{0.7911} & \textbf{0.7832} & \textbf{1.2934} \\ \bottomrule
		\end{tabular}}
	\end{center}
\end{table}

\subsection{Sampling Strategies in Viewpoint Detector}

The viewpoint detector is employed to select viewports appealing to observers inspired by the HVS that humans are more sensitive to structural information \cite{ssim}. To validate whether the viewports selected are actually beneficial to quality prediction of 360-degree images, we adopt other two viewport sampling strategies, namely random sampling and uniform sampling. Specifically, random sampling means using random viewports in VGCN during training and testing procedures. Uniform sampling denotes selecting viewports equidistantly at fixed latitude intervals \cite{soiqe}. Table \ref{table6} illustrates the performance evaluation for different viewport sampling strategies and the model with specially selected viewports achieves the best performance. Note that we only leverage the local branch to test the performance of diverse viewport selection methods.

\begin{table}[ht]
	\begin{center}
		\captionsetup{justification=centering}
		\caption{\textsc{Performance Evaluation for Different Viewpoint Detectors in VGCN Local Branch. The Best Performing Results are Highlighted in Bold.}}
		\label{table6}
		\scalebox{0.87}{
			\begin{tabular}{@{}c|ccc|ccc@{}}
				\toprule
				& \multicolumn{3}{c|}{OIQA Database}                  & \multicolumn{3}{c}{CVIQD Database}                  \\ \midrule
				Viewpoint Detector & PLCC            & SROCC           & RMSE            & PLCC            & SROCC           & RMSE            \\ \midrule
				Random             & 0.9409          & 0.9277          & 0.7076          & 0.9443          & 0.9310          & 4.6941          \\
				Uniform            & 0.9424          & 0.9317          & 0.6990          & 0.9484          & 0.9320          & 4.5215          \\
				Proposed           & \textbf{0.9529} & \textbf{0.9444} & \textbf{0.6340} & \textbf{0.9597} & \textbf{0.9539} & \textbf{3.9220} \\ \bottomrule
		\end{tabular}}
	\end{center}
\end{table}

\begin{figure}[t]
	\centerline{\includegraphics[width=8cm]{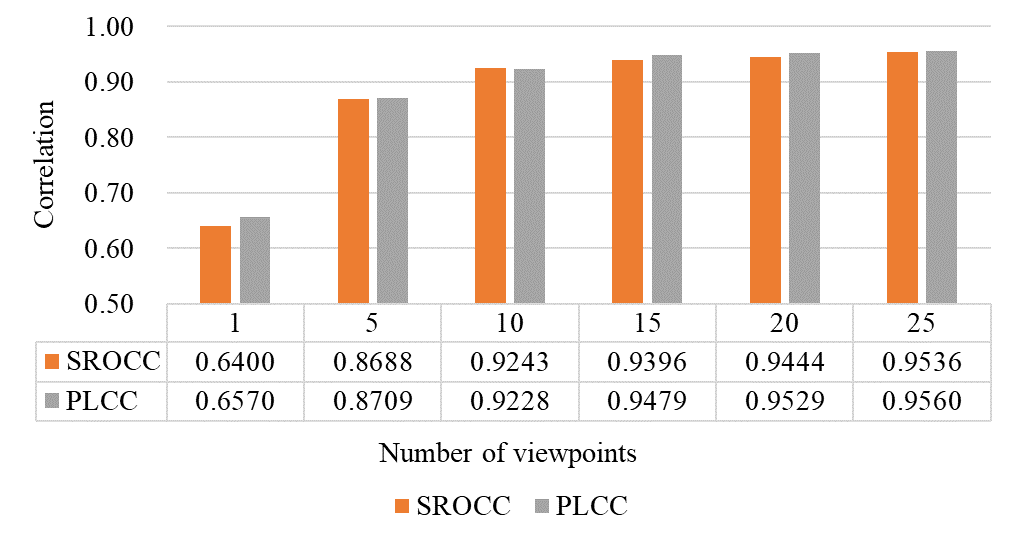}}
	\caption{The influence of the number of viewpoints in viewpoint detector.}
	\centering
	\label{fig:fig7}
\end{figure}

Moreover, as shown in Fig. \ref{fig:fig7}, we conduct an experiment to show the influence of the number of viewpoints in viewpoint detector. The SROCC and PLCC performances have been improved on OIQA database as more viewpoints are included. To keep a balance between the computation efficiency and correlation with subjective ratings, $N$ equaling to 20 will be a good choice as is done in \cite{VCNN}.

\subsection{Structure of Viewport Descriptor}

In our implementation, ResNet-18 is chosen as viewport descriptor. It is interesting to explore how viewport descriptor architecture affects the overall performance, thus we replace ResNet-18 with other models and use the local branch of VGCN to test the performance for simplicity. Table \ref{table7} lists the performance of different architectures and the results on OIQA and CVIQD database show that ResNet has better capability to extract discriminative features for measuring various distortion types and levels. Besides, ResNet-34 and ResNet-18 exhibit similar performance but ResNet-18 has fewer parameters than ResNet-34. As a result, ResNet-18 is preferable owing to its low complexity and high performance. With the fast development of deep learning technologies, either the viewpoint detector, viewport descriptor or the graph CNN can be easily replaced by other state-of-the-art algorithms and models, which can further boost the performance of current VGCN.

\begin{table}[ht]
	\begin{center}
		\captionsetup{justification=centering}
		\caption{\textsc{Performance Evaluation for Different Viewport Descriptors in VGCN Local Branch. The Best Performing Results are Highlighted in Bold.}}
		\label{table7}
		\scalebox{0.87}{
			\begin{tabular}{@{}c|ccc|ccc@{}}
				\toprule
				& \multicolumn{3}{c|}{OIQA Database}                  & \multicolumn{3}{c}{CVIQD Database}                  \\ \midrule
				Viewport Descriptor & PLCC            & SROCC           & RMSE            & PLCC            & SROCC           & RMSE            \\ \midrule
				VGG-16 \cite{vgg}   & 0.9127          & 0.9067          & 0.8538          & 0.9504          & 0.9363          & 4.4356          \\
				ResNet-18 \cite{resnet} & \textbf{0.9529} & \textbf{0.9444} & \textbf{0.6340} & 0.9597          & 0.9539          & 3.9220          \\
				ResNet-34 \cite{resnet} & 0.9516          & 0.9430          & 0.6420          & \textbf{0.9641} & \textbf{0.9573} & \textbf{3.7882} \\ \bottomrule
		\end{tabular}}
	\end{center}
\end{table}

\subsection{Ablation Study}

We conduct the ablation study to prove the effectiveness of each component in VGCN. The details are described as follows:

\subsubsection{\textbf{Spatial Viewport Graph}}

\begin{figure}[h]
	\centering
	\subfigure[]{
		\includegraphics[height=3.5cm]{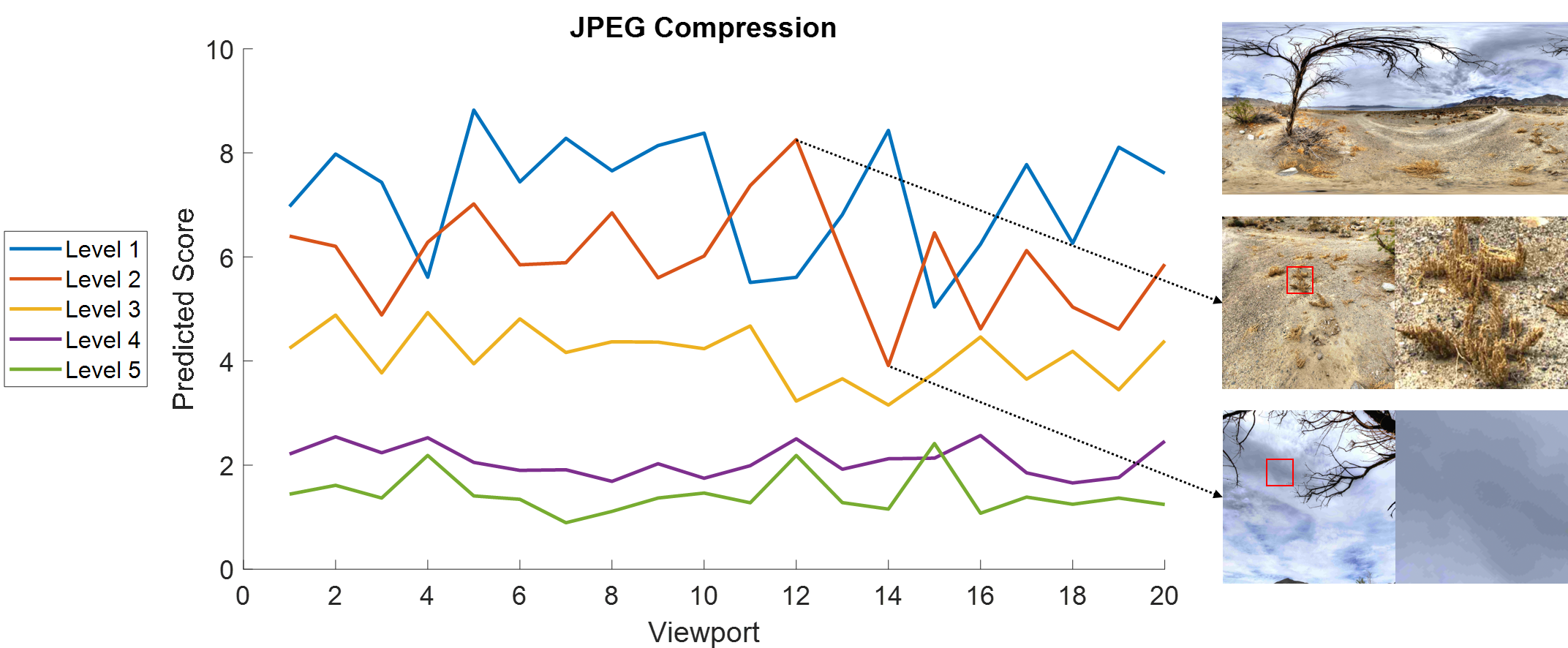}
	}
	\subfigure[]{
		\includegraphics[height=3.5cm]{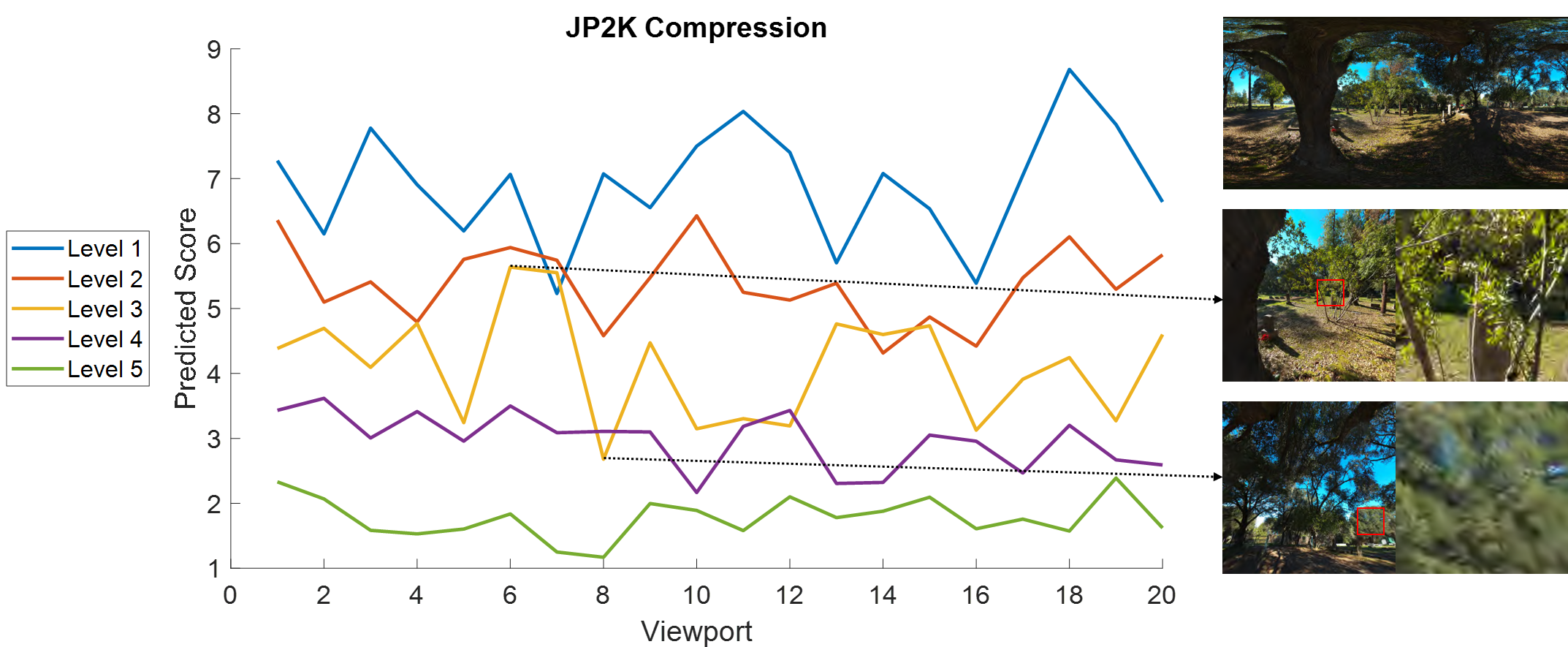}
	}
	\subfigure[OIQA Database]{
		\includegraphics[height=3.5cm]{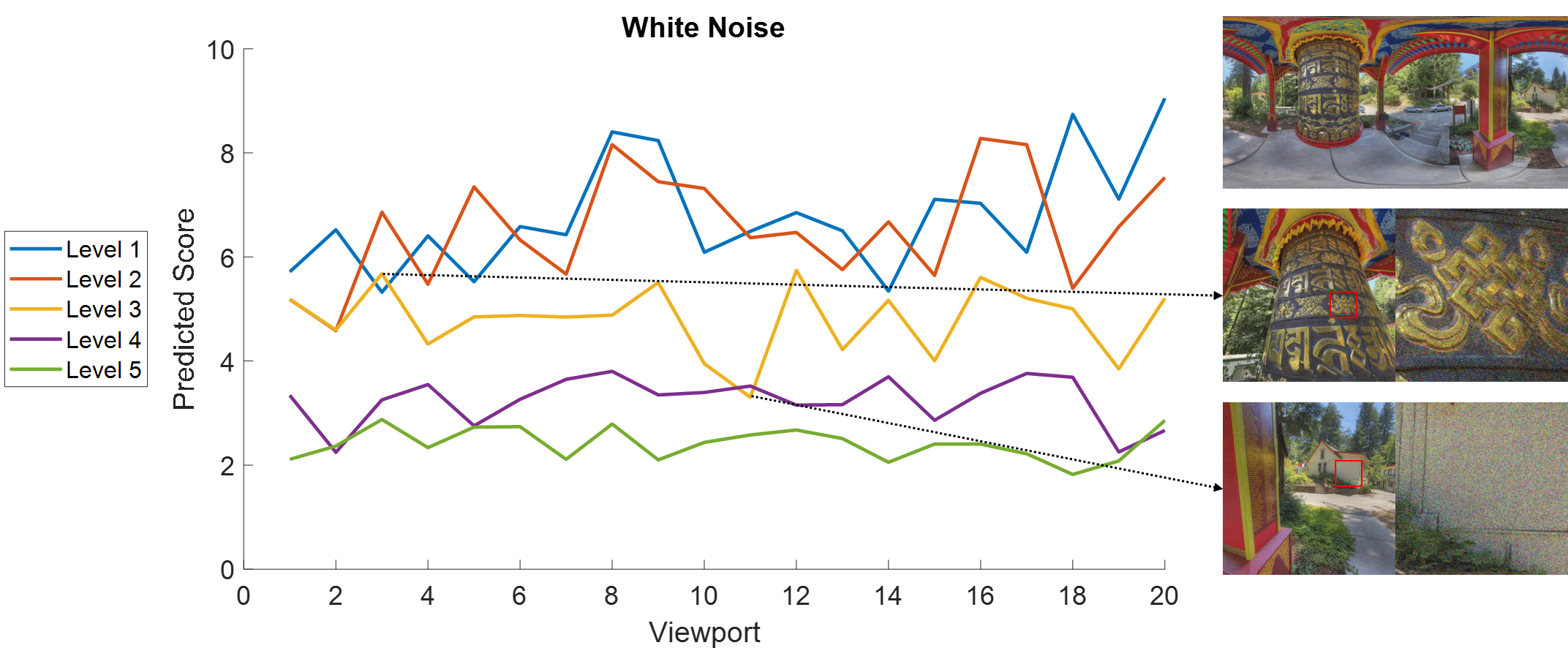}
	}
	\subfigure[]{
		\includegraphics[height=3.5cm]{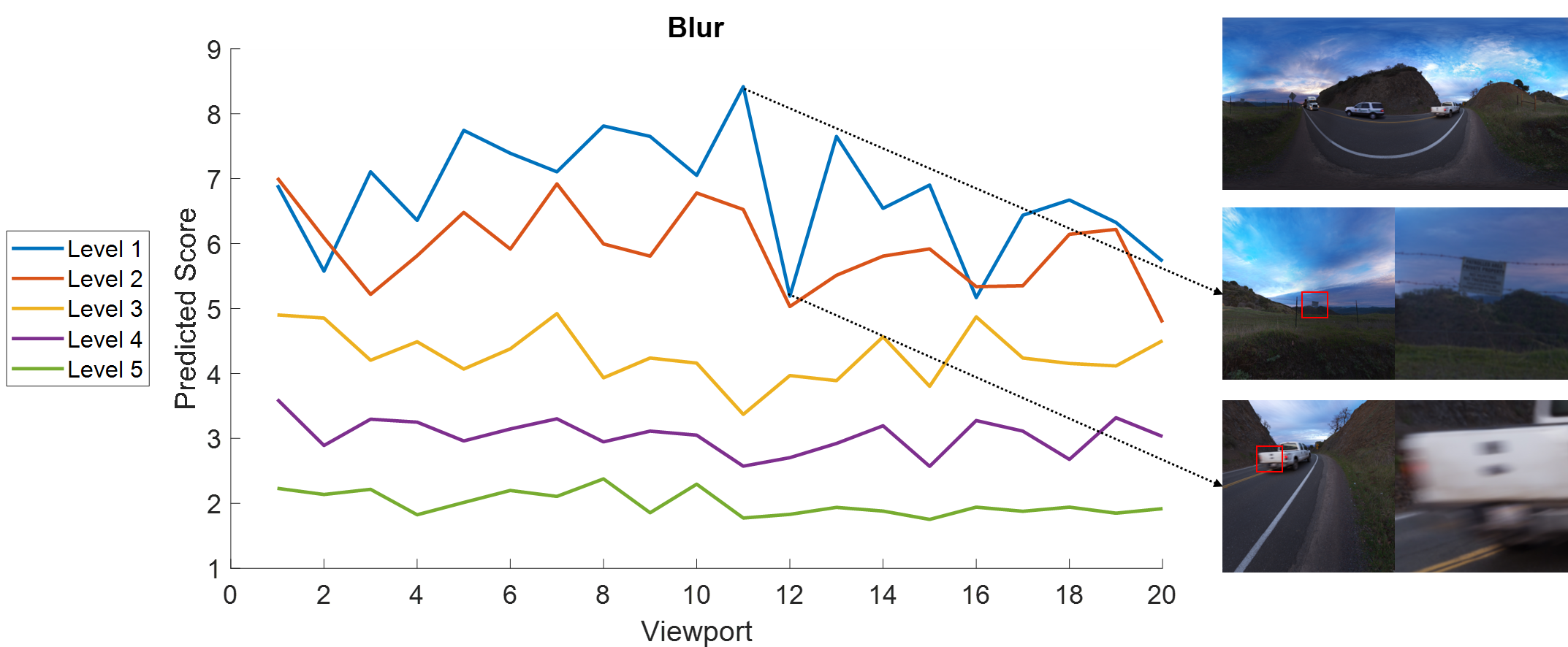}
	}
	\caption{Predicted scores of the selected viewports for different distortion types and levels by VGCN local branch.}
	\label{fig:fig8}
\end{figure}

We build a spatial viewport graph based on the selected viewports and apply the graph convolutional network to capture interactions between different viewports. To verify the effectiveness of the proposed method, we utilize five fully connected layers instead of graph convolutional layers in the viewport quality aggregator, and test the performance with the local branch. Besides, we use the viewport descriptor and fully-connected layer to estimate the local quality of the given viewports, and weight the local quality with different averaging strategies. Specifically, we weight the viewport qualities with their content and location. For content weight, we use the viewport feature vector output by the viewport descriptor to encode the viewport content and then use the channel attention mechanism \cite{Scacnn} to assign weight for each viewport. For location weight, the latitude of the viewpoint \cite{WSPSNR} is leveraged to compute the viewport weighing factor. The experimental results in Table \ref{table8} demonstrate that content and location weight can boost the performance to some extent,
but modeling the mutual dependency of viewports via GCN will bring huge improvement to the network performance due to the well consideration of viewport information interaction and aggregation. 

\begin{table}[ht]
	\begin{center}
		\captionsetup{justification=centering}
		\caption{\textsc{Performance Evaluation for Different Viewport Quality Aggregators in VGCN Local Branch. The Best Performing Results are Highlighted in Bold.}}
		\label{table8}
		\scalebox{0.76}{
			\begin{tabular}{@{}c|ccc|ccc@{}}
				\toprule
				& \multicolumn{3}{c|}{OIQA Database}                  & \multicolumn{3}{c}{CVIQD Database}                  \\ \midrule
				Viewport Quality Aggregator & PLCC            & SROCC           & RMSE            & PLCC            & SROCC           & RMSE            \\ \midrule
				Fully Connected (FC) Layer       & 0.8830          & 0.8551          & 0.9807          & 0.8610          & 0.8526          & 7.2551          \\
				FC Layer with Content Weight   & 0.8925          & 0.8772          & 0.9427          & 0.9428          & 0.9280          & 4.7531          \\
				FC Layer with Location Weight   & 0.8983          & 0.8821          & 0.9182          & 0.9430          & 0.9278          & 4.7461          \\
				Graph Convolutional Layer   & \textbf{0.9529} & \textbf{0.9444} & \textbf{0.6340} & \textbf{0.9597} & \textbf{0.9539} & \textbf{3.9220} \\ \bottomrule
		\end{tabular}}
	\end{center}
\end{table}

We visualize the predict scores of different viewports after the last activation layer of spatial viewport graph in Fig. \ref{fig:fig8}. As is shown, VGCN can distinguish different distortion levels as well as perceive the quality fluctuation of the viewports in one omnidirectional image. For JPEG compression artifacts and white noise in Fig. \ref{fig:fig8} (a) and (c), viewports with more texture information are able to mask the artifacts/noise to some extent, while distortions in smooth regions are more noticeable, thus those viewports tend to have lower predicted quality score. In Fig. \ref{fig:fig8} (b) and (d), VGCN give lower scores for the viewports with salient blurry object.

\subsubsection{\textbf{Global Branch}}

We adopt the global quality estimator to predict the perceptual quality using the entire omnidirectional image without viewport sampling. It is the supplementary of the local branch according to Fig. \ref{fig:fig1}. We compare the performance with or without global branch to demonstrate the validity of it. The results are listed in Table \ref{table9} and we can conclude that global quality estimation of 360-degree images can further boost the performance of VGCN model, since after browsing several viewports and obtaining local quality, observers will `reconstruct' the whole scenery to acquire global quality for assisting perceptual quality assessment of omnidirectional contents as illustrated in Fig. \ref{fig:fig1}. 

\begin{table}[ht]
	\begin{center}
		\captionsetup{justification=centering}
		\caption{\textsc{Performance Evaluation with or without Global Branch. The Best Performing Results are Highlighted in Bold.}}
		\label{table9}
		\scalebox{0.85}{
			\begin{tabular}{@{}c|ccc|ccc@{}}
				\toprule
				& \multicolumn{3}{c|}{OIQA Database}                  & \multicolumn{3}{c}{CVIQD Database}                  \\ \midrule
				Architecture & PLCC            & SROCC           & RMSE            & PLCC            & SROCC           & RMSE            \\ \midrule
				Without Global Branch       & 0.9529          & 0.9444          & 0.6340          & 0.9597          & 0.9539          & 3.9220          \\
				With Global Branch   & \textbf{0.9584} & \textbf{0.9515} & \textbf{0.5967} & \textbf{0.9651} & \textbf{0.9639} & \textbf{3.6573} \\ \bottomrule
		\end{tabular}}
	\end{center}
\end{table}

%\subsection{Computation Complexity}

\section{Conclusion}

In this paper, motivated by the HVS and the viewing process of omnidirectional images, we develop a blind OIQA framework containing local and global branches. Specifically, the local branch imitates the viewport information interaction and aggregation to predict perceptual quality when browsing the scenery, and the global branch simulates the subjective rating of `reconstructed' scenery in the hallucination. The viewpoint detector and viewport descriptor in the local branch aim to select attractive viewports and extract discriminative features for quality estimation. The viewport quality aggregator is designed to model the mutual dependencies of different viewports via graph convolutional networks. According to the experimental results, the proposed VGCN achieves state-of-the-art performance and exhibits a promising generalization capability for various distortion types. In the future, we will convert the viewpoint detector into a learning-based model to select more accurate viewports with reinforcement learning. Then, it can be optimized in an end-to-end manner. Besides, the temporal relationship of different viewports could be explored to improve the proposed method. In addition, we will conduct experiments on different HMDs when more omnidirectional databases with various HMDs are established in the future.

\bibliographystyle{IEEEtran}
\bibliography{references}

\begin{thebibliography}{10}
\providecommand{\url}[1]{#1}
\csname url@samestyle\endcsname
\providecommand{\newblock}{\relax}
\providecommand{\bibinfo}[2]{#2}
\providecommand{\BIBentrySTDinterwordspacing}{\spaceskip=0pt\relax}
\providecommand{\BIBentryALTinterwordstretchfactor}{4}
\providecommand{\BIBentryALTinterwordspacing}{\spaceskip=\fontdimen2\font plus
\BIBentryALTinterwordstretchfactor\fontdimen3\font minus
  \fontdimen4\font\relax}
\providecommand{\BIBforeignlanguage}[2]{{%
\expandafter\ifx\csname l@#1\endcsname\relax
\typeout{** WARNING: IEEEtran.bst: No hyphenation pattern has been}%
\typeout{** loaded for the language `#1'. Using the pattern for}%
\typeout{** the default language instead.}%
\else
\language=\csname l@#1\endcsname
\fi
#2}}
\providecommand{\BIBdecl}{\relax}
\BIBdecl

\bibitem{background1}
\BIBentryALTinterwordspacing
S.~Alan, B.~Mike, K.~Kris, and W.~Geof, ``{VR/AR} association white paper:
  Virtual and augmented reality for business.'' [Online]. Available:
  \url{https://artillry.co/wp-content/uploads/2018/04/vrara-whitepaper-enterprise-final-draft-2018.pdf}
\BIBentrySTDinterwordspacing

\bibitem{background2}
\BIBentryALTinterwordspacing
{China Academy of Information and Communications Technology (CAICT)} and
  {Huawei Technologies Co., Ltd}, ``Virtual reality/augmented reality white
  paper.'' [Online]. Available:
  \url{https://www-file.huawei.com/-/media/CORPORATE/PDF/ilab/vr-ar-en.pdf}
\BIBentrySTDinterwordspacing

\bibitem{importance}
Z.~Wang, A.~C. Bovik, and L.~Lu, ``Why is image quality assessment so
  difficult?'' in \emph{2002 IEEE International Conference on Acoustics,
  Speech, and Signal Processing}, vol.~4.\hskip 1em plus 0.5em minus
  0.4em\relax IEEE, 2002, pp. IV--3313.

\bibitem{overview360}
M.~Xu, C.~Li, S.~Zhang, and P.~Le~Callet, ``State-of-the-art in 360 video/image
  processing: Perception, assessment and compression,'' \emph{IEEE Journal of
  Selected Topics in Signal Processing}, 2020.

\bibitem{subjective&objective}
K.~Seshadrinathan, R.~Soundararajan, A.~C. Bovik, and L.~K. Cormack, ``Study of
  subjective and objective quality assessment of video,'' \emph{IEEE
  Transactions on Image Processing}, vol.~19, no.~6, pp. 1427--1441, 2010.

\bibitem{OIQADatabse}
H.~Duan, G.~Zhai, X.~Min, Y.~Zhu, Y.~Fang, and X.~Yang, ``Perceptual quality
  assessment of omnidirectional images,'' in \emph{2018 IEEE International
  Symposium on Circuits and Systems (ISCAS)}.\hskip 1em plus 0.5em minus
  0.4em\relax IEEE, 2018, pp. 1--5.

\bibitem{SOLIDDatabase}
J.~Xu, C.~Lin, W.~Zhou, and Z.~Chen, ``Subjective quality assessment of
  stereoscopic omnidirectional image,'' in \emph{Pacific Rim Conference on
  Multimedia}.\hskip 1em plus 0.5em minus 0.4em\relax Springer, 2018, pp.
  589--599.

\bibitem{SAMPVIQ}
B.~Zhang, J.~Zhao, S.~Yang, Y.~Zhang, J.~Wang, and Z.~Fei, ``Subjective and
  objective quality assessment of panoramic videos in virtual reality
  environments,'' in \emph{2017 IEEE International Conference on Multimedia \&
  Expo Workshops (ICMEW)}.\hskip 1em plus 0.5em minus 0.4em\relax IEEE, 2017,
  pp. 163--168.

\bibitem{exploration}
K.~Ma, Z.~Duanmu, Q.~Wu, Z.~Wang, H.~Yong, H.~Li, and L.~Zhang, ``Waterloo
  exploration database: New challenges for image quality assessment models,''
  \emph{IEEE Transactions on Image Processing}, vol.~26, no.~2, pp. 1004--1016,
  2016.

\bibitem{IQA&VQAsurvey}
U.~Engelke and H.-J. Zepernick, ``Perceptual-based quality metrics for image
  and video services: A survey,'' in \emph{NGI}.\hskip 1em plus 0.5em minus
  0.4em\relax IEEE, 2007.

\bibitem{StatisticalEvaluation}
H.~R. Sheikh, M.~F. Sabir, and A.~C. Bovik, ``A statistical evaluation of
  recent full reference image quality assessment algorithms,'' \emph{IEEE
  Transactions on image processing}, vol.~15, no.~11, pp. 3440--3451, 2006.

\bibitem{WaDIQaM}
S.~Bosse, D.~Maniry, K.-R. M{\"u}ller, T.~Wiegand, and W.~Samek, ``Deep neural
  networks for no-reference and full-reference image quality assessment,''
  \emph{IEEE Transactions on Image Processing}, vol.~27, no.~1, pp. 206--219,
  2017.

\bibitem{RR&NR}
Z.~Wang and A.~C. Bovik, ``Reduced-and no-reference image quality assessment,''
  \emph{IEEE Signal Processing Magazine}, vol.~28, no.~6, pp. 29--40, 2011.

\bibitem{IFC}
H.~R. Sheikh, A.~C. Bovik, and G.~De~Veciana, ``An information fidelity
  criterion for image quality assessment using natural scene statistics,''
  \emph{IEEE Transactions on image processing}, vol.~14, no.~12, pp.
  2117--2128, 2005.

\bibitem{HVS}
D.~J. Granrath, ``The role of human visual models in image processing,''
  \emph{Proceedings of the IEEE}, vol.~69, no.~5, pp. 552--561, 1981.

\bibitem{ssim}
Z.~Wang, A.~C. Bovik, H.~R. Sheikh, E.~P. Simoncelli \emph{et~al.}, ``Image
  quality assessment: from error visibility to structural similarity,''
  \emph{IEEE Transactions on image processing}, vol.~13, no.~4, pp. 600--612,
  2004.

\bibitem{msssim}
Z.~Wang, E.~P. Simoncelli, and A.~C. Bovik, ``Multiscale structural similarity
  for image quality assessment,'' in \emph{The Thrity-Seventh Asilomar
  Conference on Signals, Systems \& Computers, 2003}, vol.~2.\hskip 1em plus
  0.5em minus 0.4em\relax IEEE, 2003, pp. 1398--1402.

\bibitem{iwssim}
Z.~Wang and Q.~Li, ``Information content weighting for perceptual image quality
  assessment,'' \emph{IEEE Transactions on Image Processing}, vol.~20, no.~5,
  pp. 1185--1198, 2010.

\bibitem{cwssim}
M.~P. Sampat, Z.~Wang, S.~Gupta, A.~C. Bovik, and M.~K. Markey, ``Complex
  wavelet structural similarity: A new image similarity index,'' \emph{IEEE
  Transactions on image processing}, vol.~18, no.~11, pp. 2385--2401, 2009.

\bibitem{fsim}
L.~Zhang, L.~Zhang, X.~Mou, and D.~Zhang, ``Fsim: A feature similarity index
  for image quality assessment,'' \emph{IEEE Transactions on Image Processing},
  vol.~20, no.~8, pp. 2378--2386, 2011.

\bibitem{DIIVINE}
A.~K. Moorthy and A.~C. Bovik, ``Blind image quality assessment: From natural
  scene statistics to perceptual quality,'' \emph{IEEE Transactions on Image
  Processing}, vol.~20, no.~12, pp. 3350--3364, 2011.

\bibitem{WPDSE}
T.-J. Liu and K.-H. Liu, ``No-reference image quality assessment by
  wide-perceptual-domain scorer ensemble method,'' \emph{IEEE Transactions on
  Image Processing}, vol.~27, no.~3, pp. 1138--1151, 2017.

\bibitem{NRSharpness}
R.~Ferzli and L.~J. Karam, ``A no-reference objective image sharpness metric
  based on the notion of just noticeable blur (jnb),'' \emph{IEEE Transactions
  on image processing}, vol.~18, no.~4, pp. 717--728, 2009.

\bibitem{NRBlur}
L.~Li, W.~Lin, X.~Wang, G.~Yang, K.~Bahrami, and A.~C. Kot, ``No-reference
  image blur assessment based on discrete orthogonal moments,'' \emph{IEEE
  Transactions on cybernetics}, vol.~46, no.~1, pp. 39--50, 2015.

\bibitem{NRJPEG1}
Z.~Wang, A.~C. Bovik, and B.~L. Evan, ``Blind measurement of blocking artifacts
  in images,'' in \emph{Proceedings 2000 International Conference on Image
  Processing (Cat. No. 00CH37101)}, vol.~3.\hskip 1em plus 0.5em minus
  0.4em\relax IEEE, 2000, pp. 981--984.

\bibitem{NRJPEG2}
Z.~Wang, H.~R. Sheikh, and A.~C. Bovik, ``No-reference perceptual quality
  assessment of jpeg compressed images,'' in \emph{Proceedings. International
  Conference on Image Processing}, vol.~1.\hskip 1em plus 0.5em minus
  0.4em\relax IEEE, 2002, pp. I--I.

\bibitem{ringing}
H.~Liu, N.~Klomp, and I.~Heynderickx, ``A no-reference metric for perceived
  ringing artifacts in images,'' \emph{IEEE Transactions on Circuits and
  Systems for Video Technology}, vol.~20, no.~4, pp. 529--539, 2009.

\bibitem{NIQE}
A.~Mittal, R.~Soundararajan, and A.~C. Bovik, ``Making a `completely blind'
  image quality analyzer,'' \emph{IEEE Signal Processing Letters}, vol.~20,
  no.~3, pp. 209--212, 2012.

\bibitem{BRISQUE}
A.~Mittal, A.~K. Moorthy, and A.~C. Bovik, ``No-reference image quality
  assessment in the spatial domain,'' \emph{IEEE Transactions on image
  processing}, vol.~21, no.~12, pp. 4695--4708, 2012.

\bibitem{BLINDS}
M.~A. Saad, A.~C. Bovik, and C.~Charrier, ``Blind image quality assessment: A
  natural scene statistics approach in the dct domain,'' \emph{IEEE
  Transactions on Image Processing}, vol.~21, no.~8, pp. 3339--3352, 2012.

\bibitem{wu2015blind}
Q.~Wu, H.~Li, F.~Meng, K.~N. Ngan, B.~Luo, C.~Huang, and B.~Zeng, ``Blind image
  quality assessment based on multichannel feature fusion and label transfer,''
  \emph{IEEE Transactions on Circuits and Systems for Video Technology},
  vol.~26, no.~3, pp. 425--440, 2015.

\bibitem{CORNIA}
P.~Ye, J.~Kumar, L.~Kang, and D.~Doermann, ``Unsupervised feature learning
  framework for no-reference image quality assessment,'' in \emph{2012 IEEE
  conference on computer vision and pattern recognition}.\hskip 1em plus 0.5em
  minus 0.4em\relax IEEE, 2012, pp. 1098--1105.

\bibitem{HOSA}
J.~Xu, P.~Ye, Q.~Li, H.~Du, Y.~Liu, and D.~Doermann, ``Blind image quality
  assessment based on high order statistics aggregation,'' \emph{IEEE
  Transactions on Image Processing}, vol.~25, no.~9, pp. 4444--4457, 2016.

\bibitem{wu2017blind}
Q.~Wu, H.~Li, K.~N. Ngan, and K.~Ma, ``Blind image quality assessment using
  local consistency aware retriever and uncertainty aware evaluator,''
  \emph{IEEE Transactions on Circuits and Systems for Video Technology},
  vol.~28, no.~9, pp. 2078--2089, 2017.

\bibitem{NIMA}
H.~Talebi and P.~Milanfar, ``Nima: Neural image assessment,'' \emph{IEEE
  Transactions on Image Processing}, vol.~27, no.~8, pp. 3998--4011, 2018.

\bibitem{DIQA}
J.~Kim, A.-D. Nguyen, and S.~Lee, ``Deep cnn-based blind image quality
  predictor,'' \emph{IEEE Transactions on neural networks and learning
  systems}, vol.~30, no.~1, pp. 11--24, 2018.

\bibitem{CNNIQA}
L.~Kang, P.~Ye, Y.~Li, and D.~Doermann, ``Convolutional neural networks for
  no-reference image quality assessment,'' in \emph{Proceedings of the IEEE
  conference on computer vision and pattern recognition}, 2014, pp. 1733--1740.

\bibitem{MEON}
K.~Ma, W.~Liu, K.~Zhang, Z.~Duanmu, Z.~Wang, and W.~Zuo, ``End-to-end blind
  image quality assessment using deep neural networks,'' \emph{IEEE
  Transactions on Image Processing}, vol.~27, no.~3, pp. 1202--1213, 2017.

\bibitem{DBCNN}
W.~Zhang, K.~Ma, J.~Yan, D.~Deng, and Z.~Wang, ``Blind image quality assessment
  using a deep bilinear convolutional neural network,'' \emph{IEEE Transactions
  on Circuits and Systems for Video Technology}, 2018.

\bibitem{upenik2017performance}
E.~Upenik, M.~Rerabek, and T.~Ebrahimi, ``On the performance of objective
  metrics for omnidirectional visual content,'' in \emph{2017 Ninth
  International Conference on Quality of Multimedia Experience (QoMEX)}.\hskip
  1em plus 0.5em minus 0.4em\relax IEEE, 2017, pp. 1--6.

\bibitem{SPSNR}
M.~Yu, H.~Lakshman, and B.~Girod, ``A framework to evaluate omnidirectional
  video coding schemes,'' in \emph{2015 IEEE International Symposium on Mixed
  and Augmented Reality}.\hskip 1em plus 0.5em minus 0.4em\relax IEEE, 2015,
  pp. 31--36.

\bibitem{WSPSNR}
Y.~Sun, A.~Lu, and L.~Yu, ``Weighted-to-spherically-uniform quality evaluation
  for omnidirectional video,'' \emph{IEEE signal processing letters}, vol.~24,
  no.~9, pp. 1408--1412, 2017.

\bibitem{CPPPSNR}
V.~Zakharchenko, K.~P. Choi, and J.~H. Park, ``Quality metric for spherical
  panoramic video,'' in \emph{Optics and Photonics for Information Processing
  X}, vol. 9970.\hskip 1em plus 0.5em minus 0.4em\relax International Society
  for Optics and Photonics, 2016, p. 99700C.

\bibitem{CPPCNR}
M.~Xu, C.~Li, Z.~Chen, Z.~Wang, and Z.~Guan, ``Assessing visual quality of
  omnidirectional videos,'' \emph{IEEE Transactions on Circuits and Systems for
  Video Technology}, 2018.

\bibitem{WSSSIM}
Y.~Zhou, M.~Yu, H.~Ma, H.~Shao, and G.~Jiang, ``Weighted-to-spherically-uniform
  ssim objective quality evaluation for panoramic video,'' in \emph{2018 14th
  IEEE International Conference on Signal Processing (ICSP)}.\hskip 1em plus
  0.5em minus 0.4em\relax IEEE, 2018, pp. 54--57.

\bibitem{SSSIM}
S.~Chen, Y.~Zhang, Y.~Li, Z.~Chen, and Z.~Wang, ``Spherical structural
  similarity index for objective omnidirectional video quality assessment,'' in
  \emph{2018 IEEE International Conference on Multimedia and Expo
  (ICME)}.\hskip 1em plus 0.5em minus 0.4em\relax IEEE, 2018, pp. 1--6.

\bibitem{ssim360}
\BIBentryALTinterwordspacing
Facebook, ``Quality assessment of 360 video view sessions.'' [Online].
  Available:
  \url{https://engineering.fb.com/video-engineering/quality-assessment-of-360-video-view-sessions/}
\BIBentrySTDinterwordspacing

\bibitem{ling2018no}
S.~Ling, G.~Cheung, and P.~Le~Callet, ``No-reference quality assessment for
  stitched panoramic images using convolutional sparse coding and compound
  feature selection,'' in \emph{2018 IEEE International Conference on
  Multimedia and Expo (ICME)}.\hskip 1em plus 0.5em minus 0.4em\relax IEEE,
  2018, pp. 1--6.

\bibitem{DeepVRIQA1}
H.~G. Kim, H.-t. Lim, and Y.~M. Ro, ``Deep virtual reality image quality
  assessment with human perception guider for omnidirectional image,''
  \emph{IEEE Transactions on Circuits and Systems for Video Technology}, 2019.

\bibitem{DeepVRIQA2}
H.-T. Lim, H.~G. Kim, and Y.~M. Ra, ``Vr iqa net: Deep virtual reality image
  quality assessment using adversarial learning,'' in \emph{2018 IEEE
  International Conference on Acoustics, Speech and Signal Processing
  (ICASSP)}.\hskip 1em plus 0.5em minus 0.4em\relax IEEE, 2018, pp. 6737--6741.

\bibitem{VQA-ODV}
C.~Li, M.~Xu, X.~Du, and Z.~Wang, ``Bridge the gap between vqa and human
  behavior on omnidirectional video: A large-scale dataset and a deep learning
  model,'' pp. 932--940, 2018.

\bibitem{xu2019quality}
J.~Xu, Z.~Luo, W.~Zhou, W.~Zhang, and Z.~Chen, ``Quality assessment of
  stereoscopic 360-degree images from multi-viewports,'' in \emph{2019 Picture
  Coding Symposium}.\hskip 1em plus 0.5em minus 0.4em\relax IEEE, 2019.

\bibitem{VCNN}
C.~Li, M.~Xu, L.~Jiang, S.~Zhang, and X.~Tao, ``Viewport proposal cnn for
  360deg video quality assessment,'' in \emph{Proceedings of the IEEE
  Conference on Computer Vision and Pattern Recognition}, 2019, pp.
  10\,177--10\,186.

\bibitem{mc360iqa1}
W.~Sun, W.~Luo, X.~Min, G.~Zhai, X.~Yang, K.~Gu, and S.~Ma, ``{MC360IQA}: The
  multi-channel cnn for blind 360-degree image quality assessment,'' in
  \emph{2019 IEEE International Symposium on Circuits and Systems
  (ISCAS)}.\hskip 1em plus 0.5em minus 0.4em\relax IEEE, 2019, pp. 1--5.

\bibitem{mc360iqa2}
W.~Sun, X.~Min, G.~Zhai, K.~Gu, H.~Duan, and S.~Ma, ``{MC360IQA}: A
  multi-channel cnn for blind 360-degree image quality assessment,'' \emph{IEEE
  Journal of Selected Topics in Signal Processing}, 2019.

\bibitem{surf}
H.~Bay, T.~Tuytelaars, and L.~Van~Gool, ``Surf: Speeded up robust features,''
  in \emph{European conference on computer vision}.\hskip 1em plus 0.5em minus
  0.4em\relax Springer, 2006, pp. 404--417.

\bibitem{dhp}
M.~Xu, Y.~Song, J.~Wang, M.~Qiao, L.~Huo, and Z.~Wang, ``Predicting head
  movement in panoramic video: A deep reinforcement learning approach,''
  \emph{IEEE Transactions on pattern analysis and machine intelligence}, 2018.

\bibitem{resnet}
K.~He, X.~Zhang, S.~Ren, and J.~Sun, ``Deep residual learning for image
  recognition,'' in \emph{Proceedings of the IEEE conference on computer vision
  and pattern recognition}, 2016, pp. 770--778.

\bibitem{kipf2016semi}
T.~N. Kipf and M.~Welling, ``Semi-supervised classification with graph
  convolutional networks,'' in \emph{International Conference on Learning
  Representations}, 2017.

\bibitem{yan2018spatial}
S.~Yan, Y.~Xiong, and D.~Lin, ``Spatial temporal graph convolutional networks
  for skeleton-based action recognition,'' in \emph{Thirty-Second AAAI
  Conference on Artificial Intelligence}, 2018.

\bibitem{glorot2011deep}
X.~Glorot, A.~Bordes, and Y.~Bengio, ``Deep sparse rectifier neural networks,''
  in \emph{Proceedings of the fourteenth international conference on artificial
  intelligence and statistics}, 2011, pp. 315--323.

\bibitem{wang2018videos}
X.~Wang and A.~Gupta, ``Videos as space-time region graphs,'' in
  \emph{Proceedings of the European Conference on Computer Vision (ECCV)},
  2018, pp. 399--417.

\bibitem{vgg}
K.~Simonyan and A.~Zisserman, ``Very deep convolutional networks for
  large-scale image recognition,'' \emph{arXiv preprint arXiv:1409.1556}, 2014.

\bibitem{pennec2006riemannian}
X.~Pennec, P.~Fillard, and N.~Ayache, ``A riemannian framework for tensor
  computing,'' \emph{International Journal of computer vision}, vol.~66, no.~1,
  pp. 41--66, 2006.

\bibitem{CVIQDDatabase}
W.~Sun, K.~Gu, S.~Ma, W.~Zhu, N.~Liu, and G.~Zhai, ``A large-scale compressed
  360-degree spherical image database: From subjective quality evaluation to
  objective model comparison,'' in \emph{2018 IEEE 20th International Workshop
  on Multimedia Signal Processing (MMSP)}.\hskip 1em plus 0.5em minus
  0.4em\relax IEEE, 2018, pp. 1--6.

\bibitem{imagenet}
J.~Deng, W.~Dong, R.~Socher, L.-J. Li, K.~Li, and L.~Fei-Fei, ``Imagenet: A
  large-scale hierarchical image database,'' in \emph{2009 IEEE conference on
  computer vision and pattern recognition}.\hskip 1em plus 0.5em minus
  0.4em\relax IEEE, 2009, pp. 248--255.

\bibitem{pascal}
M.~Everingham, L.~Van~Gool, C.~K. Williams, J.~Winn, and A.~Zisserman, ``The
  pascal visual object classes (voc) challenge,'' \emph{International journal
  of computer vision}, vol.~88, no.~2, pp. 303--338, 2010.

\bibitem{live}
H.~Sheikh, ``Live image quality assessment database release 2,''
  \emph{http://live. ece. utexas. edu/research/quality}, 2005.

\bibitem{PerformanceMeasure}
V.~Q.~E. Group \emph{et~al.}, ``Final report from the video quality experts
  group on the validation of objective models of video quality assessment,
  phase ii,'' \emph{2003 VQEG}, 2003.

\bibitem{krasula2016accuracy}
L.~Krasula, K.~Fliegel, P.~Le~Callet, and M.~Kl{\'\i}ma, ``On the accuracy of
  objective image and video quality models: New methodology for performance
  evaluation,'' in \emph{2016 Eighth International Conference on Quality of
  Multimedia Experience (QoMEX)}.\hskip 1em plus 0.5em minus 0.4em\relax IEEE,
  2016, pp. 1--6.

\bibitem{aldahdooh2018improved}
A.~Aldahdooh, E.~Masala, O.~Janssens, G.~Van~Wallendael, M.~Barkowsky, and
  P.~Le~Callet, ``Improved performance measures for video quality assessment
  algorithms using training and validation sets,'' \emph{IEEE Transactions on
  Multimedia}, vol.~21, no.~8, pp. 2026--2041, 2018.

\bibitem{DeepQA}
J.~Kim and S.~Lee, ``Deep learning of human visual sensitivity in image quality
  assessment framework,'' in \emph{Proceedings of the IEEE conference on
  computer vision and pattern recognition}, 2017, pp. 1676--1684.

\bibitem{BMPRI}
X.~Min, G.~Zhai, K.~Gu, Y.~Liu, and X.~Yang, ``Blind image quality estimation
  via distortion aggravation,'' \emph{IEEE Transactions on Broadcasting},
  vol.~64, no.~2, pp. 508--517, 2018.

\bibitem{soiqe}
Z.~Chen, J.~Xu, C.~Lin, and W.~Zhou, ``Stereoscopic omnidirectional image
  quality assessment based on predictive coding theory,'' \emph{IEEE Journal of
  Selected Topics in Signal Processing}, 2020.

\bibitem{Scacnn}
L.~Chen, H.~Zhang, J.~Xiao, L.~Nie, J.~Shao, W.~Liu, and T.-S. Chua, ``Sca-cnn:
  Spatial and channel-wise attention in convolutional networks for image
  captioning,'' in \emph{Proceedings of the IEEE conference on computer vision
  and pattern recognition}, 2017, pp. 5659--5667.

\end{thebibliography}

\begin{IEEEbiography}[{\includegraphics[width=1in,height=1.25in,clip,keepaspectratio]{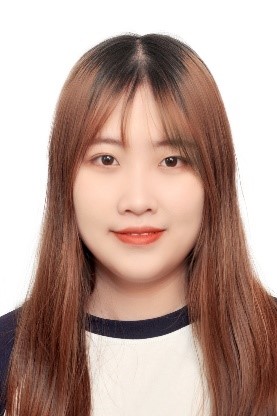}}]{Jiahua Xu}
	received the B.Sc. degree in information engineering from Nanjing University of Aeronautics and Astronautics in 2018. She is currently pursuing the M.Sc. degree with the Department of Electronic Engineer and Information Science, University of Science and Technology of China. She was a visiting student with the Korea Advanced Institute of Science and Technology, Daejeon, South Korea, in 2016. Her current research interests include image/video quality assessment, computer vision and deep learning.
\end{IEEEbiography}

\begin{IEEEbiography}[{\includegraphics[width=1in,height=1.25in,clip,keepaspectratio]{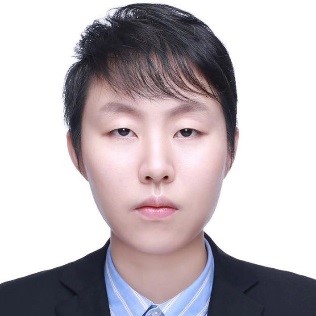}}]{Wei Zhou}
	(S'19) is a Ph.D. candidate with the Department of Electronic Engineer and Information Science at University of Science and Technology of China, and was a visiting student in the National Institute of Informatics, Tokyo in 2017.
\end{IEEEbiography}

\begin{IEEEbiography}[{\includegraphics[width=1in,height=1.25in,clip,keepaspectratio]{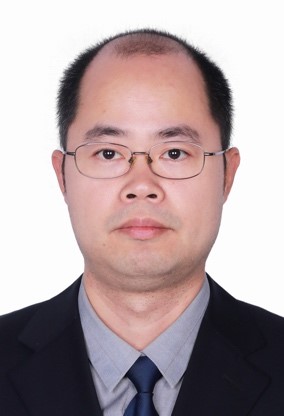}}]{Zhibo Chen}
	(M'01-SM'11) received the B. Sc., and Ph.D. degree from Department of Electrical Engineering Tsinghua University in 1998 and 2003, respectively. He is now a professor in University of Science and Technology of China. His research interests include image and video compression, visual quality of experience assessment, immersive media computing and intelligent media computing. He has more than 100 publications and more than 50 granted EU and US patent applications. He is IEEE senior member, member of IEEE Visual Signal Processing and Communications Committee, and member of IEEE Multimedia System and Applications Committee. He was TPC chair of IEEE PCS 2019 and organization committee member of ICIP 2017 and ICME 2013, served as TPC member in IEEE ISCAS and IEEE VCIP.
\end{IEEEbiography}


%\end{IEEEbiography}
%% insert where needed to balance the two columns on the last page with
%%\newpage


%\vfill


\end{document}